\documentclass[fleqn,twocolumn,amsmath,amssymb,longbibliography]{revtex4-1}

\usepackage{xifthen}% provides \isempty test

\newcommand{\refAppendix}[6]{#1
  \ifthenelse{\isempty{#2}}%
    {}% if #2 is empty
    {\protect\cite{#2}}% if #1 is not empty
    #3\protect\ref{#4}#5#6\xspace
}

\usepackage{forloop,ifthen} % for loops and if statements
\usepackage{multirow} % merge rows and columns
\usepackage{ esint }
\usepackage{amsmath} % need for subequations

    \usepackage{graphicx}
    \usepackage{grffile}
    \usepackage{transparent}
    \usepackage{fancybox}  
    \usepackage{color}
    \usepackage{dcolumn}
    \usepackage{mathtools} % needed for dcases
    \usepackage{bm}
    \usepackage{xspace}
     \usepackage[colorlinks=true,     linkcolor={red!70!black},
     citecolor={green!50!black}, urlcolor={blue!50!black}, pdfborder={0 0 0}]{hyperref}

    \usepackage{calc}

    \newcommand{\CO}[1]{}%SuppressText
    \newcommand{\emphLabel}[1]{\textbf{({#1})}}

\usepackage[usenames,dvipsnames]{xcolor}
\usepackage{ amssymb }

\newcommand{\schr}{Schr\"odinger\xspace}
\newcommand{\ps}{phase space\xspace}
\newcommand{\gs}{comb-state\xspace}
\newcommand{\gss}{comb-states\xspace}

\newcommand{\suppress}[1]{}%SuppressText

\newcommand{\emphCaption}[1]{{\bf{#1}}}

\begin{document}

\title{On the Formation of Lines in Quantum Phase Space}

\author{Ole Steuernagel${}^{1}$, Popo Yang$^{2}$, and Ray-Kuang Lee$^{2, 3, 4, 5}$}

\affiliation{${}^{1}${Department of Physics,~Astronomy~and~Mathematics,~University~of~Hertfordshire,~Hatfield, AL10 9AB, UK}\\
${}^{2}${Institute of Photonics Technologies, National Tsing Hua University, Hsinchu 30013, Taiwan}\\
$^{3}${Department of Physics, National Tsing Hua University, Hsinchu 30013, Taiwan}\\
$^{4}${Physics Division, National Center for Theoretical Sciences, Taipei 10617, Taiwan}\\
$^{5}${Center for Quantum Technology, Hsinchu 30013, Taiwan}}

\date{\today}

\begin{abstract}
  We theoretically study the formation of lines in \ps using Wigner's distribution $W.$ In trapped quantum systems
  such lines form generically, crisscrossing \ps and they can have astonishing extent, reaching
  across the entire state.  In classical systems this does not happen. We show that the formation of
  such straight line patterns is due to the formation of `randomized \gss'. We establish their
  stability to perturbations, and that they are tied to coherences in configuration space. We
  additionally identify generic higher-order `eye' patterns in \ps which occur less often since they
  arise from more specific symmetric \gss; we show that the perturbation of eye patterns through
  their randomization tends to deform them into lines. Lines in \ps should give rise to large probability
  peaks in measurements.
\end{abstract}

\maketitle

\section{Introduction\label{sec:Introduction}}

Quantum waves frequently form long lines in \ps. This has not been reported
before~\cite{Korsch_PD81,Dragoman_OC97,TorresVega_PRA98,Zurek_NAT01,Gao_OC14,Martins_AHEP20}, and is
astonishing when viewed from the perspective of classical \ps densities.

To study \ps behaviour we map the quantum waves~$\psi$ onto their associated Wigner distribution,
$W$~\cite{Wigner_PR32}. While time evolves, $W$ forms lines and does so repeatedly. The lines
crisscross~$W$, often in such a way that they reach across the entire distribution. This trend, to
form lines in \ps, is enhanced by attractive and suppressed by repulsive nonlinear interactions
of~$\psi$.

We note that such straight lines should create significant peaks detectable in (rotated quadrature)
measurements, as used in quantum~\cite{Hofheinz_NAT09} or atom optical~\cite{Kurtsiefer_NAT97}
experi\-ments measuring projections of~$W$.

Formally, we study one-dimensional single-particle quantum waves~$\psi(x,t)$, in position $x$ and time $t$,
whose evolution obeys linear or nonlinear Schr\"odinger equations (NLSEs)~\footnote{We use a unit-free
  description, setting $\hbar=1$ and particle mass $M=1$. For details see, e.g.,
  Ref.~\cite{Oliva_Shear_19}. Eq.~(\ref{eq:_NLSE}) is norm conserving~\protect{\cite{Tao_BAMS09}}.} of the form
\begin{eqnarray}
   \label{eq:_NLSE}
  i \frac{\partial\psi}{\partial t}%\partial_t\psi
  =-\frac{1}{2} \frac{\partial^2 \psi}{\partial x^2} %\partial^2_x\psi
  + V(x) \psi - \gamma(t) |\psi|^\epsilon \psi \; .
\end{eqnarray}
We always assume either the conservative potential~$V(x)$ to be trapping, or the nonlinear (energy
conserving) interactions to be self-attracting ($\gamma>0$).

Such attractive nonlinear interactions describe multi-particle or field phenomena which
can lead to the formation, stabilization and interaction of pulses in
plasmas~\cite{Zabusky_Kruskal_PRL65}, nonlinear optics~\cite{Kivshar_Book03} or dilute ultracold
clouds of atoms, and the generation of rogue waves~\cite{SotoCrespo_PRL16}, tidal bores, dam break
scenarios~\cite{Marcucci_NC19} and many other nonlinear wave phenomena~\footnote{The associated
  energy expression is~\protect{\cite{Tao_BAMS09}}
\begin{eqnarray}
  \label{eq:_NLSE_Energy}
  {\cal H} = \int dx \left[ \frac{1}{2} \left| \frac{\partial\psi}{\partial x}\right|^2 +
  V |\psi^2| -\frac{2 \gamma}{\epsilon+2} |\psi(x,t)|^{\epsilon+2} \right] ,
\end{eqnarray}
our description is unit-free }. For order $\epsilon=2$~\footnote{We investigate NLSEs with
$\epsilon$ varying from 0.5 to 3.5, for $\epsilon=2$ this is the time-dependent Gross–Pitaevskii
equation, for large nonlinearities blowup instabilities can occur~\protect{\cite{Tao_BAMS09}}.},
Eq.~(\ref{eq:_NLSE}) is also known as the Gross-Pitaevskii equation.

Few analytical solutions for NLSEs~(\ref{eq:_NLSE}) are known and generally little is established
about the generic behaviour of solutions for arbitrary initial states and in the presence of
external potentials. We numerically investigate their \ps behaviour, showing that they
often form straight lines crisscrossing \ps and also `eye' patterns for a large variety of
different scenarios, different initial states, different confining potentials and different classes
of NLSEs of varying order~$\epsilon+1$ and strength~$\gamma$ of their nonlinearity.

In this work, after we remind ourselves of the behaviour of classical systems, in the next
paragraph, we introduce Wigner's distribution in Section~\ref{sec:WignerDist}, then we will
concentrate on the dynamics of the linear \schr equation of quantum mechanics for a trapped system
in Section~\ref{sec:TrappedLSE}. In Section~\ref{sec:TheoryRandomGridStates}, we show that the
formation of (positive) straight lines in \ps is due to the formation of randomized \gss, whereas
eye patterns are due to more symmetrical \gss with locally concave or convex arrangements of the
weights of their peaks. Finally, we consider nonlinear systems without trapping potential in
Section~\ref{sec:FreeNLSE} followed by nonlinear systems with trapping potential in
Section~\ref{sec:TrappedNLSE}.

%\subsection
\paragraph*{Classical systems: \label{subsec:Classical}}
spread-out states subjected to con\-ser\-va\-tive hamiltonian time evolution in classical \ps, typically,
form delicate folded patterns on ever smaller scales as the hamiltonian flow stretches and folds their
(initially concentrated, but non-singular) distributions. Similar whorl patterns can also form
in the quantum case~\cite{Korsch_PD81}, at least temporarily, but they are limi\-ted by a minimum
scale first identified by Zurek~\cite{Zurek_NAT01,Oliva_Shear_19}.

Lines in classical \ps can arise for free particles with distributions initially spatially
concentrated as their nonzero momentum spread over time induces an unlimited affine shear in
\ps~\cite{Kurtsiefer_NAT97}. Systems \emph{isomorphic} to free particles, namely, when $V(x)$ forms
linear ramps or harmonic traps induce purely classical
transport~\cite{Oliva_PhysA17,Ole_FreeHOSC_14}. This can result in displacements, rotations and
shearing but does not at all change $W$'s interference patterns in phase
space~\cite{Oliva_PhysA17,Ole_FreeHOSC_14}. Being in this sense tri\-vi\-al we will not discuss such
cases any further.

% \onecolumngrid  
\begin{widetext}
%\vspace{\columnsep}
\begin{figure}[t]
   \hspace{-0.2cm}
   \begin{minipage}[h]{2\columnwidth}
 %%  "b" to have captions on the same line
\includegraphics[height=0.217\columnwidth]{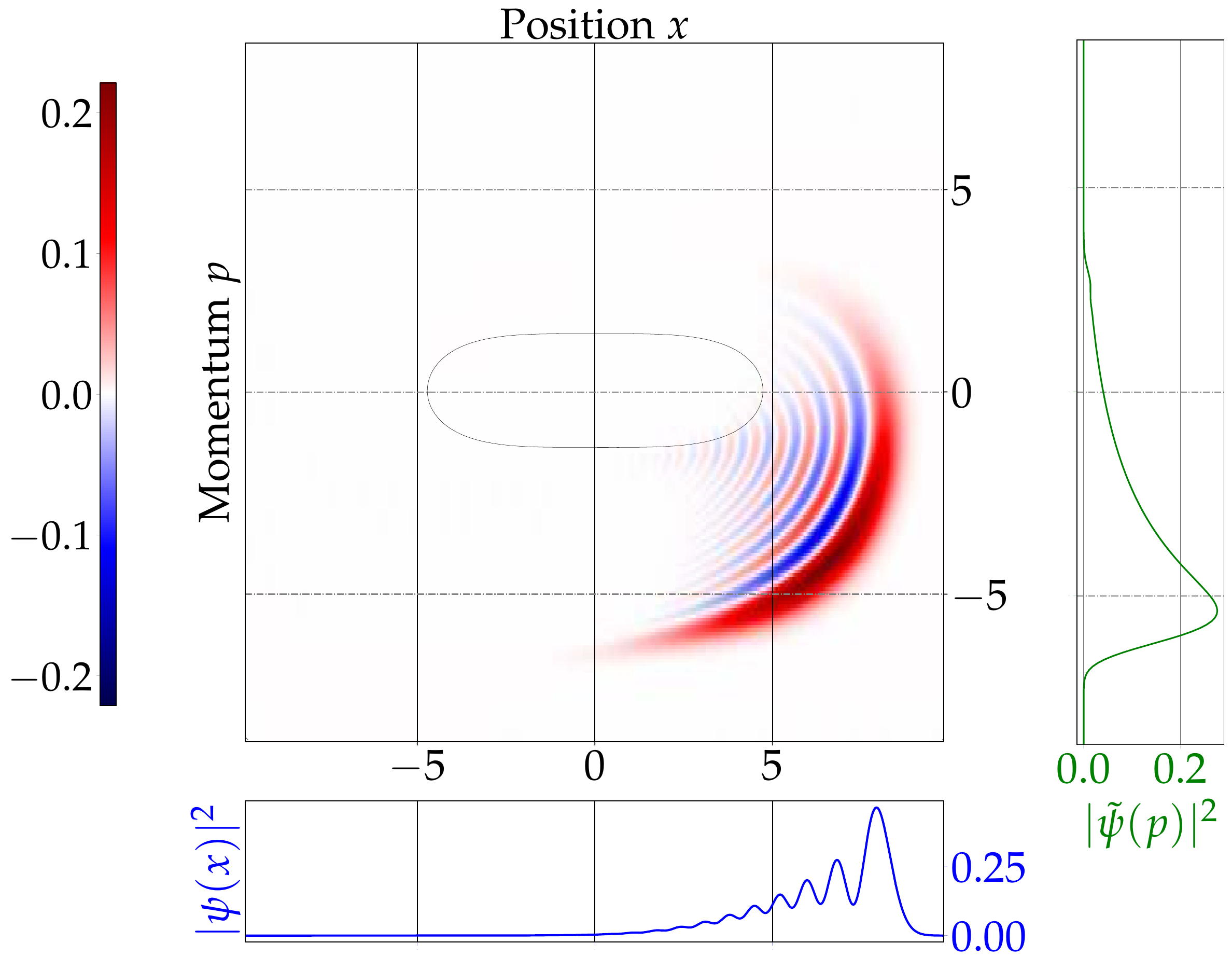}
\includegraphics[height=0.217\columnwidth]{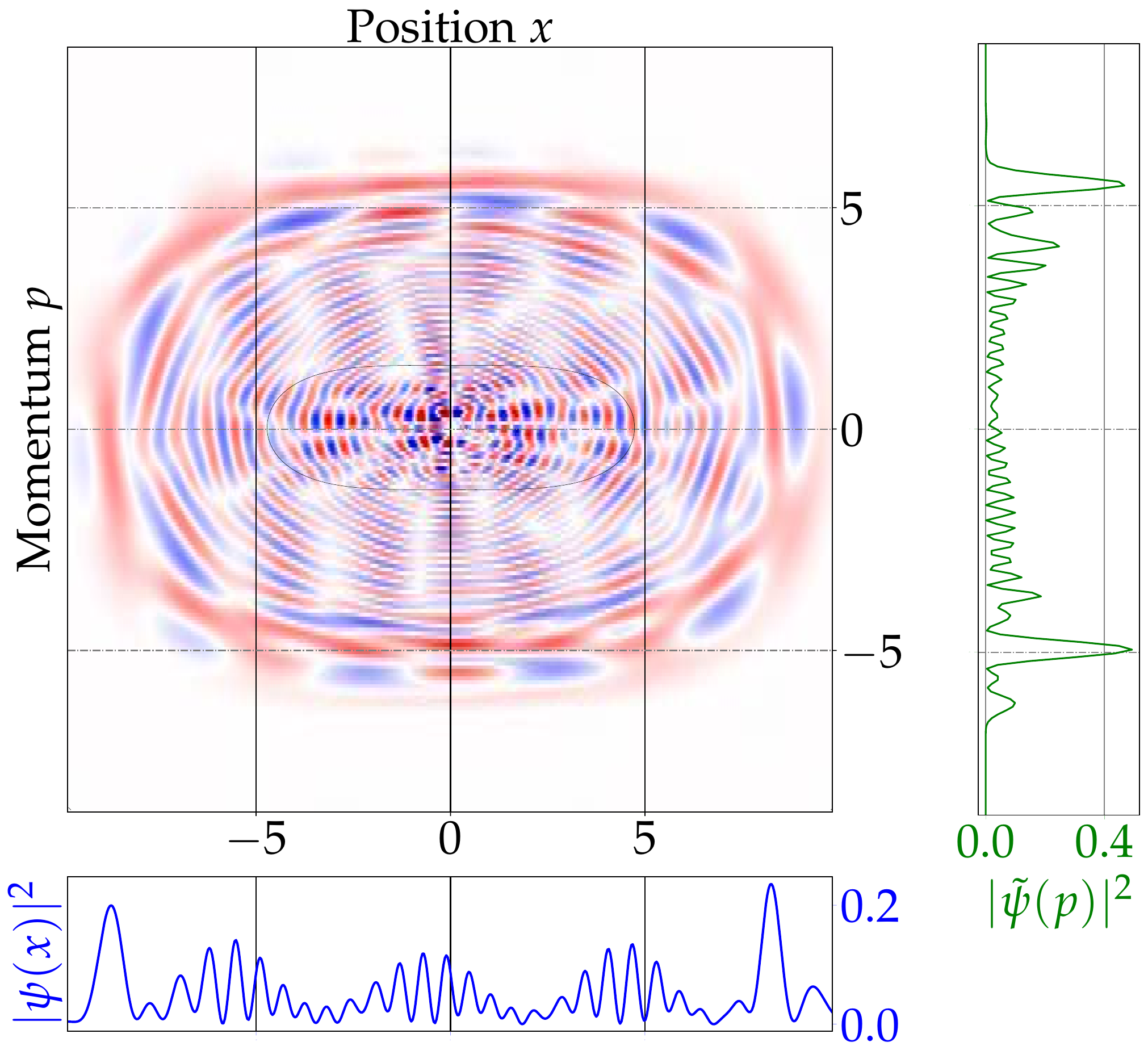}
\includegraphics[height=0.217\columnwidth]{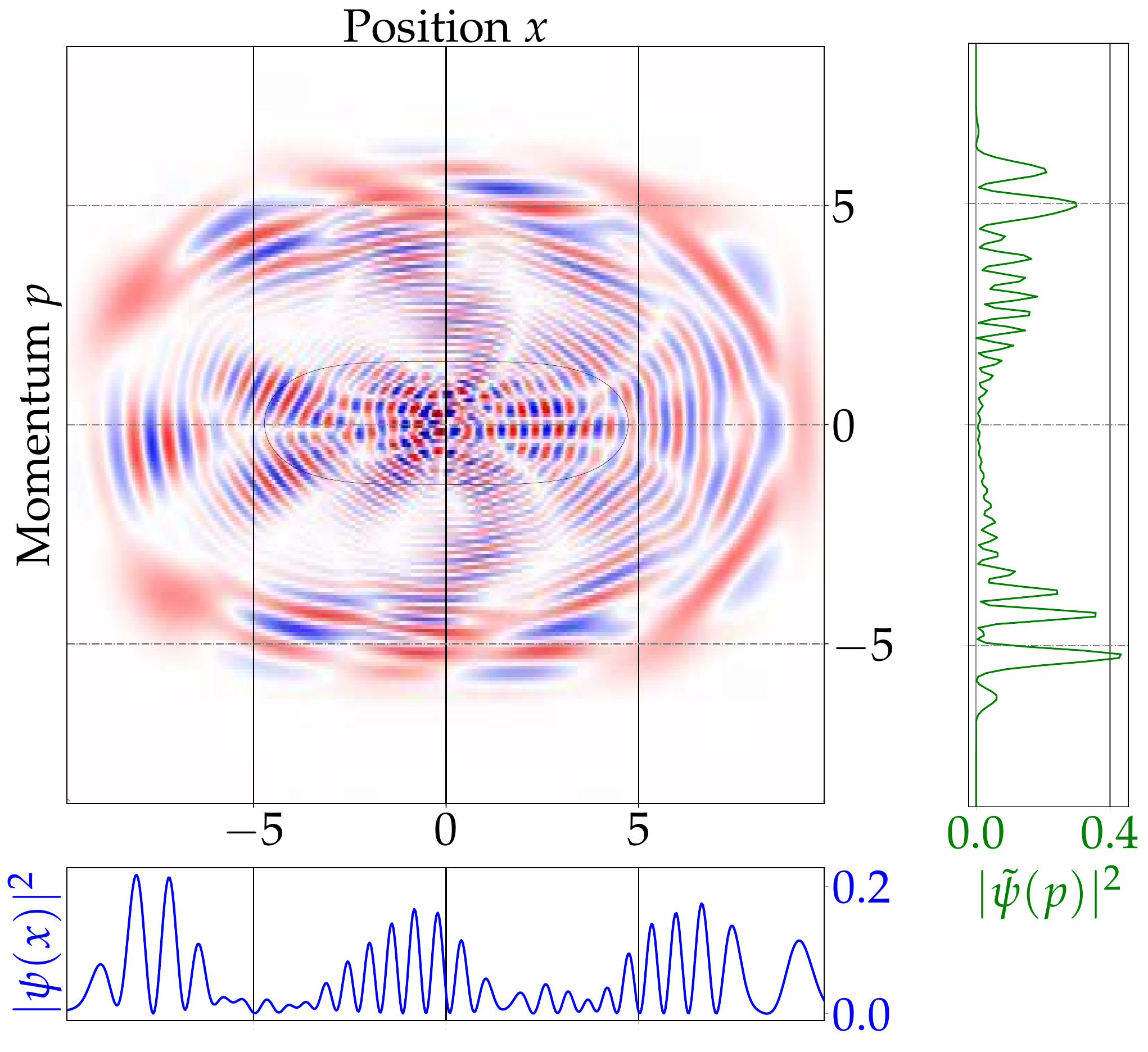}
\includegraphics[height=0.217\columnwidth]{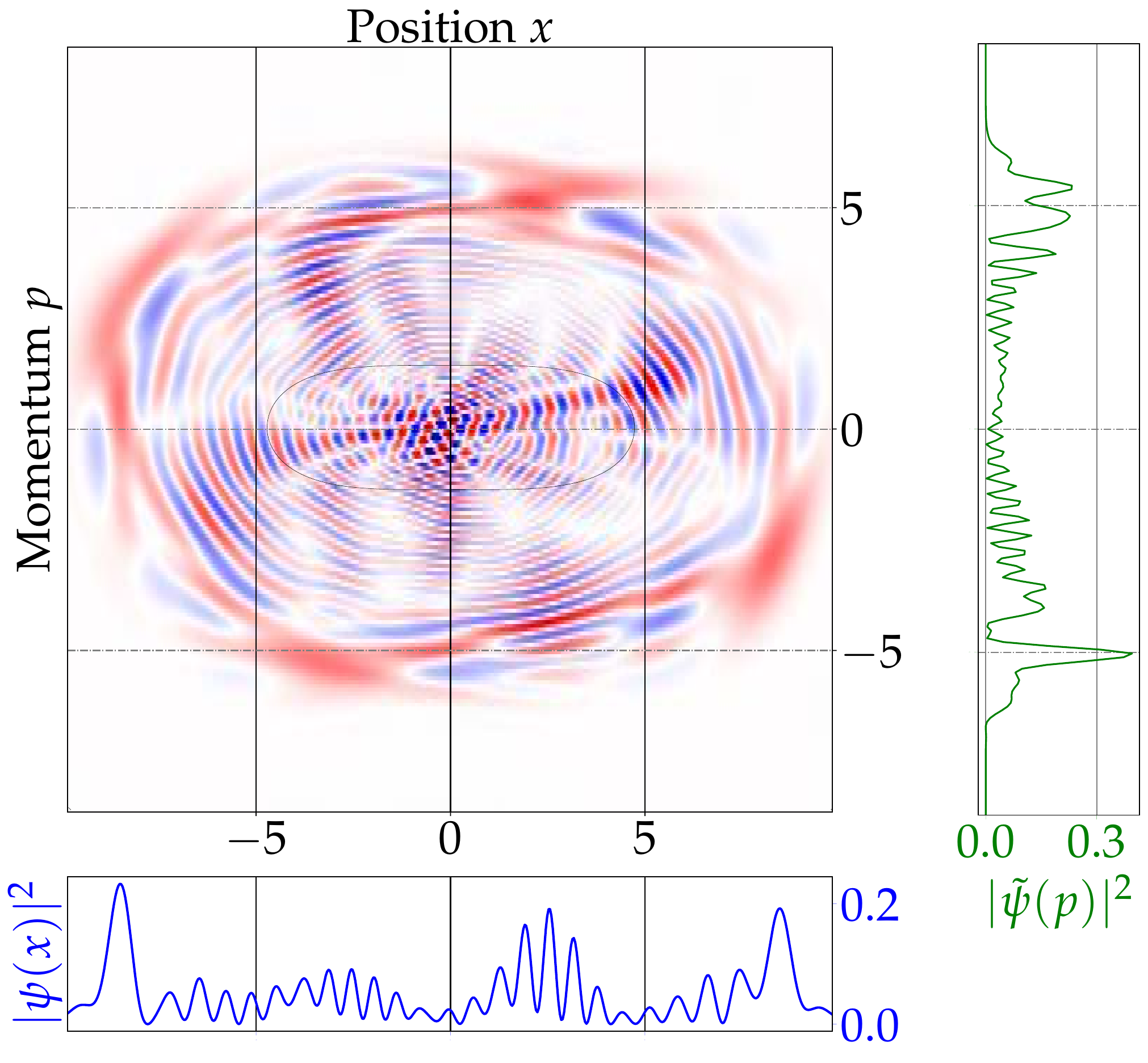}
     \put(-434,95){\rotatebox{0}{\emphLabel{A} $t=10$}}
     \put(-324,95){\rotatebox{0}{\emphLabel{B} $t=77$}}
     \put(-212,95){\rotatebox{0}{\emphLabel{C} $t=110$}}
     \put(-102,95){\rotatebox{0}{\emphLabel{D} $t=192$}}
\\
\includegraphics[height=0.216\columnwidth]{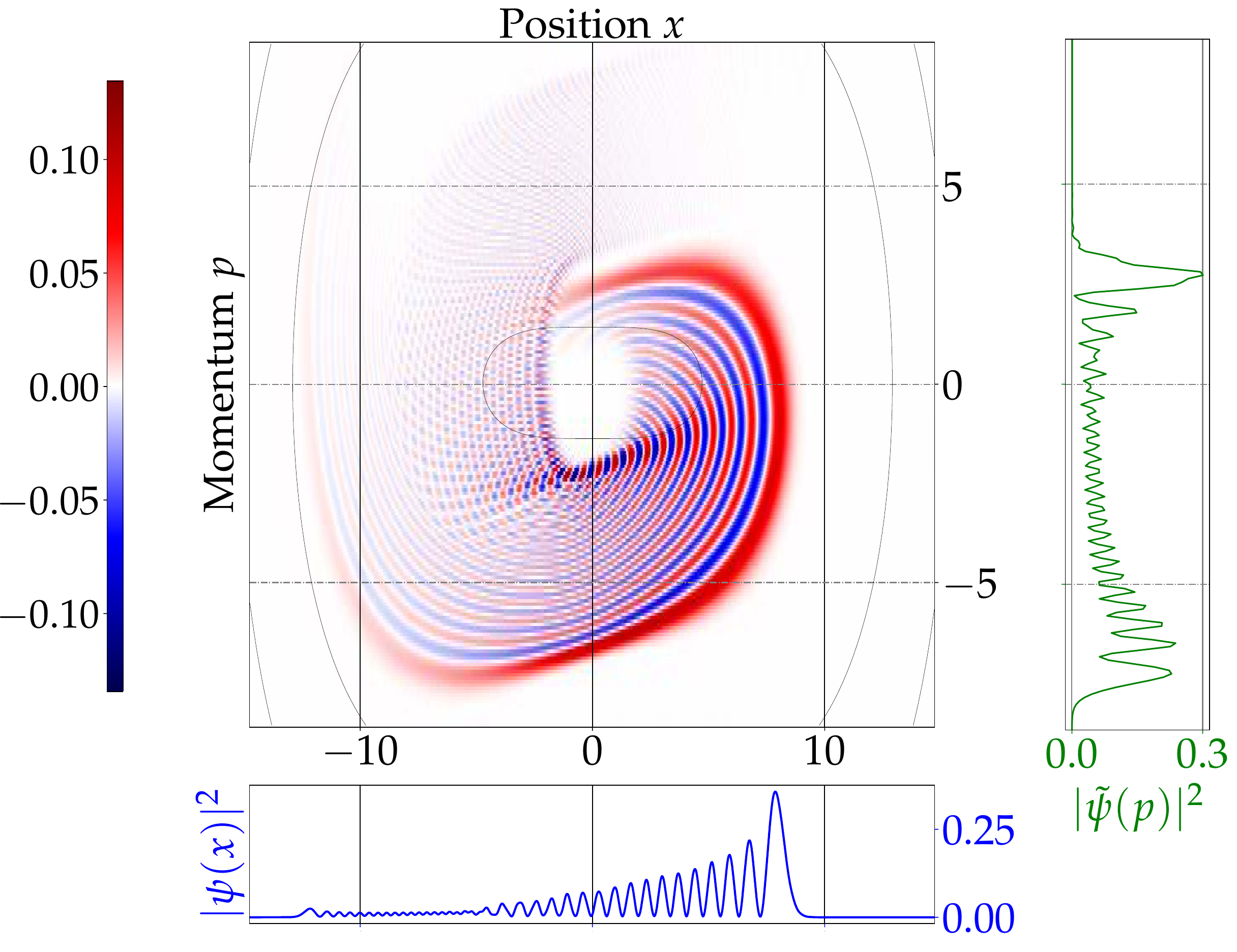}
\includegraphics[height=0.216\columnwidth]{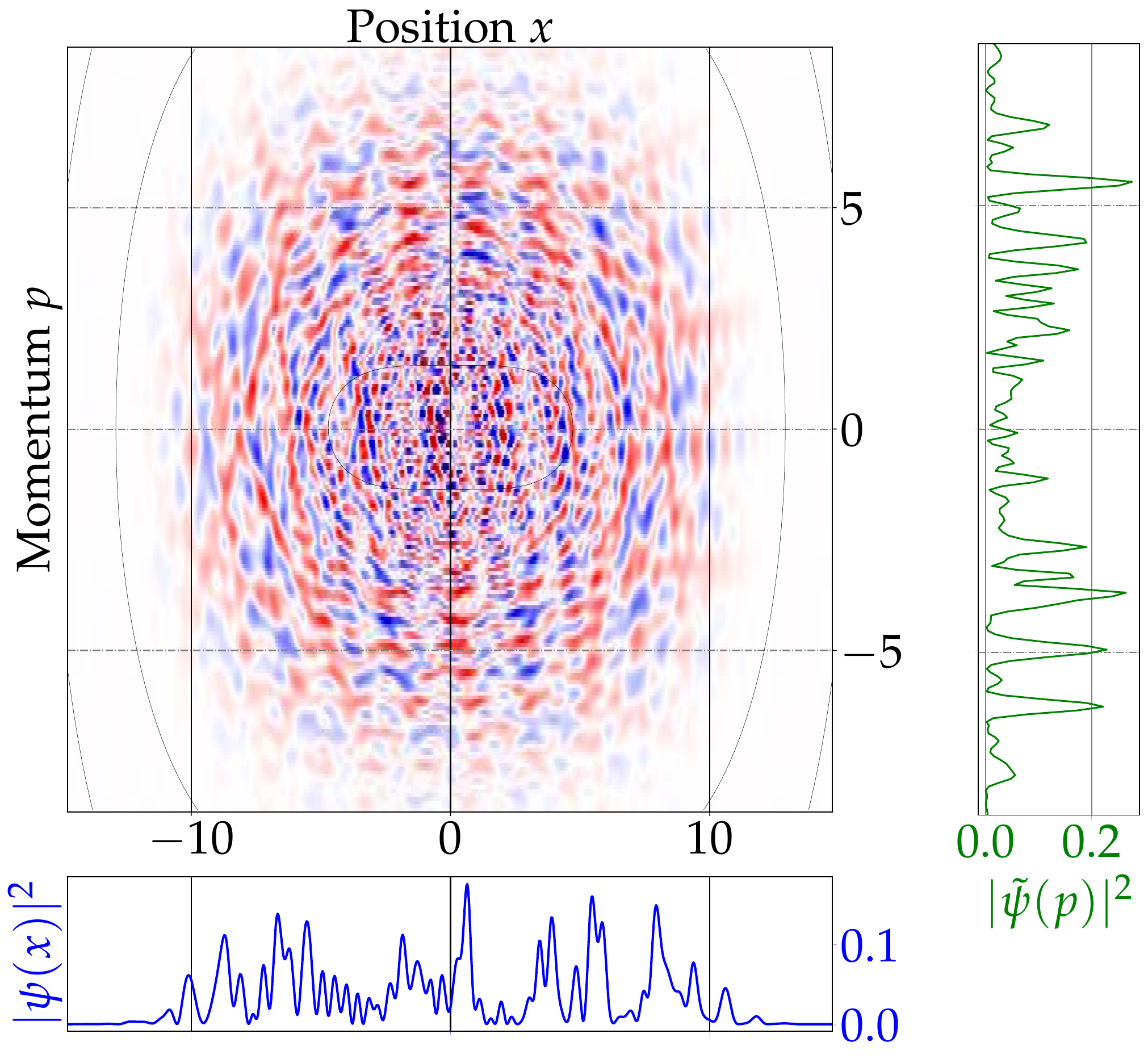}
\includegraphics[height=0.216\columnwidth]{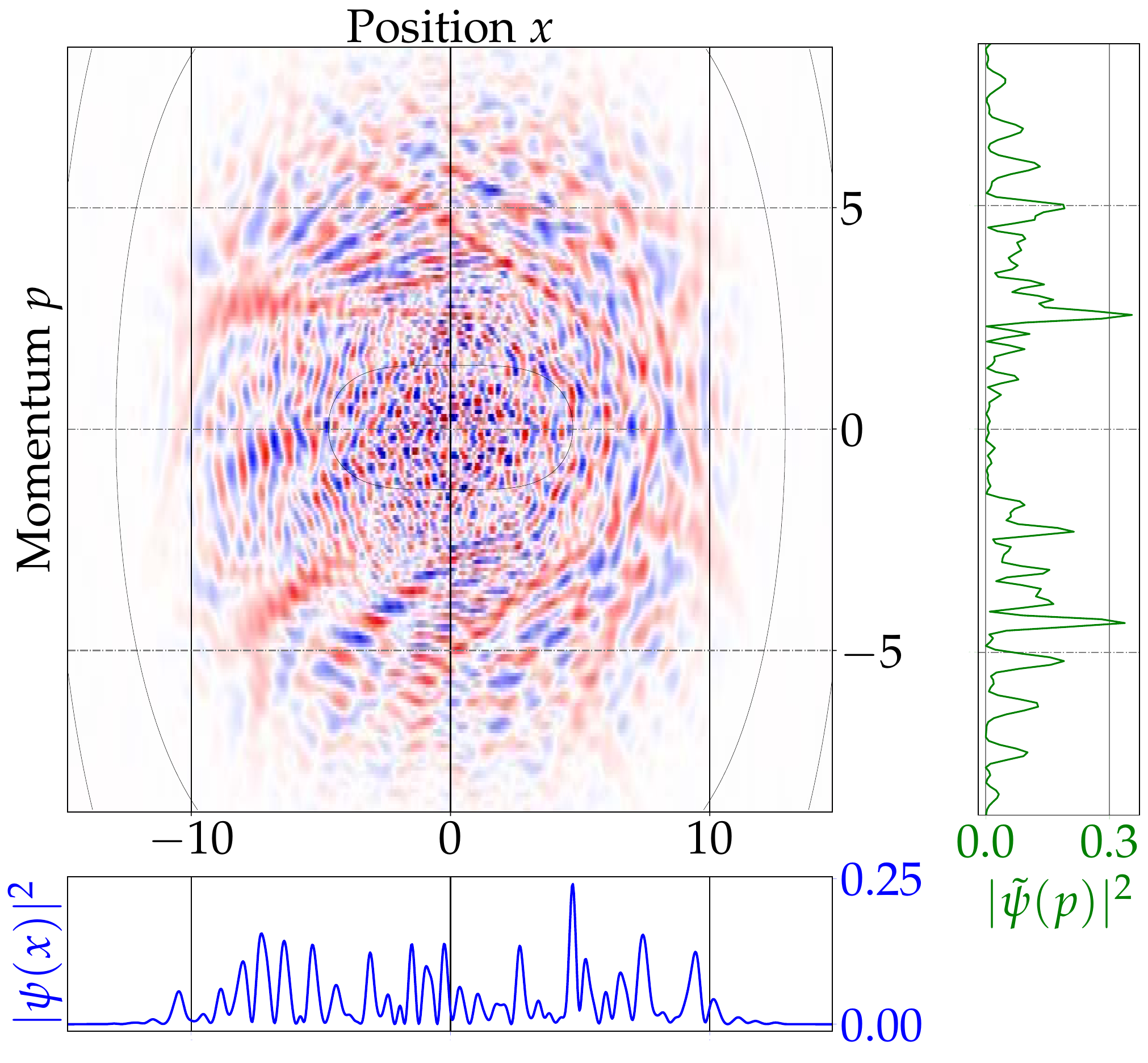}
\includegraphics[height=0.216\columnwidth]{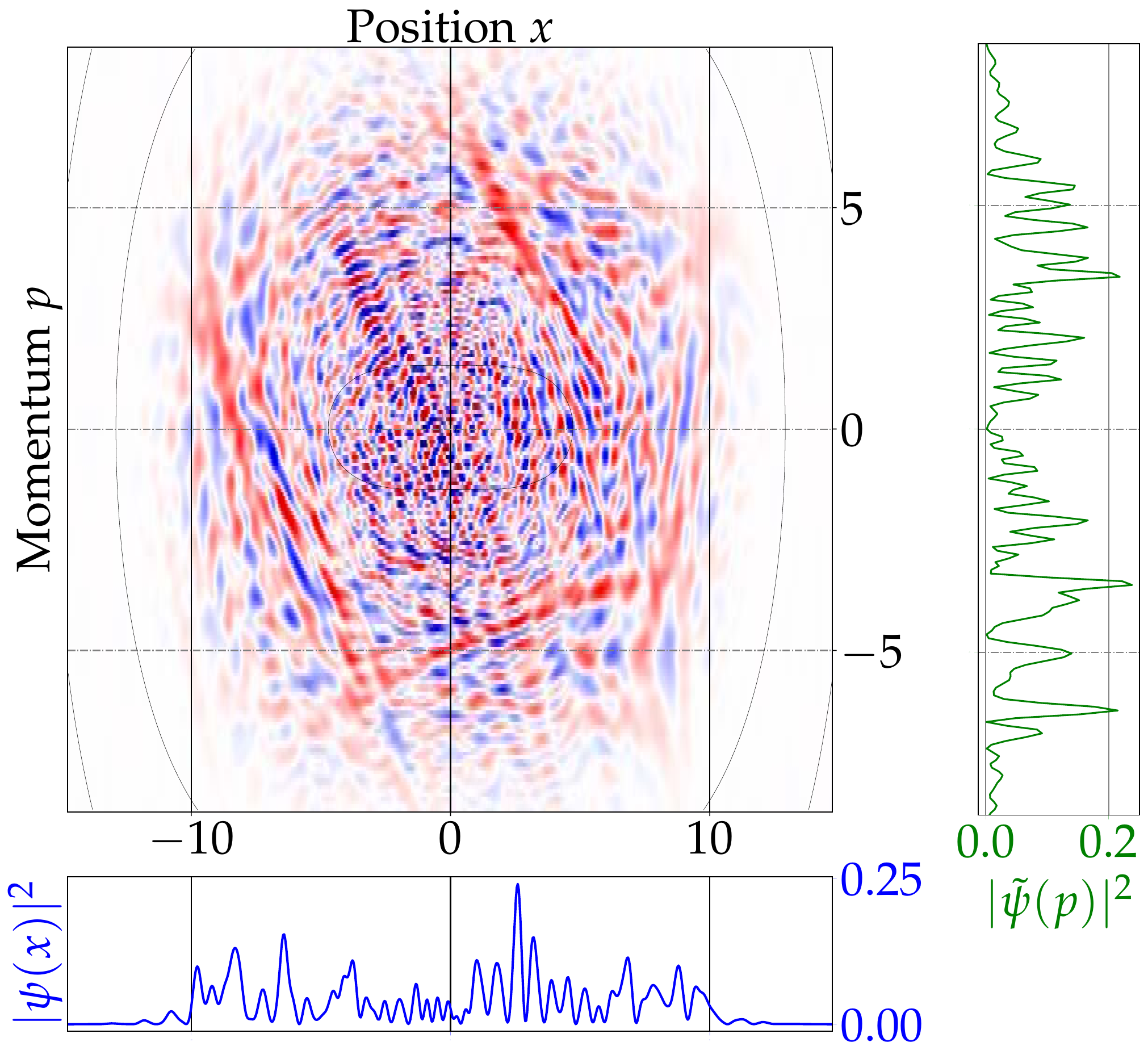}
     \put(-434,95){\rotatebox{0}{\emphLabel{a} $t=10$}}
     \put(-324,95){\rotatebox{0}{\emphLabel{b} $t=77$}}
     \put(-212,95){\rotatebox{0}{\emphLabel{c} $t=110$}}
     \put(-102,95){\rotatebox{0}{\emphLabel{d} $t=192$}}
     \caption{\emphCaption{The Wigner distribution evolved in a linear system with potential
         $V(x)=x^4/500$}, has a recurrence time~\cite{Averbukh_PLA89,Robinett_PR04}
       $T_r \approx 750$ at which the evolved state roughly reforms~\cite{Oliva_Shear_19}. For the
       top row, \emphLabel{A}-\emphLabel{D}, the initial state
       $\psi_0= (2/\pi)^{1/4} \exp[-(x-9)^2]$ is used; for the bottom row,
       \emphLabel{a}-\emphLabel{d}, the initial state is more squeezed in
       $p$:~$\psi_0= (2/9\pi)^{1/4} \exp[-(x-9)^2/9]$. For short times \emphLabel{A} and
       \emphLabel{a} the state forms a fringed crescent. At greater times the distribution covers
       its energy corridor in \ps and around $t=77$, \emphLabel{B} and \emphLabel{b},
       straight lines first appear. Such straight lines tend to have larger extent in cases where
       the state covers a larger area in \ps, \emphLabel{c} versus \emphLabel{C}.  At time
       $t = 192 \approx T_r /4$, \emphLabel{D} and \emphLabel{d}, fractional revivals of the initial
       state with approximately fourfold symmetry form~\cite{Oliva_Shear_19,Averbukh_PLA89,Robinett_PR04}.}
     \label{fig:Linear_x4_0.002_gauss9}
   \end{minipage}
\end{figure}
\end{widetext}

For systems confined by a non-harmonic trapping potential, lines do not form in classical phase
space unless one starts out with special initial states (back-propagated line states). Such lines
then form once but not again.

In the quantum case we find the dynamics creates lines in many different scenarios, in view of the
classical behaviours this is puzzling and demands an explanation. We will show that these lines in
\ps are created by the coherences of randomized \gss.  More symmetrical \gss can create ringed
`eye' patterns, upon perturbation such eye patterns morph into lines.

\section{Wigner's distribution \label{sec:WignerDist}}

Here, we do not assume periodic boundary conditions thus avoiding quantization of momentum into discrete
momentum modes~\cite{Infeld_PRL81,Trillo_OL91}.

To study \ps behaviour for Eq.~(\ref{eq:_NLSE}) we determine~$\psi$'s Wigner
distribution~$W$~\cite{Hillery_PR84} associated with pure states $\psi(x,t)$~\footnote{We use a
  unit-free description, setting $\hbar=1$ and particle mass $M=1$. For details see, e.g.,
  Ref.~\cite{Oliva_Shear_19}.}, namely

\begin{flalign}
  \label{eq:W}
  W(x,p,t) = \frac{1}{\pi} \hspace{-0.065cm} \int_{-\infty}^{\infty} & \hspace{-0.2cm} dy
  \; \psi(x+y,t) \psi^*(x-y,t)  e^{-2i p y}. &
\end{flalign}
$W$ is a function of $x,$ $t$ and momentum~$p$ and known to fully represent all information
contained in $\psi$. By construction $W$ is nonlocal (through~$y$) and normalized:
$\int dp \int dx \;W(x,p,t)=1$. Unlike~$\psi$, $W$ is always real-valued but features nega\-tive
regions~\cite{Wigner_PR32} and is thus considered a distribution featuring
`quasi-probabilities'~\cite{Zachos_book_05}.

Here, we consider 1D problems and always assume wave functions to be normalized
$\int |\psi(x,t)|^2 dx =1$.

The projections of $W$ yield the densities in
position~$P(x,t) = |\psi(x,t)|^2 = \int dp \; W(x,p,t)$ and in
momentum~$\tilde P(p,t) = |\tilde \psi(p,t)|^2 = \int dx \; W(x,p,t)$, respectively. Thus,
long straight lines, reaching across entire distributions, can only form when they have
\emph{positive} values.

\section{%Line formation in
  Trapped Linear Systems \label{sec:TrappedLSE}}

\begin{widetext}  
\begin{figure}[t]
   \hspace{-0.2cm}
   \begin{minipage}[h]{2\columnwidth}
\includegraphics[height=0.215\columnwidth]{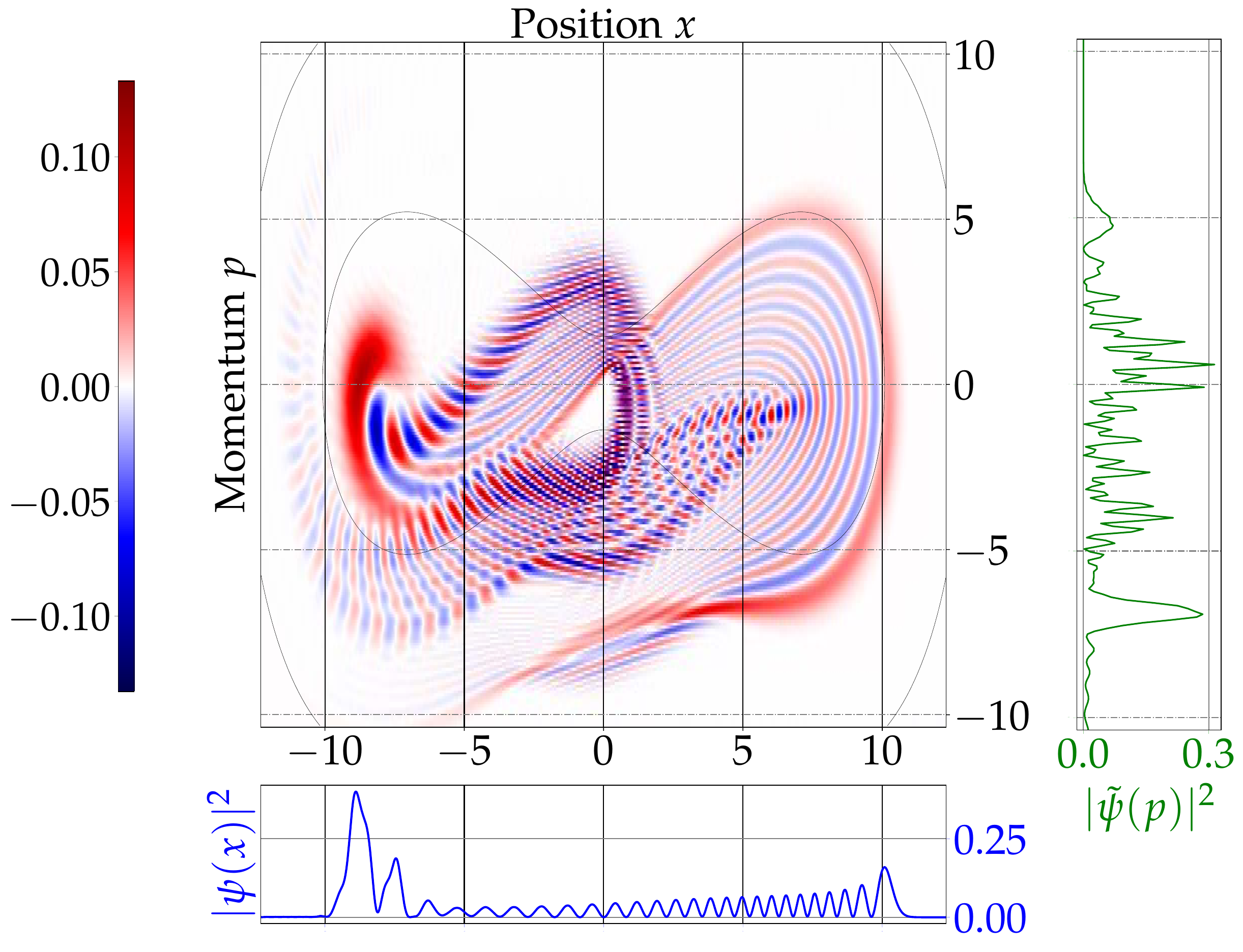} \includegraphics[height=0.215\columnwidth]{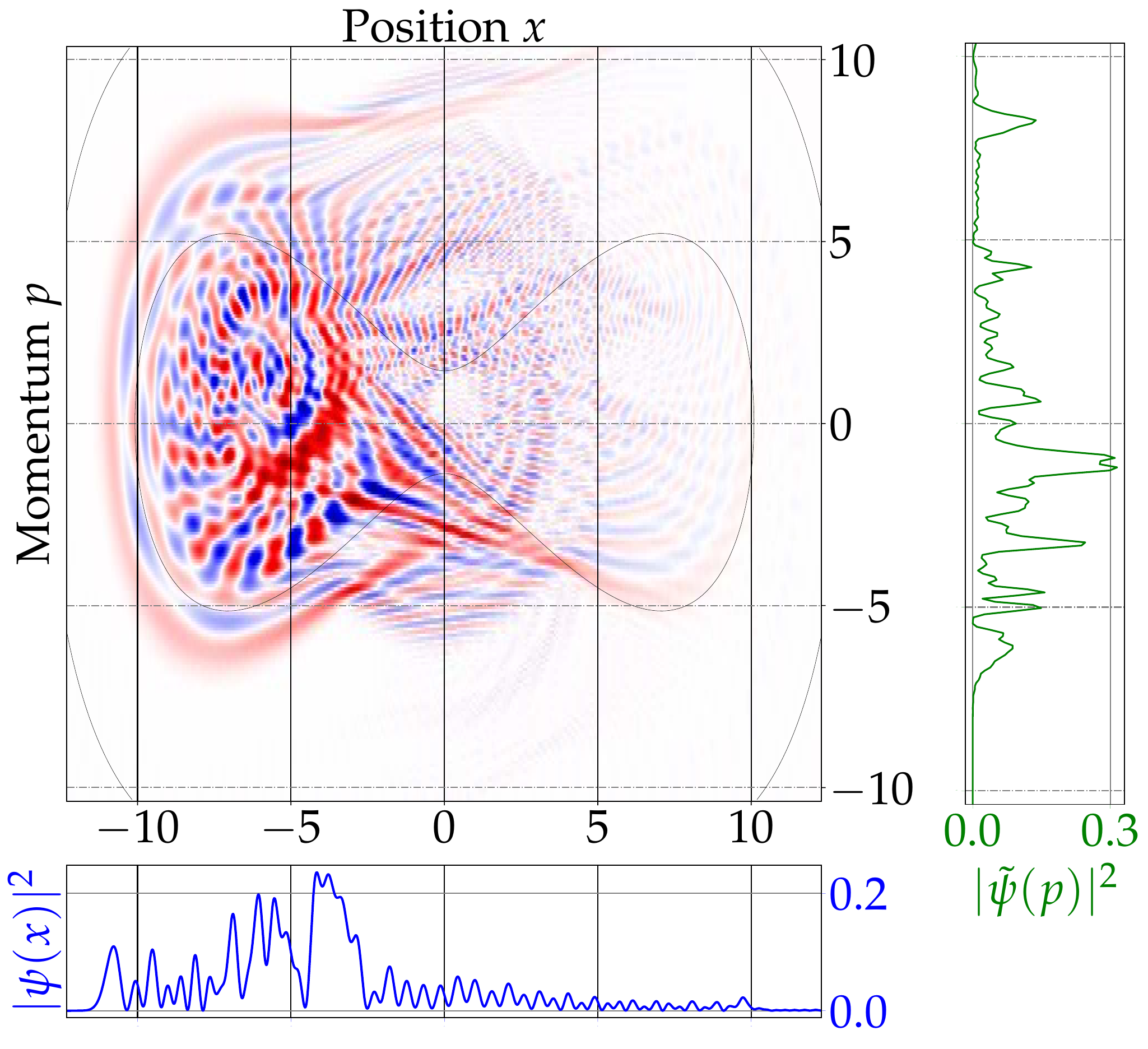} \includegraphics[height=0.215\columnwidth]{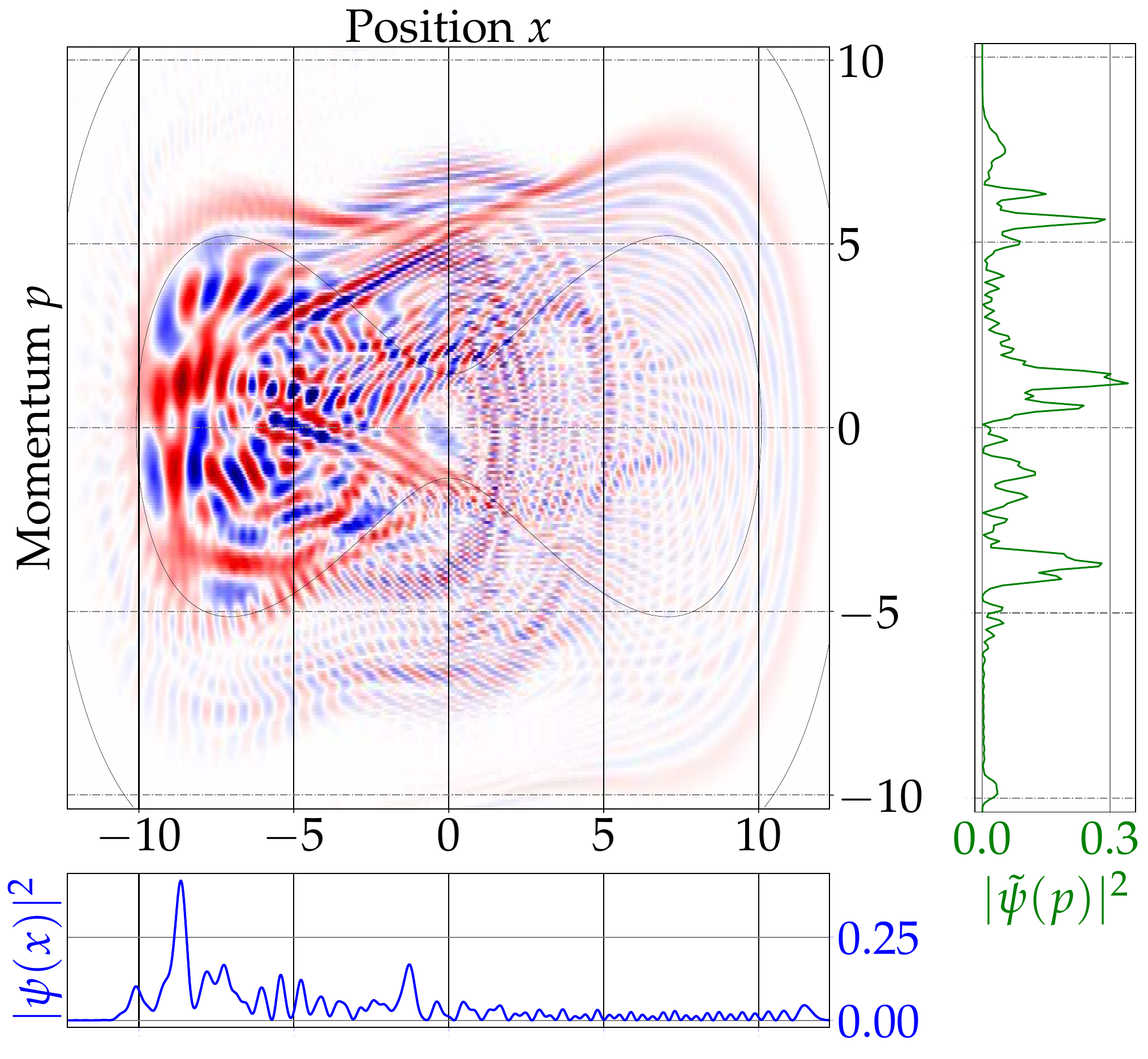}
\includegraphics[height=0.215\columnwidth]{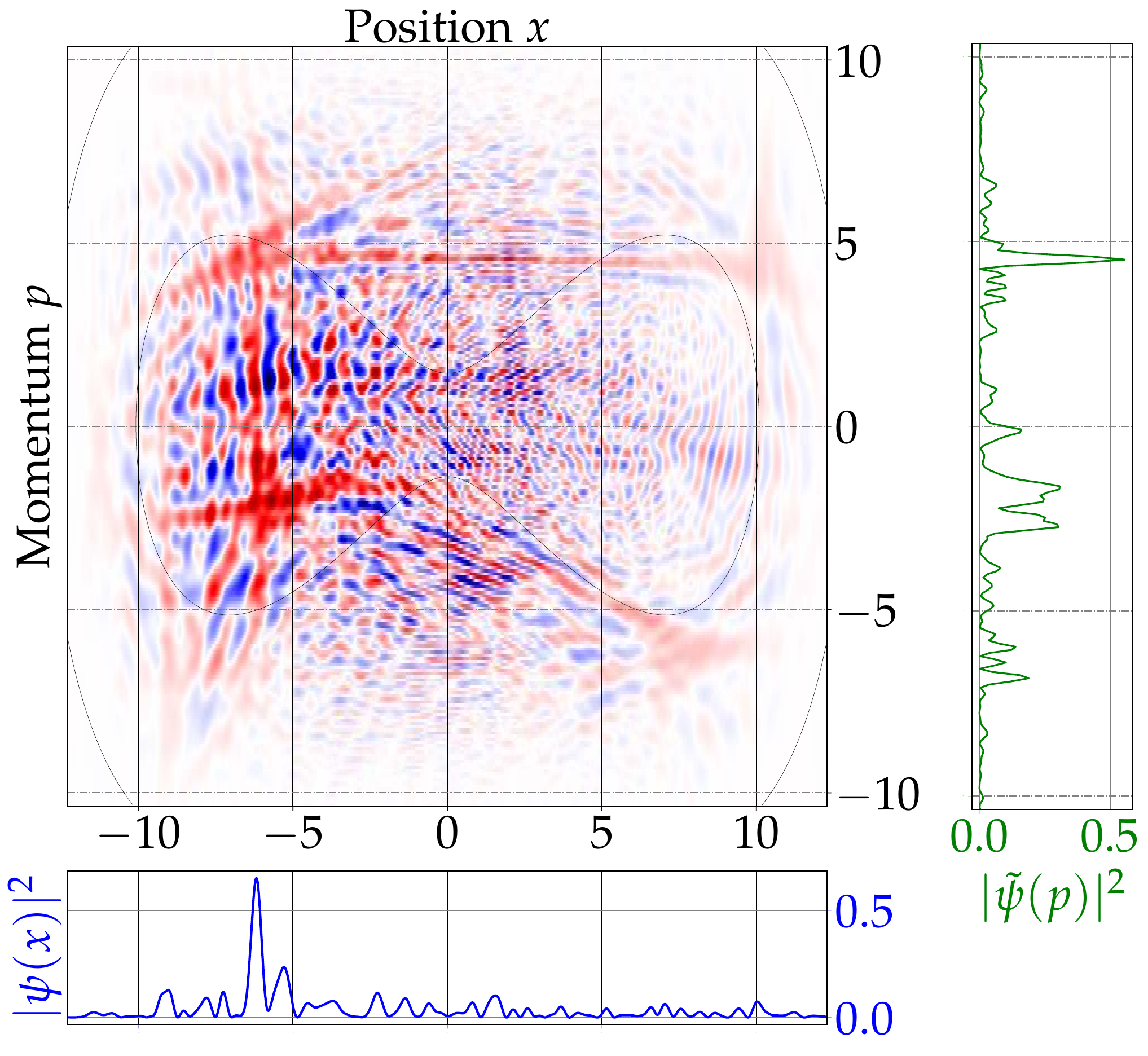}
     \put(-434,95){\rotatebox{0}{\emphLabel{A} $t=5$}}
     \put(-324,95){\rotatebox{0}{\emphLabel{B} $t=7.8$}}
     \put(-212,95){\rotatebox{0}{\emphLabel{C} $t=10$}}
     \put(-102,95){\rotatebox{0}{\emphLabel{D} $t=161.4$}}
     \caption{\emphCaption{Wigner distributions evolved in double well potential
         $V(x)=x^4/200-x^2/2$}. The initial state $\psi_0= (2/5\pi)^{1/4} \exp[-(x+9.9)^2/5]$ has an
       energy that partly exceeds the central barrier as can be seen clearly at short times
       \emphLabel{A}. Subsequently, lines form quite soon \emphLabel{B} and keep reappearing
       \emphLabel{C}-\emphLabel{D}.}
     \label{fig:Linear_x4_0.005_x2_0.5_gauss5}
   \end{minipage}
\end{figure}  
%\twocolumngrid
\end{widetext}

We find that trapped systems sooner or later form lines in \ps. When we choose a spatially
concentrated initial state, the (anharmonic) potential $V(x)$ disperses the state over its energy
corridor in \ps (the black background lines in Figs.~\ref{fig:Linear_x4_0.002_gauss9}
and~\ref{fig:Linear_x4_0.005_x2_0.5_gauss5} depict energy contours). This process has to happen
first until finally the state is sufficiently dispersed to self-interfere as irregular standing
waves, i.e. form random \gss, see \emphLabel{b}-\emphLabel{d} and, to a lesser extent,
\emphLabel{B}-\emphLabel{D} in Fig.~\ref{fig:Linear_x4_0.002_gauss9}.

These random \gss are responsible for the formation of lines in \ps, see
Section~\ref{sec:TheoryRandomGridStates}.

Simple enough trapped systems can show state revivals after a `recurrence' or `revival' time~$T_r$
(at $t=T_r$ the evolved state is identical to the initial state~\cite{Oliva_Kerr_18} or very
%\onecolumngrid

\begin{figure}[b]
  \hspace{-0.2cm}
  \begin{minipage}[b]{0.7\columnwidth}
    %% "b" to have captions on the same line
    \includegraphics[width=\columnwidth,angle=0]{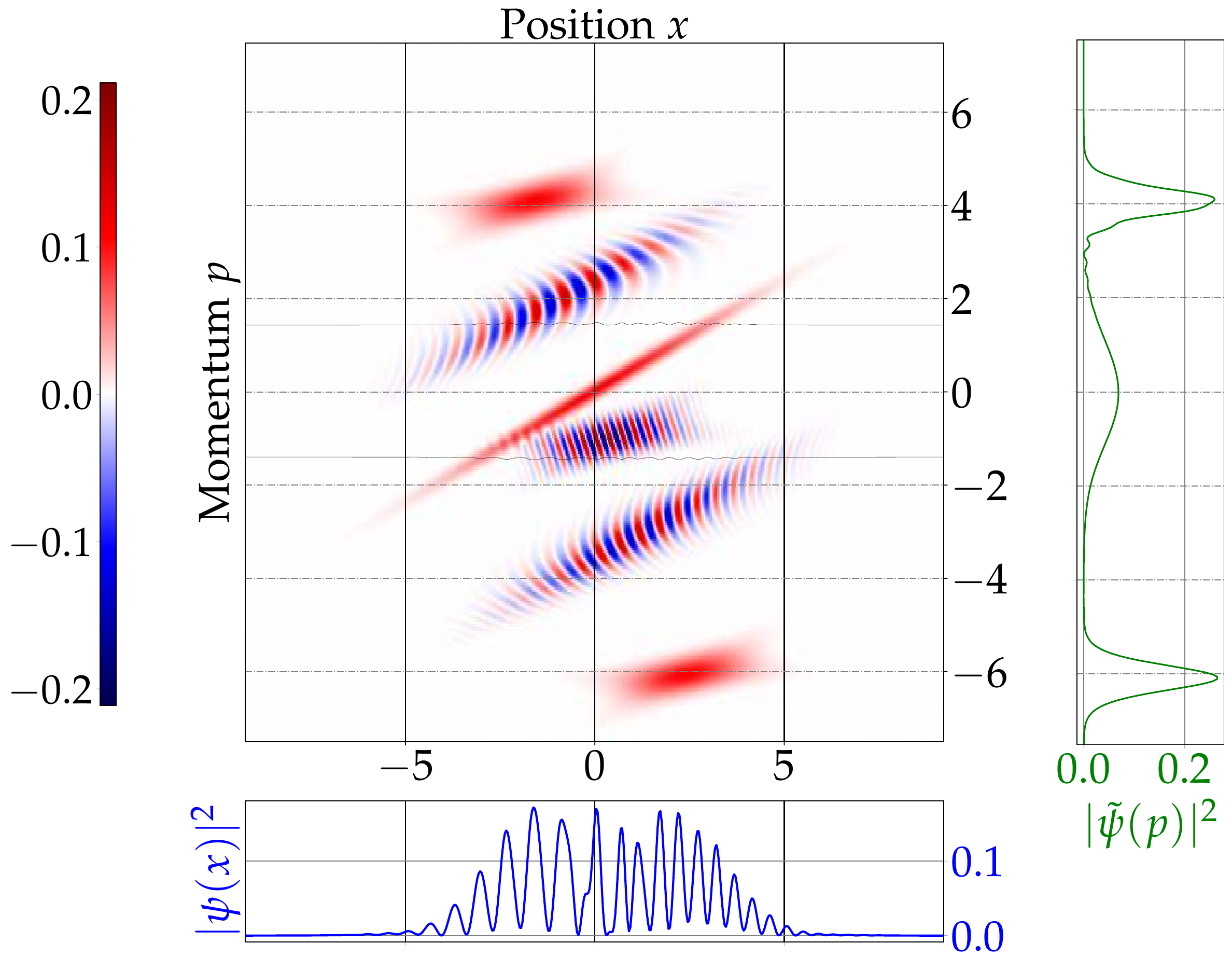}
  \end{minipage}
  \caption{\emphCaption{$W$'s elementary \ps interference
      fringes}~\cite{Zurek_NAT01,Wigner_PR32,Schleich_01} are roughly described by
    Eq.~(\ref{eq:WSchrodingersCat}) but tend to be curved when created between two peaks of unequal
    shape [see fringes centered on (-1,2) and (2,-3)] whereas peaks of equal shape create straight
    fringes [here at (0,-1)].
    \label{fig:ThreeUnequalPeaks_curvedInterference}}
\end{figure}

\noindent
similar to
it~\cite{Averbukh_PLA89,Robinett_PR04,Oliva_Shear_19}). For such, not too highly excited
linear systems~$T_r$ can be so short that we can study it numerically. At suitable fractions of
$T_r$, namely, at times of `fractional revivals' of the initial
state~\cite{Oliva_Kerr_18,Averbukh_PLA89,Robinett_PR04,Oliva_Shear_19}, we witness pronounced
formation of lines in \ps, for an example see Fig.~\ref{fig:Linear_x4_0.002_gauss9} \emphLabel{D}
and \emphLabel{d}.

\section{Comb-states have to be random to form lines in \ps \label{sec:TheoryRandomGridStates}}

The formation of (positive) straight lines in \ps is due to the formation of randomized \gss. If the
\gss are too symmetrical, they form higher order concentric ring or `eye' patterns, instead of
lines. These eye patterns are most pronounced for \gss with locally concave or convex arrangements
of the weights of their peaks. We now give numerical and semi-analytical evidence~to support these
claims.

\subsection{Interference between pairs of peaks \label{subsec:PairPeakInterference}}

To theoretically underpin that the coherences between \gs peaks gives rise to the observed
formation of lines in \ps, we will now isolate the pertinent trigonometric terms that are
responsible for the observed phenomena.

$W$ for a ``Schr\"odinger cat" state formed from two squeezed states
$G(x,p)=\pi^{-1} e^{-x^{2}/\xi^{2} - {p^{2}\xi^{2}}}$, with squeezing parameter~$\xi$, according to
Eq.~(\ref{eq:W}), has the simple form
\begin{subequations}
\begin{eqnarray}
  W(x,p) & =  \frac{
             G(x-\Delta x/2,p)+G(x+\Delta x/2,p) }{2} \;\; &
             \label{eq:compassstate}
  \\
 & + \; G(x,p)\cos\left(p \Delta x\right) \label{eq:interferenceterm}. &
 \end{eqnarray}\label{eq:WSchrodingersCat}
\end{subequations}
\indent This approximation for the description of a pair of peaks shows that they form an interference
pattern~(\ref{eq:interferenceterm}) of peak-width, halfway between peaks, with fringes whose spatial
frequency ($k \thicksim p$) is proportional~to the interpeak distance~$\Delta x$,
(see expression~(\ref{eq:interferenceterm}) and\refAppendix{ Appendix}{}{}{sec:Appendix_2peaks}{}{).}

\begin{figure}[b]
  \hspace{-0.2cm}
  \begin{minipage}[b]{1.01\columnwidth}
    %% "b" to have captions on the same line
    \includegraphics[width=\columnwidth,angle=0]{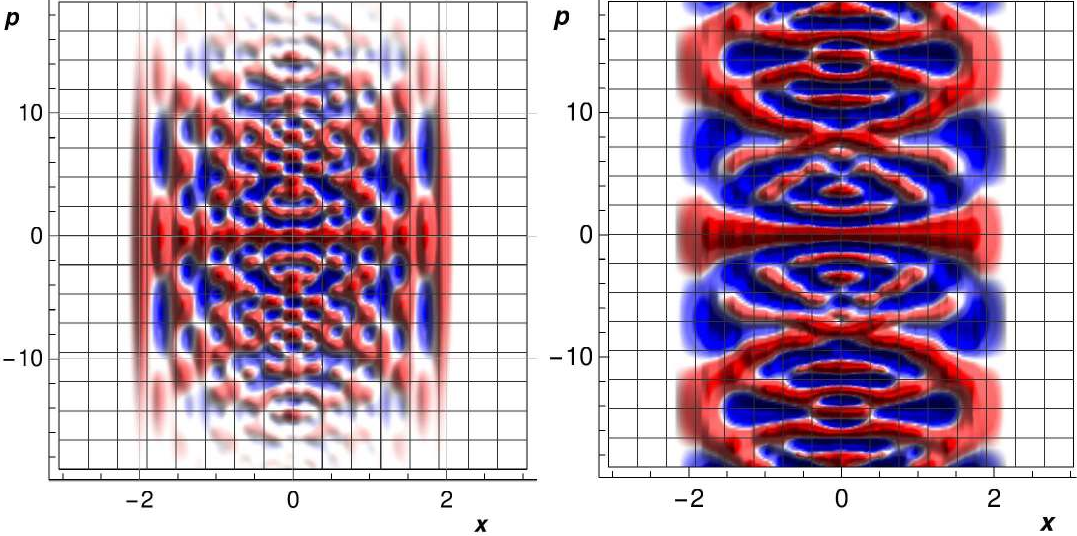}
    \put(-245,5){\rotatebox{0}{\emphLabel{A}}}
     \put(-119,5){\rotatebox{0}{\emphLabel{B}}}
  \end{minipage}
  \caption{\emphCaption{\emphLabel{A} $W$ of \gs versus \emphLabel{B} ${\cal I}$ of
      Eq.~(\ref{eq:InterferenceTrigTerms})}, with $\Lambda = 1$, with same locations and weights of
    peaks of \gs as in Fig.~\ref{fig:Qx_grid_Quadratic7}~\emphLabel{E}. Both representations yield
    similar interference patterns: we conclude that coherences between peaks in \gss are responsible for
    formation of lines.
    \label{fig:GridState_Trigonometric}}
\end{figure}

We emphasise that the interference pattern in \ps $\cos\left(p \Delta x\right)$ does not
generally have such straight line behaviour, which requires equal shapes $G$ of the constituent
peaks. In general the associated interference fringes are curved, see
Fig.~\ref{fig:ThreeUnequalPeaks_curvedInterference}.

To investigate the \gs scenario we strip out the terms~(\ref{eq:compassstate}) corresponding to the
peaks themselves and only retain the terms~(\ref{eq:interferenceterm}) describing \ps interference,
compare Fig.~\ref{fig:GridState_Trigonometric}~\emphLabel{A} with~\emphLabel{B}.

We are left with the resulting simplified expression for a random \gs's interference term
\begin{flalign}
\label{eq:InterferenceTrigTerms}
\hspace{-0.02cm}
{\cal I} = \sum_{m= 1}^{N - 1}&\sum_{n =  m + 1}^{N} \hspace{-0.065cm} \Lambda(x - \frac{X_m + X_n}{2}) \nonumber
\\ & \times \cos(p [X_m - X_n] - \phi_m+\phi_n) \, ,  %& \hspace{-0.0cm}
\end{flalign}
describing the effective overlap between peaks through~$\Lambda$. The inter-peak distances, $X_m-X_n$,
modulate the cosine-term in~(\ref{eq:InterferenceTrigTerms}) analog\-ously to
expression~(\ref{eq:interferenceterm}). Every peak at position~$X_m$ caries its own (constant)
phase~$\phi_m$.

Numerically, whenever $W$ (generated from peaks as in Fig.~\ref{fig:Qx_grid_Quadratic7}
and~\ref{fig:Grid_Random6}) forms lines, then so does a plot of
expression~(\ref{eq:InterferenceTrigTerms}), Fig.~\ref{fig:GridState_Trigonometric} is a typical
example.

Yet, we did not manage to extract an analytical expression that obviously displays the fact that
formation of straight lines with positive values is encoded in~(\ref{eq:InterferenceTrigTerms}), we
therefore now discuss the emergence of straight lines due to random \gss qualitatively instead.

\subsection{Interference in combs of peaks \label{subsec:ArtificialGridPeaksInterference}}

\begin{figure}[t]
  \hspace{-0.2cm}
  \begin{minipage}[b]{1.0\columnwidth}
    %% "b" to have captions on the same line
    \includegraphics[width=0.32\columnwidth,angle=0]{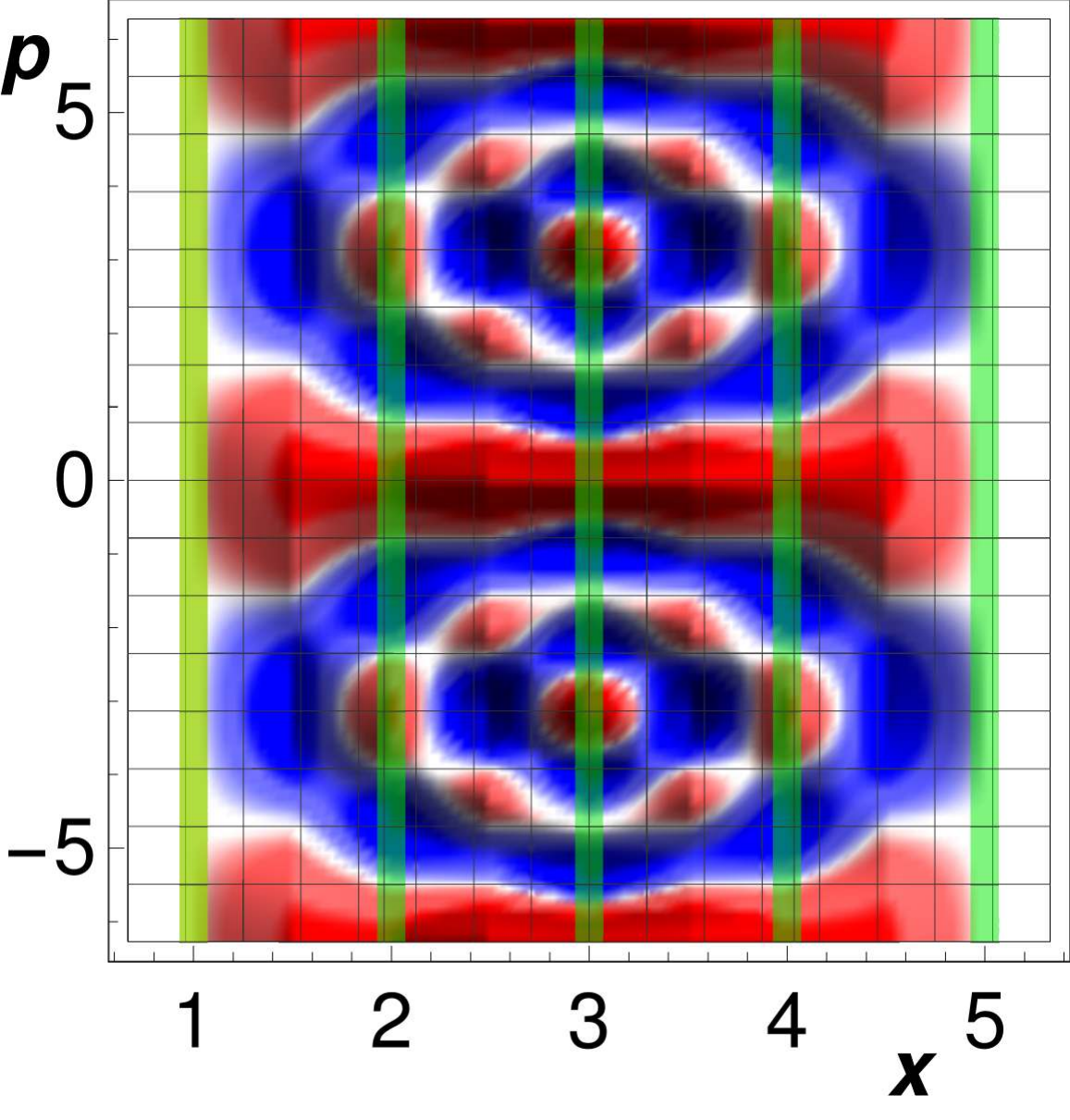}
    \includegraphics[width=0.32\columnwidth,angle=0]{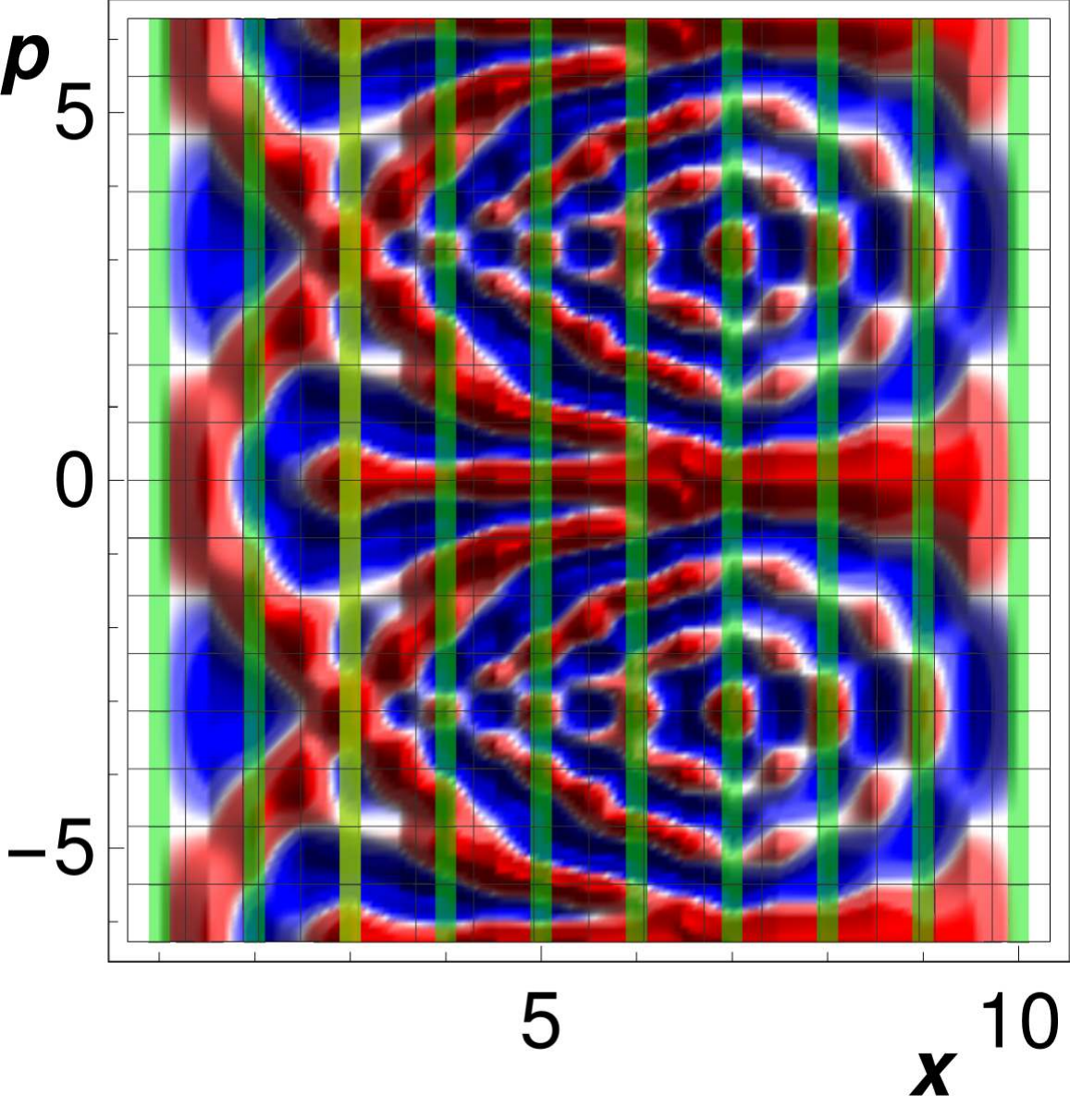}
       \includegraphics[width=0.32\columnwidth,angle=0]{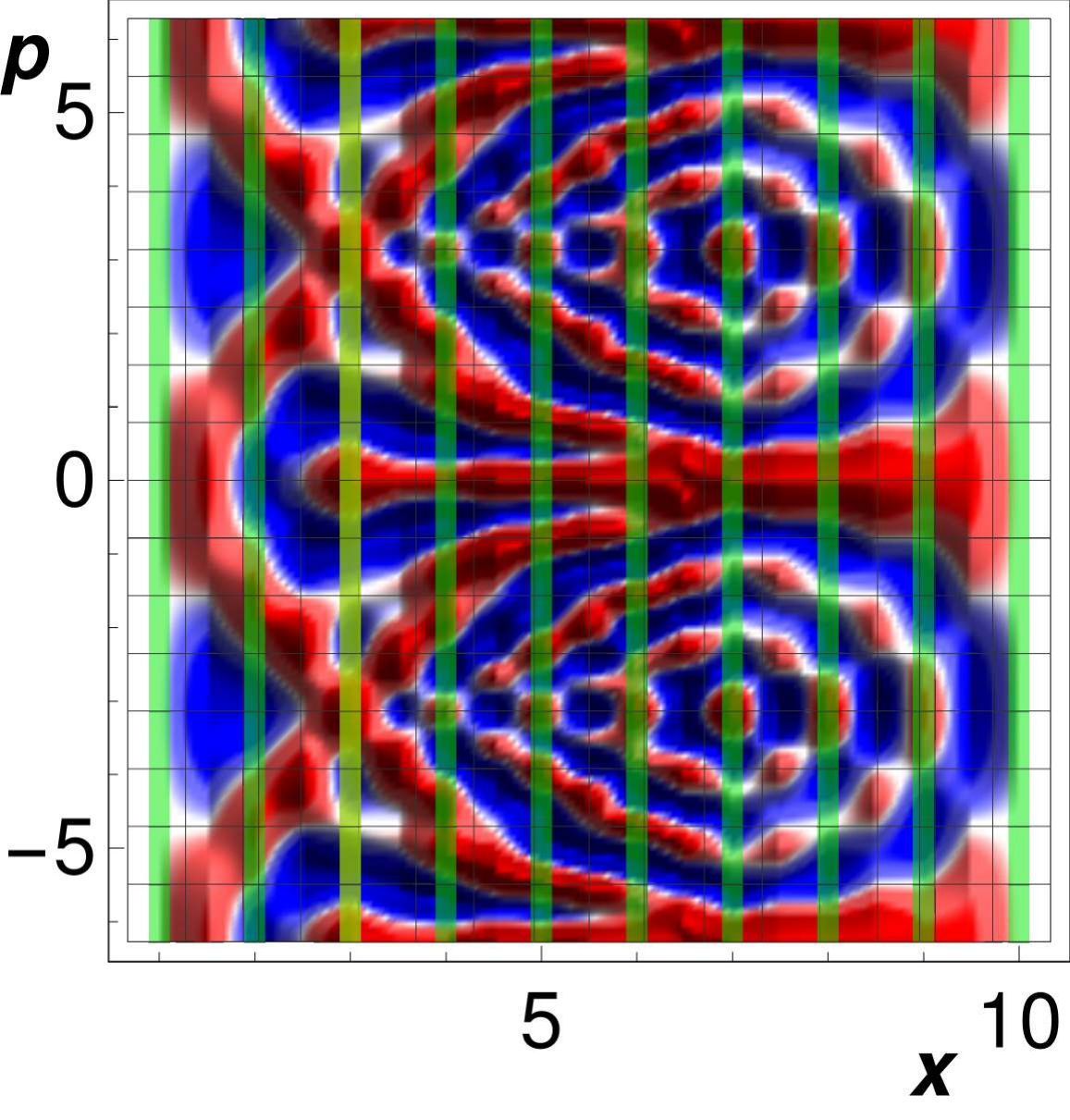}
     \put(-245,-0){\rotatebox{0}{\emphLabel{A}}}
     \put(-159,-0){\rotatebox{0}{\emphLabel{B}}}
     \put(-79,-0){\rotatebox{0}{\emphLabel{C}}}
  \end{minipage}
  \caption{\emphCaption{Plot of interference pattern $\cal I$ of
      Eq.~(\ref{eq:InterferenceTrigTerms}) for equal equidistant peaks.} The green strips indicate
     positions, $X_m$, of identical peaks. {\emphLabel{A}, all relative
      phases are zero: eye-patterns form}. \emphLabel{B}, phase of the central peak ($x=3$) is
    shifted by $\pi$ and \emphLabel{C} phase of the third peak, at position $x=3$, is shifted by
    $\pi$: these randomising phase shifts lead to the forma\-tion of lines (\emphLabel{B}--\emphLabel{C});
    also see{\protect \refAppendix{ Appendix}{}{}{sec:Appendix_Eyes}{}{.}}
    \label{fig:grid_1_phase_Pi}}
\end{figure}

In Figs.~\ref{fig:grid_1_phase_Pi}~\emphLabel{A} and~\ref{fig:Qx_grid_Quadratic7}~\emphLabel{A} we
observe that more symmetrical \gss with fixed peak-to-peak distances and fixed constant phase across
all peaks do not form straight lines in \ps, also see\refAppendix{
  Appendix}{}{}{sec:Appendix_Eyes}{}{.}

When these \gss are sufficiently randomized, however, they exhibit formation of straight lines in \ps,
see Figs.~\ref{fig:grid_1_phase_Pi}~\emphLabel{B}--\emphLabel{C}
and~\ref{fig:Qx_grid_Quadratic7}~\emphLabel{D}--\emphLabel{E}. Additionally, we observe that
imprinting completely random phases on each peak or shifting their individual momenta randomly (but
moderately) or changing their relative weights randomly (but moderately) does not destroy the
formation of straight lines in \ps (see\refAppendix{ Appendix}{}{}{sec:Appendix_SingleEye}{}{}
and\refAppendix{}{}{}{sec:Appendix_phiRand}{}{):} the formation of straight lines in \ps from
randomized \gss is a stable phenomenon.

This stability can be understood from the functional form of the interference pattern~$\cal I$.  For
example, shifts of a local phase $\phi_m$ entail a `holistic effect' since several terms in
Eq.~(\ref{eq:InterferenceTrigTerms}) are affected in a synchronized fashion (see
Fig.~\ref{fig:grid_1_phase_Pi}), thus interference patterns are modified smoothly rather than
abruptly.

We find that for randomized \gss the formation
%of lines in \ps is generic, see
%Fig.~\ref{fig:Grid_Random6}, and that the formation of eye patterns occurs most clearly when locally

\begin{widetext}
%\onecolumngrid  
\vspace{\columnsep}
 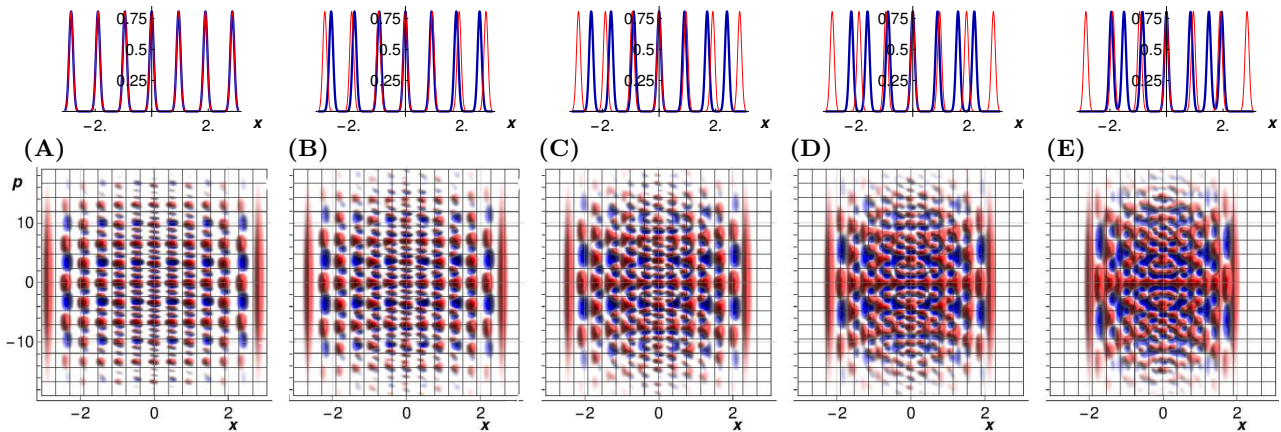
\begin{figure}[b!]
   \hspace{-0.2cm}
   \begin{minipage}[b]{1.98\columnwidth}
     \includegraphics[width=0.99\columnwidth]{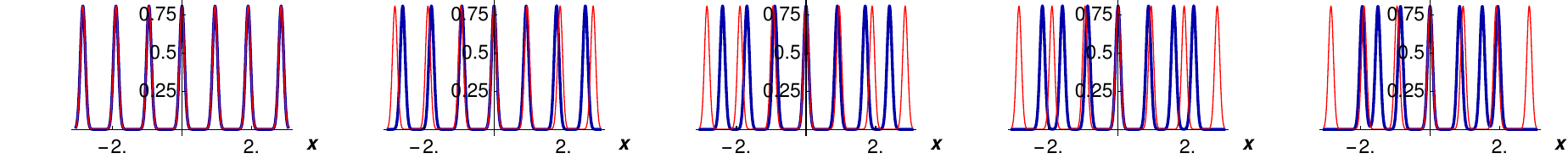}
     \\
     \includegraphics[width=0.99\columnwidth]{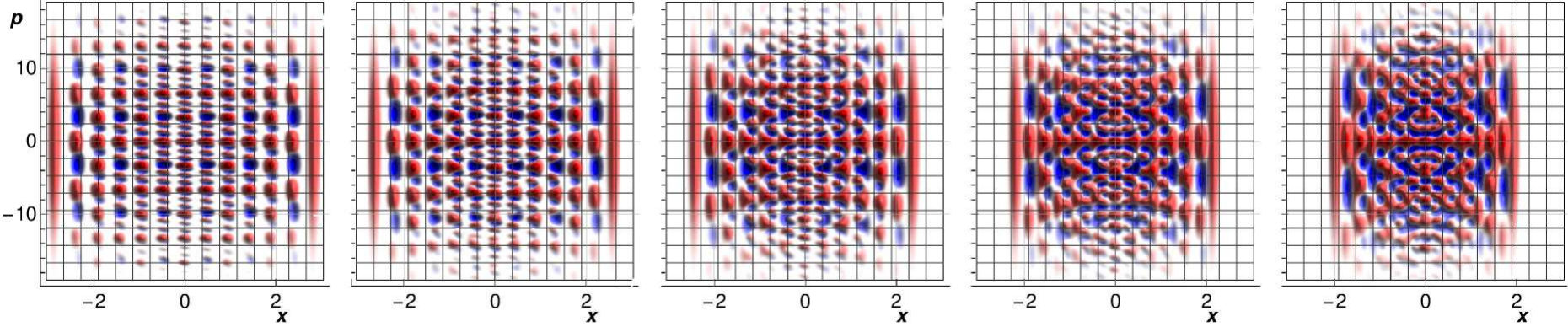}
     \put(-475,105){\rotatebox{0}{\emphLabel{A}}}
     \put(-375,105){\rotatebox{0}{\emphLabel{B}}}
     \put(-280,105){\rotatebox{0}{\emphLabel{C}}}
     \put(-185,105){\rotatebox{0}{\emphLabel{D}}}
     \put(-088,105){\rotatebox{0}{\emphLabel{E}}}
     \caption{\emphCaption{Randomized \gs and associated Wigner distributions:
       }$P(x)$ (top row) of a 7-peak state with  constant peak-to-peak spacing (red curves) gets
       increasingly squashed towards the center (\emphLabel{A}-\emphLabel{E}) reducing the
       peak-to-peak distances (blue curves). The resulting \gss, randomized in interpeak distances, form Wigner
       distributions~$W(x,p)$ that develop lines crisscrossing \ps (bottom row).
       \label{fig:Qx_grid_Quadratic7}}
 \end{minipage}
 \end{figure}
\end{widetext}
%\twocolumngrid

% We find that for randomized \gss the formation
\noindent
of lines in \ps is generic, see Fig.~\ref{fig:Grid_Random6}, and that the formation of eye patterns
occurs most clearly when locally concave \gss form, see Fig.~\ref{fig:Grid_Random6}~\emphLabel{D}
and \emphLabel{E}, Fig.~\ref{fig:8sech_straightLines}~\emphLabel{B}-\emphLabel{D},
and\refAppendix{ Appendix}{}{}{sec:Appendix_Eyes}{}{.}

\subsection{Line formation in random potentials \label{sec:AndersonLocal}}

Evolution in random potentials $V(x)$ is an obvious
candidate for the synthesis of random \gss. Here we
create $V(x)$ from random Fourier series.

\begin{figure}[t]
  \hspace{-0.2cm}
  \begin{minipage}[t]{1.01\columnwidth}
    %% "b" to have captions on the same line
    \includegraphics[width=0.49\columnwidth,angle=0]{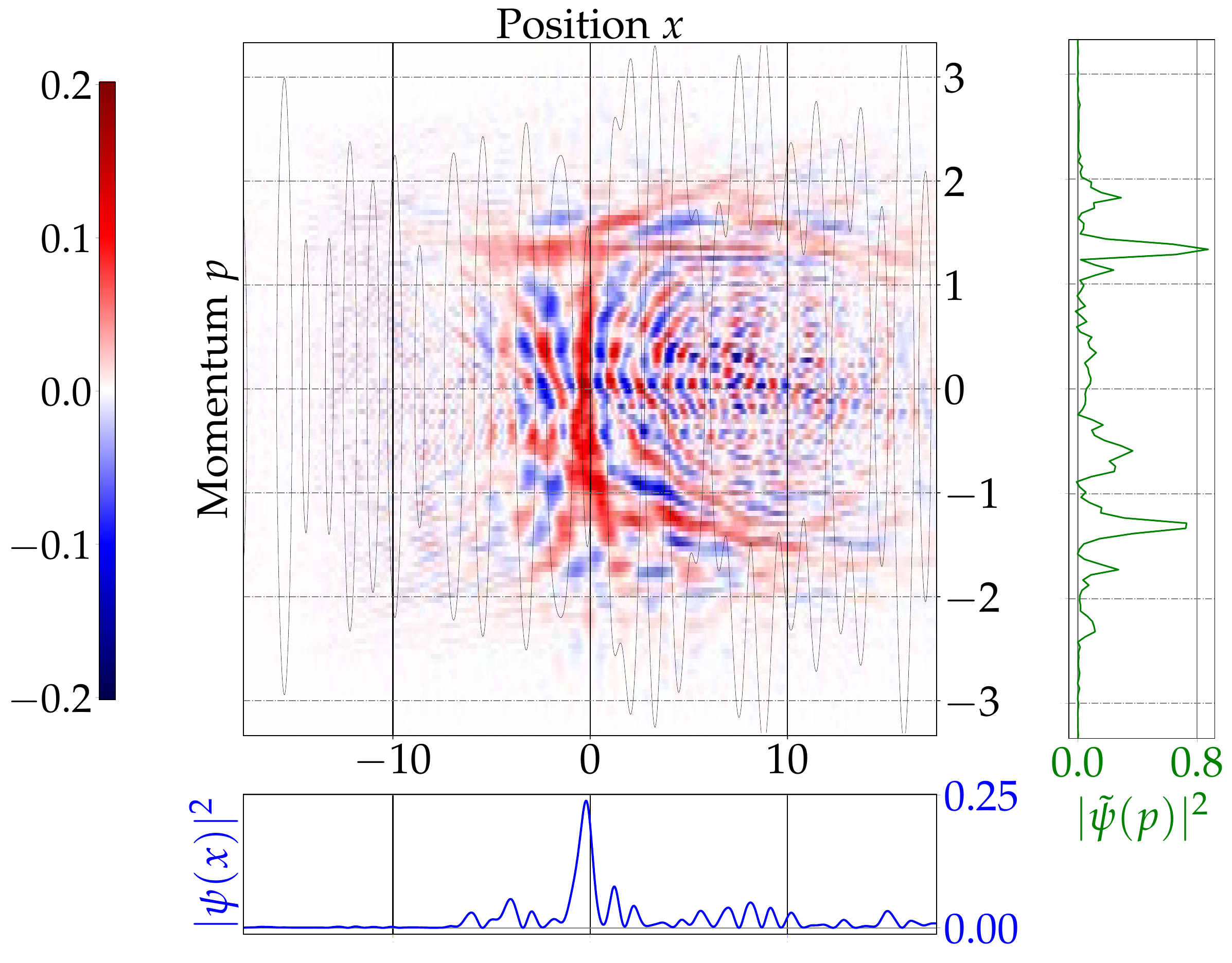}
        \includegraphics[width=0.49\columnwidth,angle=0]{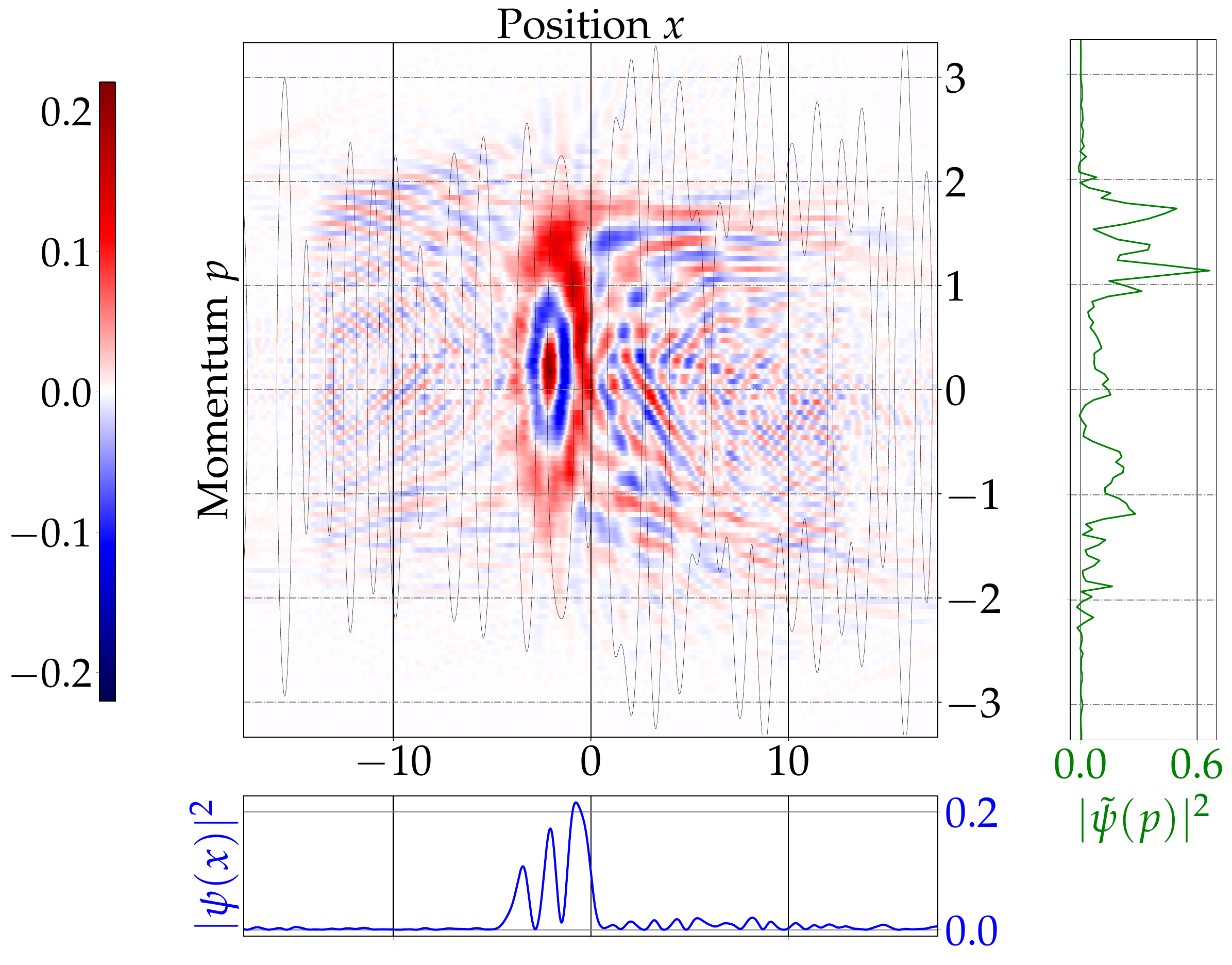}
    \put(-245,5){\rotatebox{0}{\emphLabel{A}}}
    \put(-119,5){\rotatebox{0}{\emphLabel{B}}}
  \end{minipage}
  \caption{\emphCaption{Random \gss form in a random potential} giving rise to lines across \ps
    and the formation of `eye' patterns.
    \label{fig:AndersonLineEye}}
\end{figure}

We commonly observe the formation of lines in \ps, see Fig.~\ref{fig:AndersonLineEye}~\emphLabel{A}
for a representative example.

Additionally, for a more symmetrical \gs with a convex peak-weighting distribution, eyes form
in \ps as well, see Fig.~\ref{fig:AndersonLineEye}~\emphLabel{B}.

\section{Nonlinear Systems \label{sec:NLSE}}

\subsection{Line formation in free nonlinear systems \label{sec:FreeNLSE}}

For the free ($V=0$) Schr\"odinger equation~(\ref{eq:_NLSE}) of order
three ($\epsilon =2$) with
attractive nonlinearity, $\gamma >0$, it is known that initial states

\begin{eqnarray}
  \label{eq:_InitialStateSECH}
  \hspace{-0.65cm}
  \psi(x,0) = \frac{\text{sech}(x)}{\sqrt{2}} \text{ with } \gamma = 2 N^2, N = 1,2,3,\ldots
\end{eqnarray}
give rise to breather solutions with up to $N+1$ peaks and repetition period
$T=\frac{\pi}{2}$~\cite{Schrader_IEEE95}. Their associated Wigner distributions~$W$ can display
straight lines and ringed `eye' shapes in \ps, see Fig.~\ref{fig:8sech_straightLines}.

\begin{figure}[t]
  \hspace{-0.2cm}
  \begin{minipage}[b]{\columnwidth} \includegraphics[%width=\columnwidth,
  height=0.6\columnwidth,angle=0]{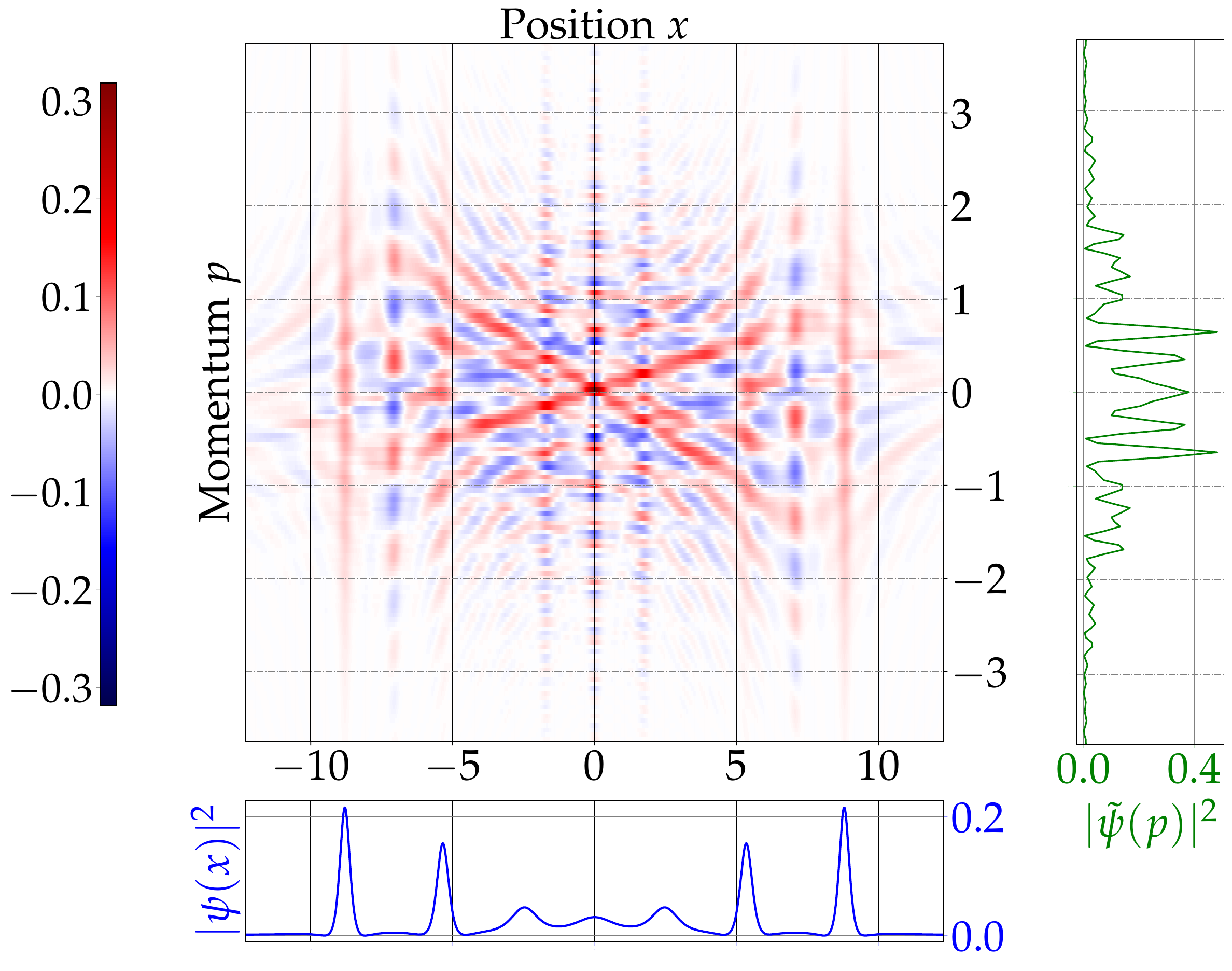}
  \end{minipage}
  \caption{\emphCaption{In a `dam break scenario' \gss form} resulting in the formation of lines in \ps.
    Here, for a linear increase in nonlinearity ($\epsilon = 2$) such that $\gamma(t)=150$ at $t=5.75$,
    starting from $\gamma(t=0)=0$, and an initial `straight-wall'
    state $\psi_0=\exp[-32 (x/5\pi)^{18}]/\sqrt{5 \pi 2^{2/3} \Gamma(19/(18))}$.
    \label{fig:DamBreak}}
\end{figure}

Straight lines also form for generic initial conditions which in the free case ($V=0$) lead to
evolution fulfilling the {``soliton resolution conjecture''}~\cite{Tao_BAMS09}. The lines only form
initially, while the radiative background and pre-solitonic peaks still overlap, see
Fig.~\ref{fig:DamBreak} for an example. This finding applies to wide classes of nonlinear \schr
equations as long as the interactions are attractive ($\gamma>0$), see
Fig.~\ref{fig:changeOrderEpsilon}.

In the case of repulsive interactions a confining potential is needed to trap the system state such
that it self-interferes, forming \gss with straight lines in \ps, see Fig.~\ref{fig:RepulsiveNLSE}.

\subsection{Line formation in trapped nonlinear systems \label{sec:TrappedNLSE}}

Straight lines form repeatedly, we believe in perpetuity, when we confine the spread of the wave
function by an
external trapping potential since it traps the radiative background~\cite{Tao_BAMS09}. For an
example see the bottom row of Fig.~\ref{fig:WithPotential}.
Lines can form for nonlinear systems with different orders $\epsilon+1$, see
Fig.~\ref{fig:changeOrderEpsilon}.

\begin{widetext}%
%  \onecolumngrid  
\vspace{\columnsep}
\begin{figure}[b]
   \hspace{-0.2cm}
   \begin{minipage}[b]{2.02\columnwidth}
 %%  "b" to have captions on the same line
     \includegraphics[height=0.219\columnwidth]{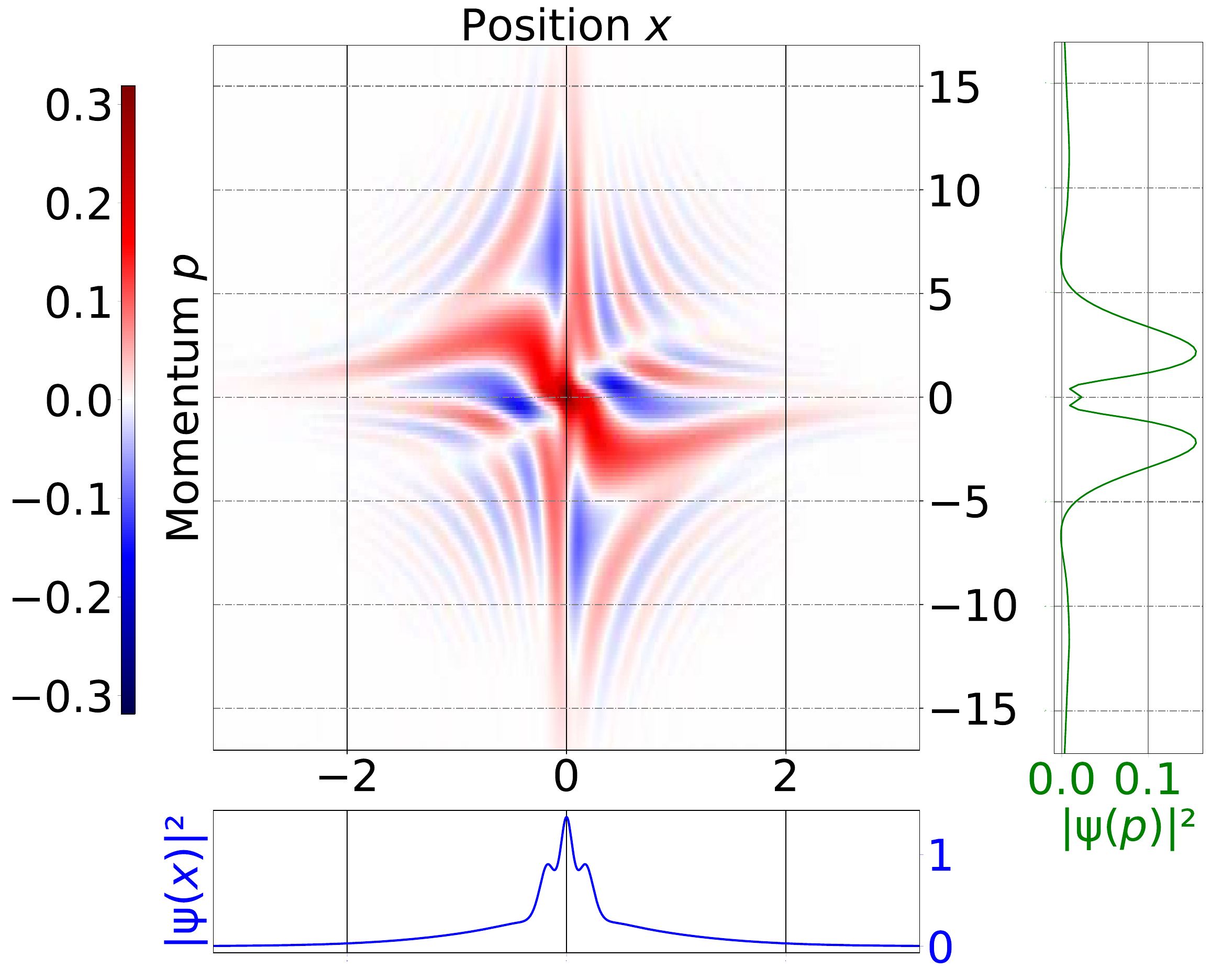}
     \includegraphics[height=0.219\columnwidth]{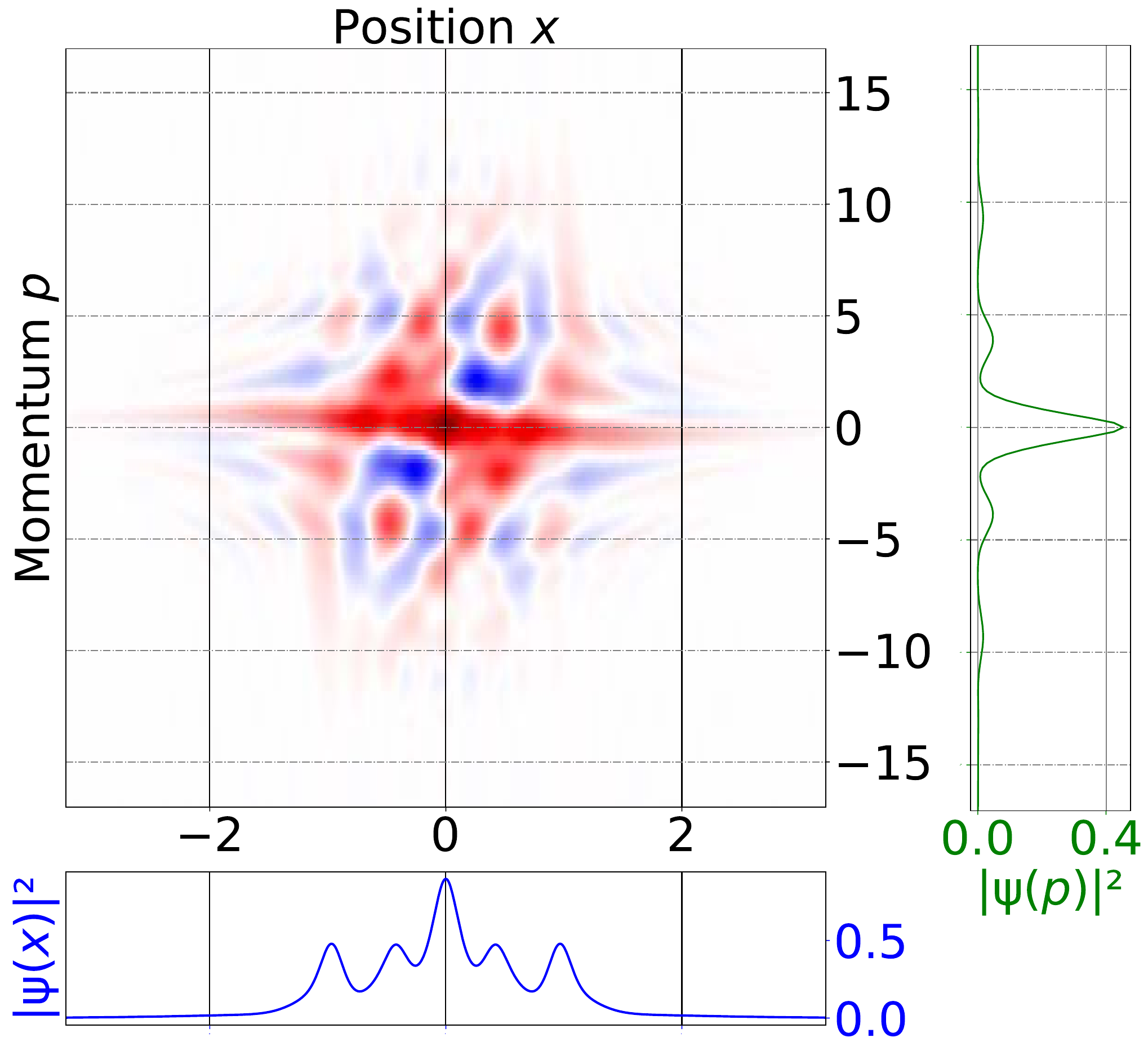}
     \includegraphics[height=0.219\columnwidth]{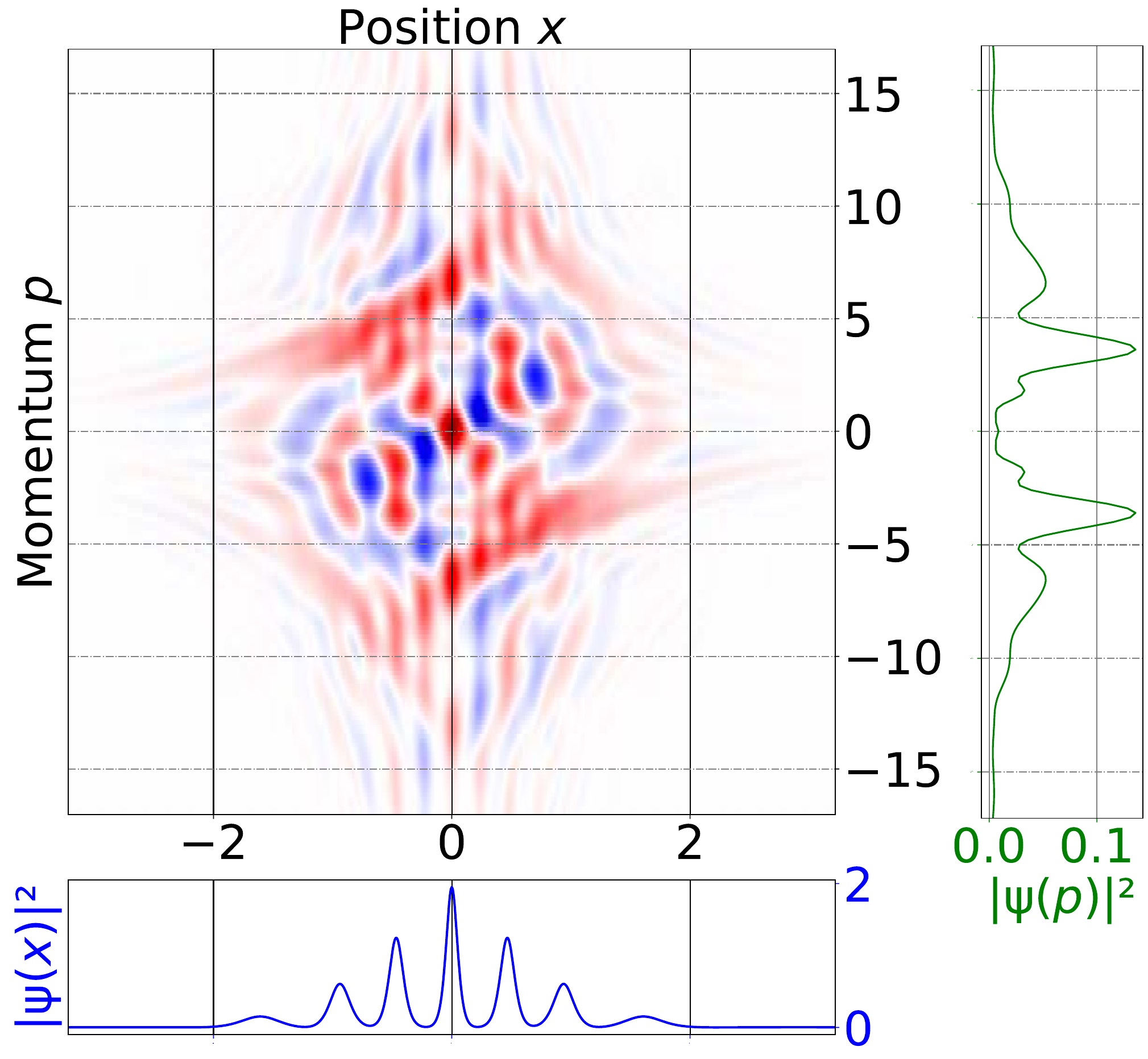}
     \includegraphics[height=0.219\columnwidth]{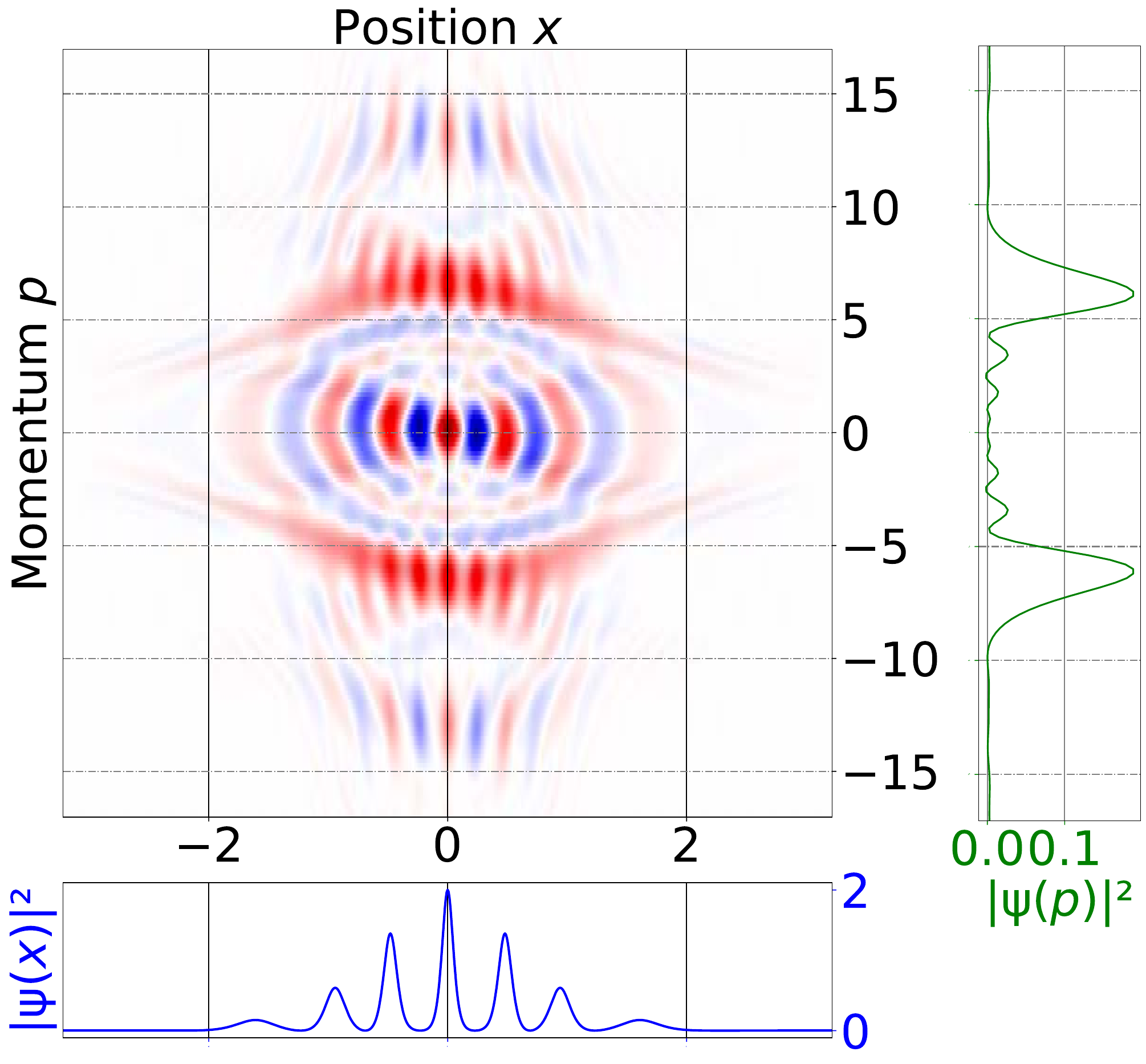}
     \put(-474,91){\rotatebox{0}{\emphLabel{A}}}
     \put(-354,91){\rotatebox{0}{\emphLabel{B}}}
     \put(-232,91){\rotatebox{0}{\emphLabel{C}}}
     \put(-112,91){\rotatebox{0}{\emphLabel{D}}}
     \caption{\emphCaption{Wigner distributions evolved from initial
         sech-state~(\protect{\ref{eq:_InitialStateSECH}}) with $\gamma=128$ ($N=8$):} \emphLabel{A}
       $t=0.10$, \emphLabel{B} $t=0.31$, \emphLabel{C} $t=0.76$, and \emphLabel{D} $t=0.79$.  Note
       the formation of zero lines (white) in \emphLabel{A} and positive (red) straight lines
       throughout.  In \emphLabel{B}, \emphLabel{C} and \emphLabel{D} a regular array of peaks
       in the position distribution~$P(x)$ which, in \emphLabel{C} and \emphLabel{D}, coincides
       with a concave momentum distribution~$\tilde P(p)$, giving rise to the formation of
       `double-eye' patterns, compare Fig.~\ref{fig:Grid_Random6}~\emphLabel{E}.}
     \label{fig:8sech_straightLines}
   \end{minipage}
\end{figure}
\end{widetext}
%\twocolumngrid

We hope the reader finds our conjecture plausible that the formation of slightly randomized peaks is
responsible for the formation of straight line patterns in \ps.

Whereas line formation is enhanced by the presence of attractive nonlinear interactions,
Fig.~\ref{fig:WithPotential}, it can be present even for repulsive interactions ($\gamma < 0$) if a
confining potential traps the state such that \gss can form, see\refAppendix{
  Appendix}{}{}{subsec:Appendix_RepulsiveNLSE}{}{.}

\subsection*{Conclusions \label{subsec:Conclusion}}

We have established that the states of many different types of quantum systems display formation of
\emph{positive} lines criss-crossing \ps; such lines cannot form in classical systems.

The formation of these lines is a robust phenomenon.

It will be interesting to see whether in higher-dimensional systems similar `pencils' form in \ps.

The presence of attractive nonlinear interactions enhances the formation of lines criss-crossing
\ps.

We moreover speculate that the formation of these lines might be able to illuminate the formation of
rogue waves in nonlinear systems~\cite{SotoCrespo_PRL16}, using the \ps perspective.

Lines can also occur in linear or repulsive systems, if the state is confined by a trapping
potential.

We expect that it should be possible to experimentally detect such lines through the detection of
large peaks in measurements of suitably rotated quadratures and in quantum state reconstruction
experiments~\cite{Hofheinz_NAT09,Kurtsiefer_NAT97}.

%\begin{widetext}
  \onecolumngrid
  \begin{figure}[b]
  \hspace{-0.2cm}
  \begin{minipage}[h]{2.02\columnwidth}
%%  "b" to have captions on the same line
    \includegraphics[height=0.216\columnwidth]{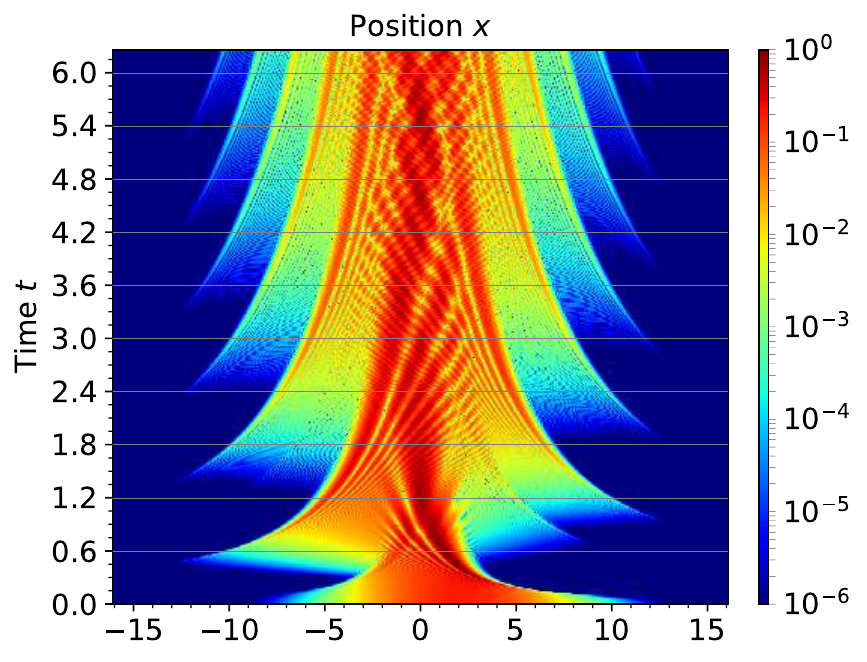}
    \includegraphics[height=0.216\columnwidth]{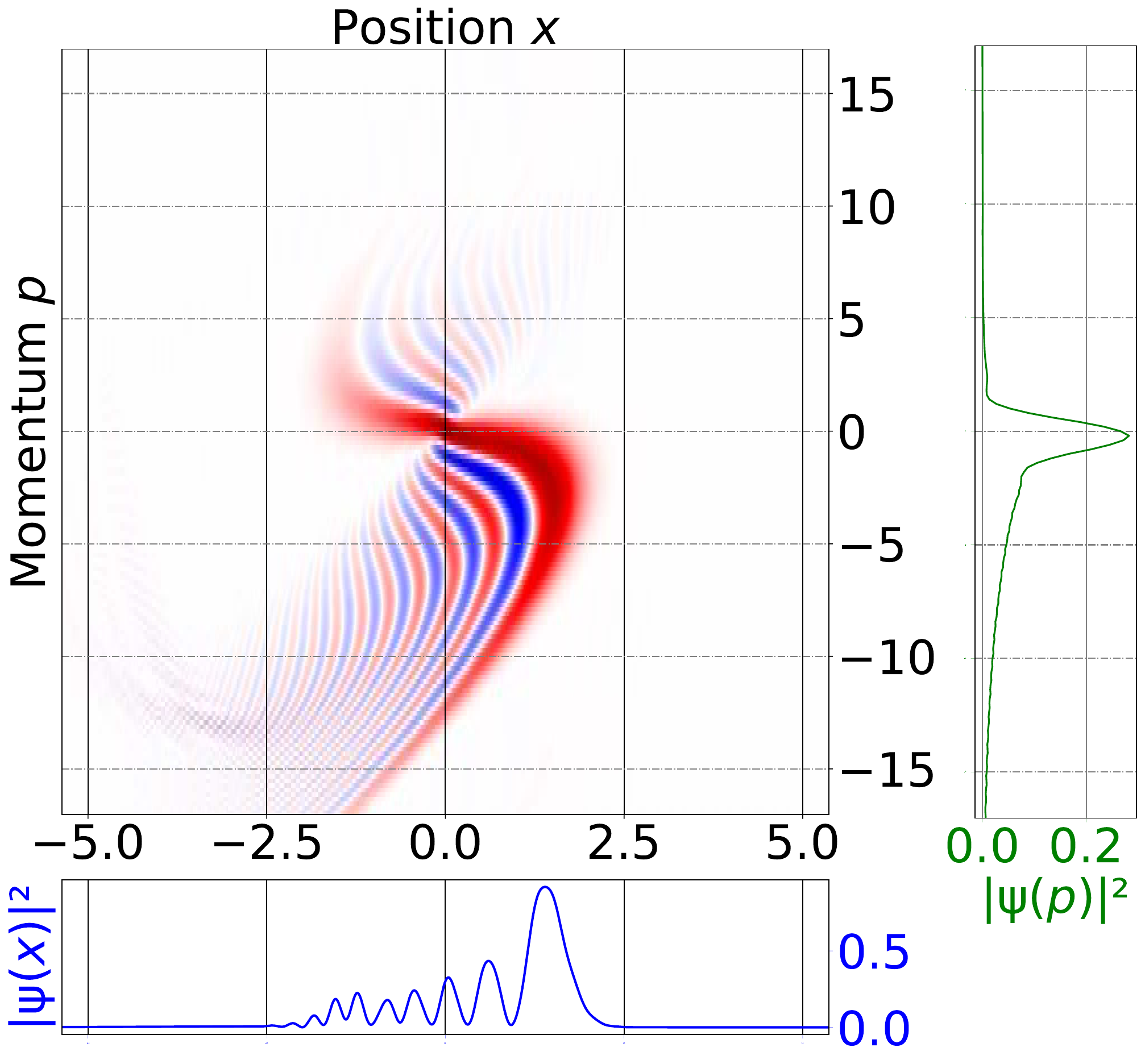}
    \includegraphics[height=0.216\columnwidth]{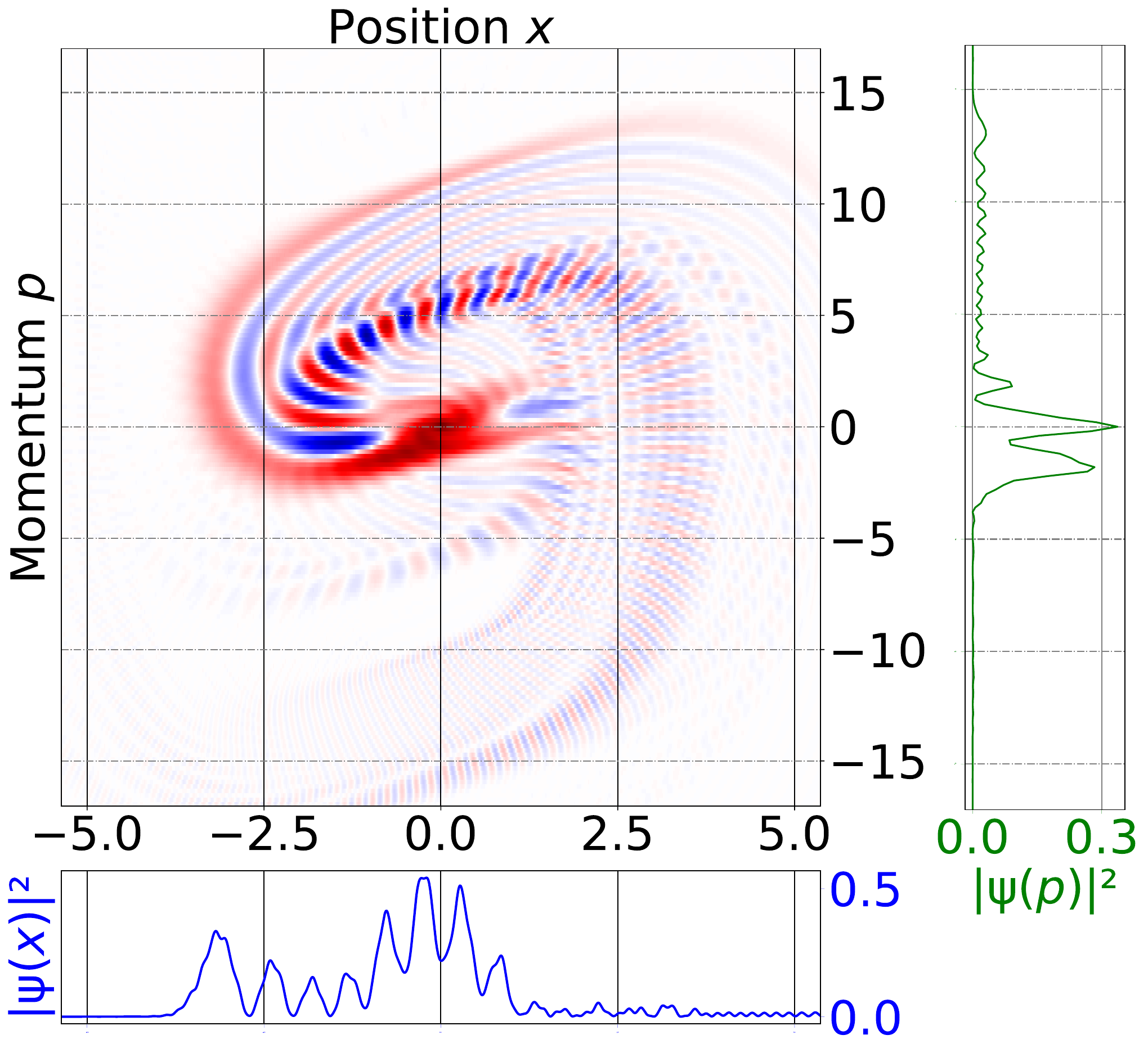}
    \includegraphics[height=0.216\columnwidth]{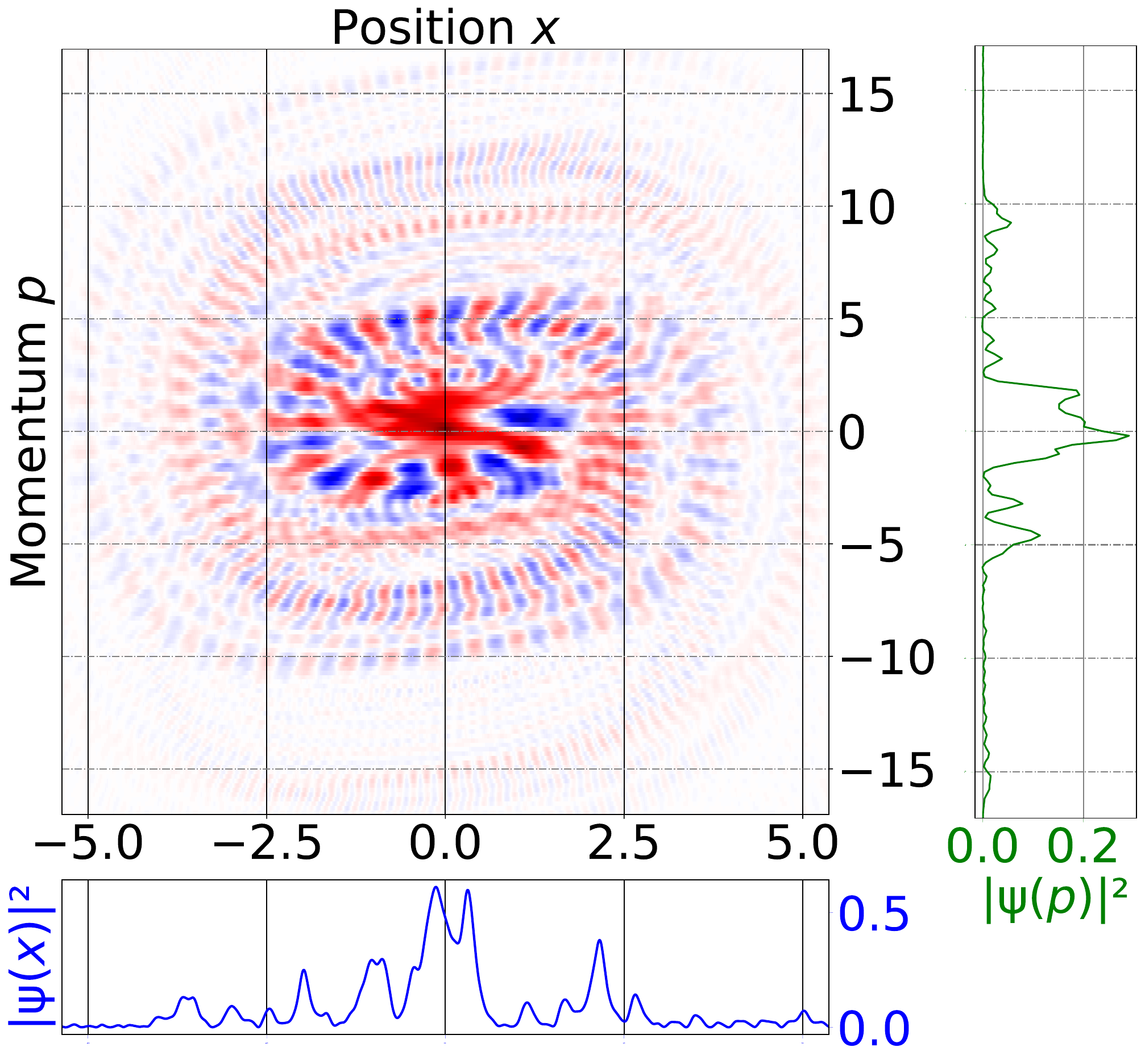}
    \put(-480,90){\rotatebox{0}{\textcolor{white}{\emphLabel{A}}}}
    \put(-477,34){\rotatebox{90}{\textcolor{white}{\boldmath{$|\psi(x,t)|^2$}}}}
    \put(-347,90){\rotatebox{0}{\emphLabel{B}}}
    \put(-229,90){\rotatebox{0}{\emphLabel{C}}}
    \put(-109,90){\rotatebox{0}{\emphLabel{D}}}
    \\
    \includegraphics[height=0.215\columnwidth]{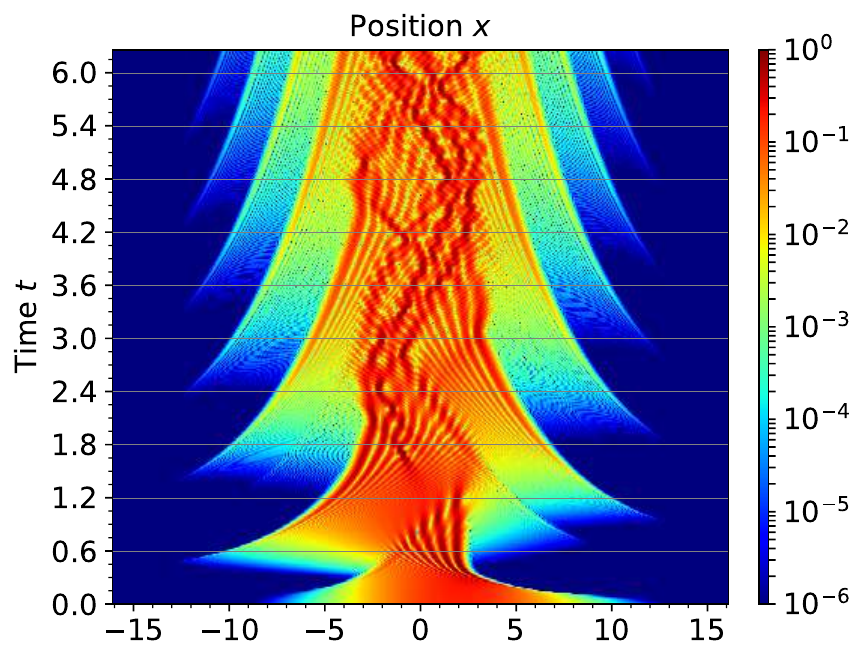}
    \includegraphics[height=0.215\columnwidth]{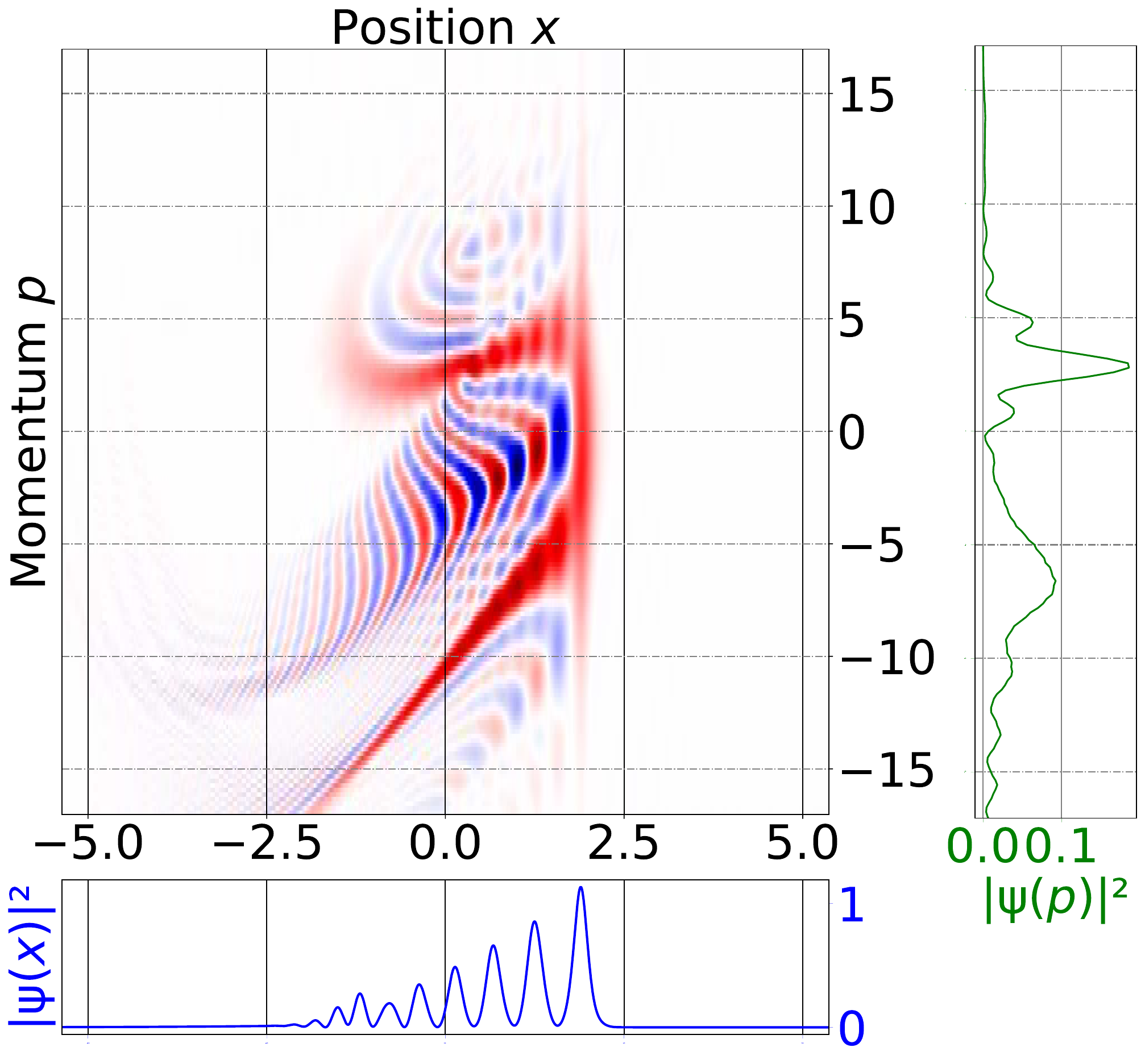}
    \includegraphics[height=0.215\columnwidth]{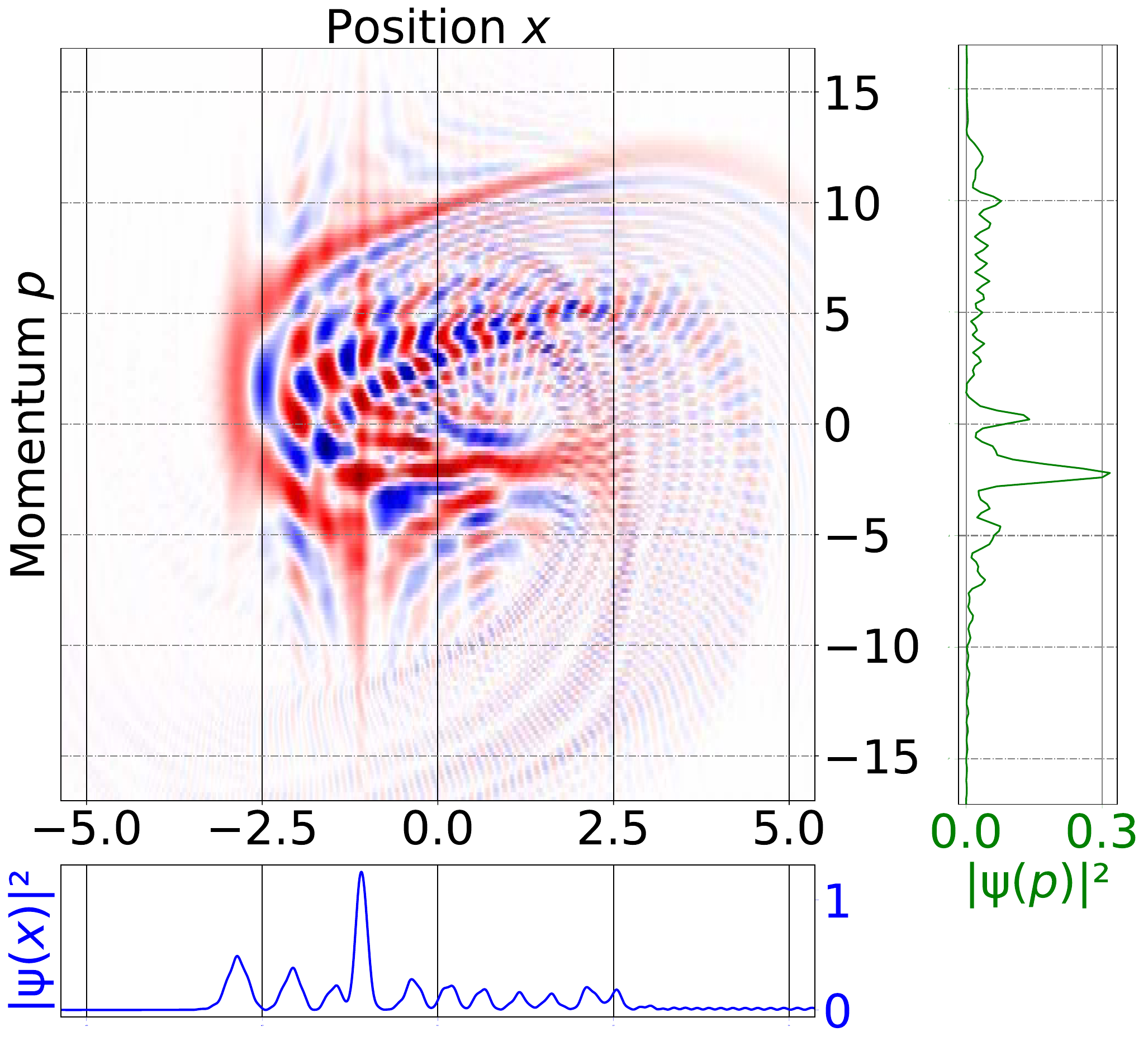}
    \includegraphics[height=0.215\columnwidth]{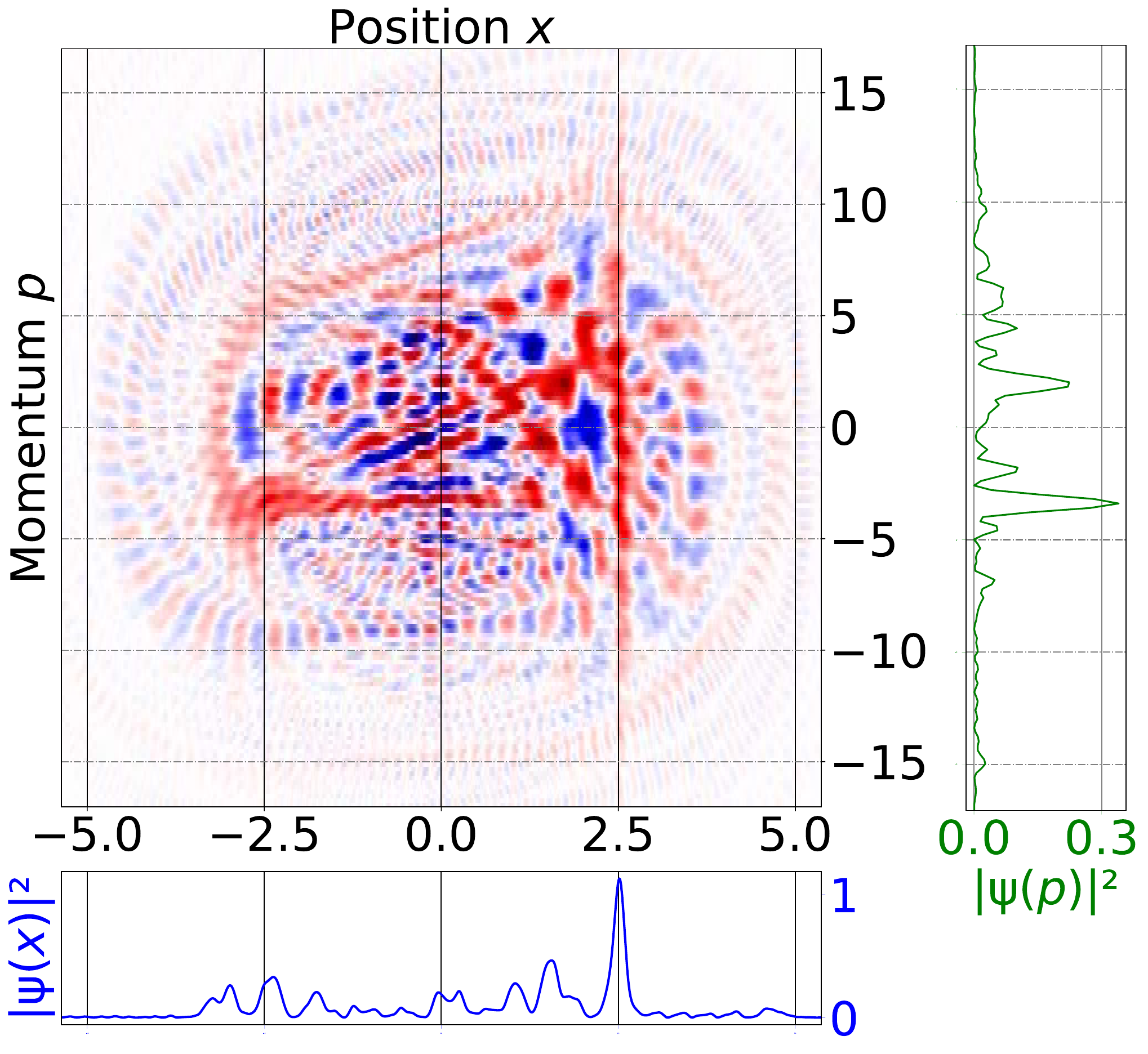}
    \put(-480,90){\rotatebox{0}{\textcolor{white}{\emphLabel{a}}}}
    \put(-478,34){\rotatebox{90}{\textcolor{white}{\boldmath{$|\psi(x,t)|^2$}}}}
    \put(-347,90){\rotatebox{0}{\emphLabel{b}}}
    \put(-229,90){\rotatebox{0}{\emphLabel{c}}}
    \put(-109,90){\rotatebox{0}{\emphLabel{d}}}
    \caption{\emphCaption{Enhanced line formation with attractive interaction: } The initial
      state $\psi_0(x) = 0.43 \exp[-(x-2)^2/19]$ evolves in the potential $V(x) = \frac{1}{10} x^4$
      without, ($\gamma =0$) top row
      (\emphLabel{A}-\emphLabel{D}), and in the presence of attractive interaction~($\gamma=50$ and $\epsilon=2$)
      bottom row (\emphLabel{a}-\emphLabel{d}).  The spatial probability density $|\psi(x,t)|^2$,
      \emphLabel{A} and \emphLabel{a}, shows dynamics dominated by oscillations due to the
      potential's confining forces. The associated Wigner distributions (using the same colouring as
      in Fig.~\ref{fig:8sech_straightLines}), at times \emphLabel{B} and \emphLabel{b} $t=0.55$,
      \emphLabel{C} and \emphLabel{c} $t=1.76$, and \emphLabel{D} and \emphLabel{d} $t=4.52$,
      prominently display straight lines crisscrossing \ps for the nonlinear case
      \emphLabel{b}-\emphLabel{d} whereas only weak lines form in the linear case, also
      see{\protect \refAppendix{ Appendix}{}{}{subsec:Appendix_RepulsiveNLSE}{}{.}}
      \label{fig:WithPotential}}
\end{minipage}
\end{figure}
\vspace{0.1cm}
%\end{widetext}

\begin{acknowledgments}
  O.S. is extremely grateful to Denys Bondar for sharing his python code on github and his patient
  explanations on how to use it. He also appreciates the hospitality of the National Center for
  Theoretical Sciences during his stay in Hsinchu. This work is partially supported by the Ministry
  of Science and Technology of Taiwan (No. 108-2923-M-007-001-MY3 and No. MOST: 110-2123-M-007-002),
  Office of Naval Research Global, US Army Research Office, and the collaborative research program
  of the Institute for Cosmic Ray Research (ICRR) at the University of Tokyo.
\end{acknowledgments}
%\end{widetext}
\twocolumngrid

%\end{widetext}

%\bibliography{Ole_Bibliography}
\bibliography{Formation_of_Lines_in_PhaseSpacePatterns.bbl}

\clearpage

\setcounter{section}{0}
\renewcommand{\thesection}{A.~\arabic{section}}
\renewcommand{\thefigure}{\thesection~\arabic{figure}}
\setcounter{equation}{0}
\renewcommand{\theequation}{\thesection~\arabic{equation}}%.\arabic{figure} 
\renewcommand{\emphLabel}[1]{\textbf{{#1}}}

\onecolumngrid
\vspace{\columnsep}

\begin{center}
{   \large \bf -- Appendix --  \\ \vspace{0.25cm}
 {On the Formation of Lines in Quantum Phase Space} }
%   --Supplement--}%al Material}
 \\ \vspace{0.25cm}
 Ole Steuernagel, Popo Yang and Ray-Kuang Lee 
 \end{center}

%\onecolumngrid  

%\onecolumngrid  
\section{Wigner distribution fringes between two-peak combinations\label{sec:Appendix_2peaks}}

\begin{figure*}[h]
\centering
\includegraphics[width=0.98\columnwidth]{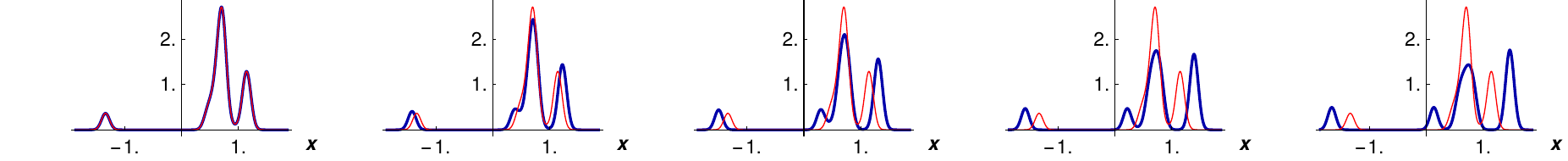}
\\
\includegraphics[width=0.98\columnwidth]{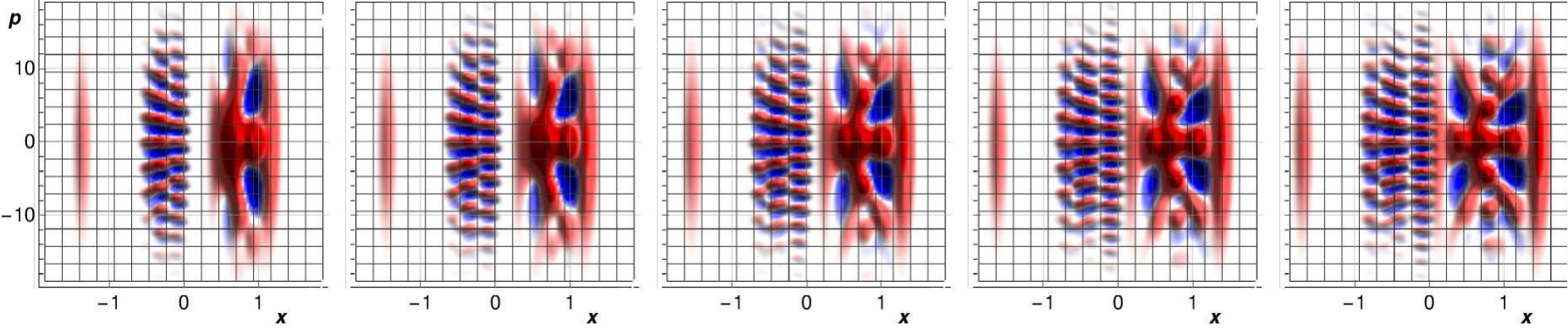}

\caption{\emphCaption{Interference fringes in \ps for a 2-peaked state }form midway between wave
  function peaks and at a spatial frequency proportional to the interpeak distance, compare
  Eq.~(\ref{eq:interferenceterm}).  Here a distribution which is roughly concentrated in two spatial
  locations ($x<0$ versus $0<x$) [see $P(x)$ top row] displays simple interference in the
  region~$-1<x<0$ [see $W(x,p)$ bottom row], which is graded (the spatial frequency increases from
  $x=-1$ to $x=0$) since $P(x)$ for positive values of $x$ is spatially spread out.
  \\
  The \ps structure for the positive region, $x>0$, arises from the coherence of the three peaks located
  in that region. Therefore, the positive region by itself provides a simple
  and illustrative case for how lines crisscrossing \ps form.
  \label{sfig:2peaks}}
\end{figure*}
\vspace{-1.45cm}
 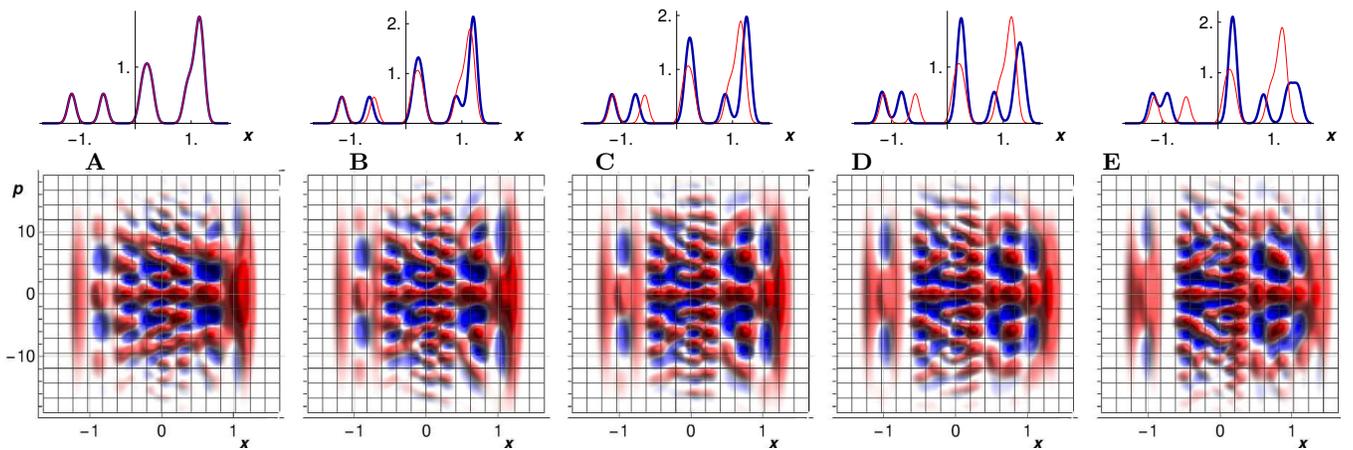
\begin{figure}[b]
   \hspace{-0.2cm}
   \begin{minipage}[b]{\columnwidth}
 %%  "b" to have captions on the same line
     \hspace{-0.5cm}
     \includegraphics[width=1.01\columnwidth]{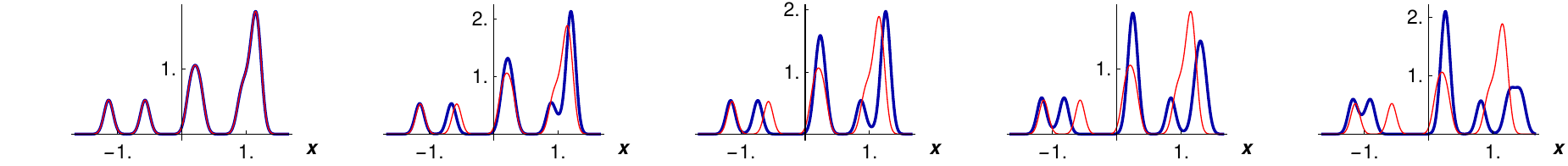}
     \\ \vspace{0cm}
     \includegraphics[width=0.99\columnwidth]{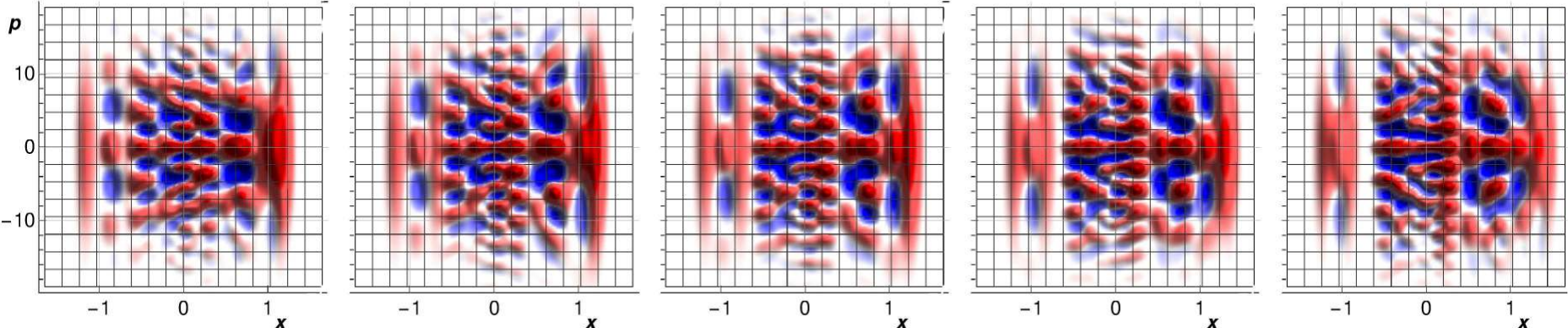}
     \put(-475,105){\rotatebox{0}{\emphLabel{A}}}
     \put(-375,105){\rotatebox{0}{\emphLabel{B}}}
     \put(-282,105){\rotatebox{0}{\emphLabel{C}}}
     \put(-185,105){\rotatebox{0}{\emphLabel{D}}}
     \put(-090,105){\rotatebox{0}{\emphLabel{E}}}
     \caption{\emphCaption{Random 7-peak \gss and associated Wigner distributions: }$P(x)$
       (top row) of states with 7 equally weighted peaks which are randomly distributed in position
       (and thus coalescing into 4 or 5 humps) are shown together with the associated Wigner
       distributions $W(x,p)$ (bottom row). $W(x,p)$ displays straight lines crisscrossing
       \ps. Panels \emphLabel{D} and \emphLabel{E} show a `double-eye' pattern (bottom row) due to
       the concave arrangement of the weights of the last three (rightmost) humps (top row).
       \label{fig:Grid_Random6}}
 \end{minipage}
 \end{figure}

\newpage
\section{NLSEs of Different Orders  \label{sec:Appendix_DifferentNLSEs}}
\vspace{-.85cm}
%\begin{widetext}
%\onecolumngrid  
\vspace{\columnsep}
 \begin{figure}[h]
   \hspace{-0.2cm}
   \begin{minipage}[b]{\columnwidth}%,\columnwidth}
 %%  "b" to have captions on the same line
     \includegraphics[height=0.216\columnwidth]{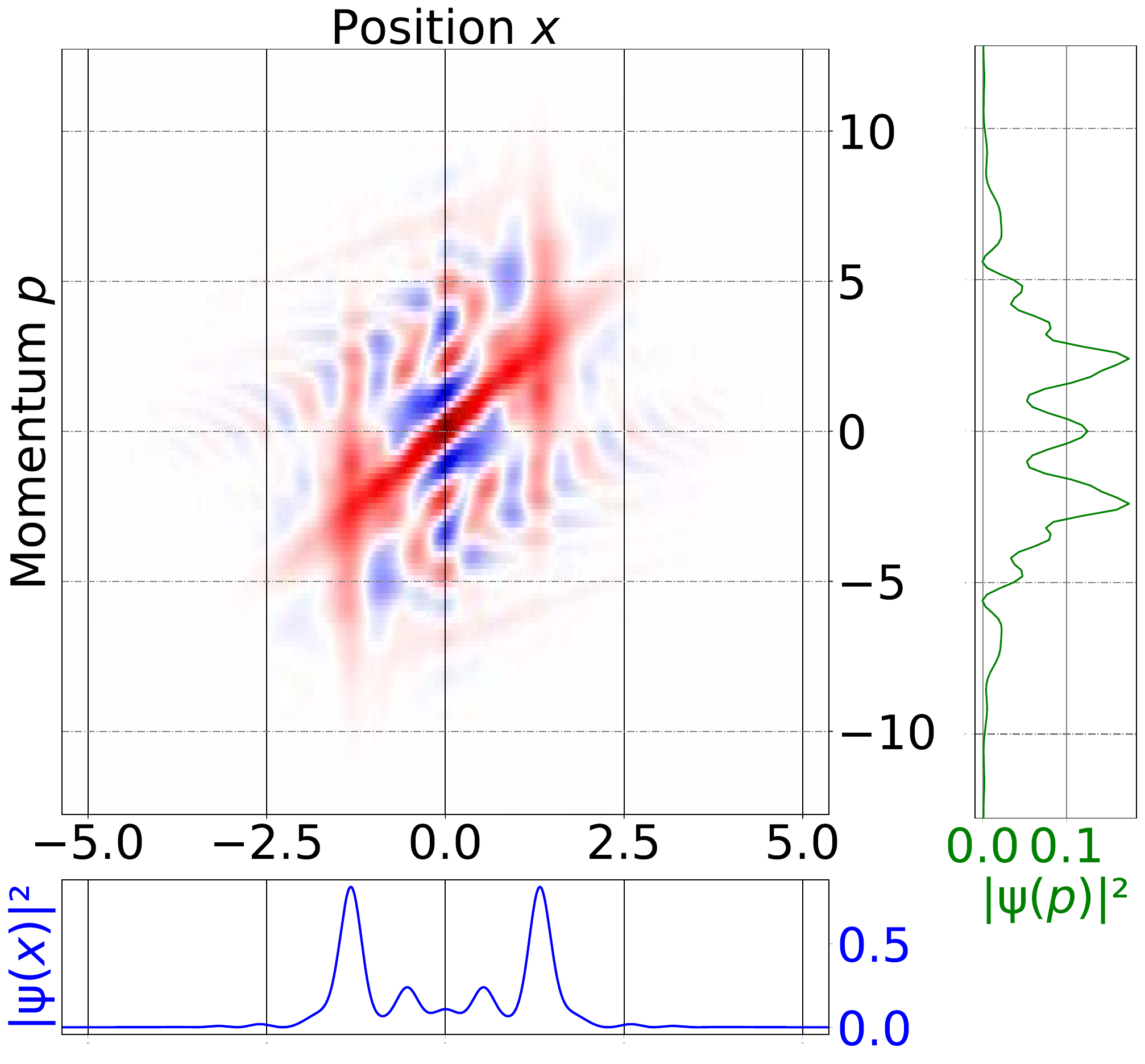}
     \includegraphics[height=0.216\columnwidth]{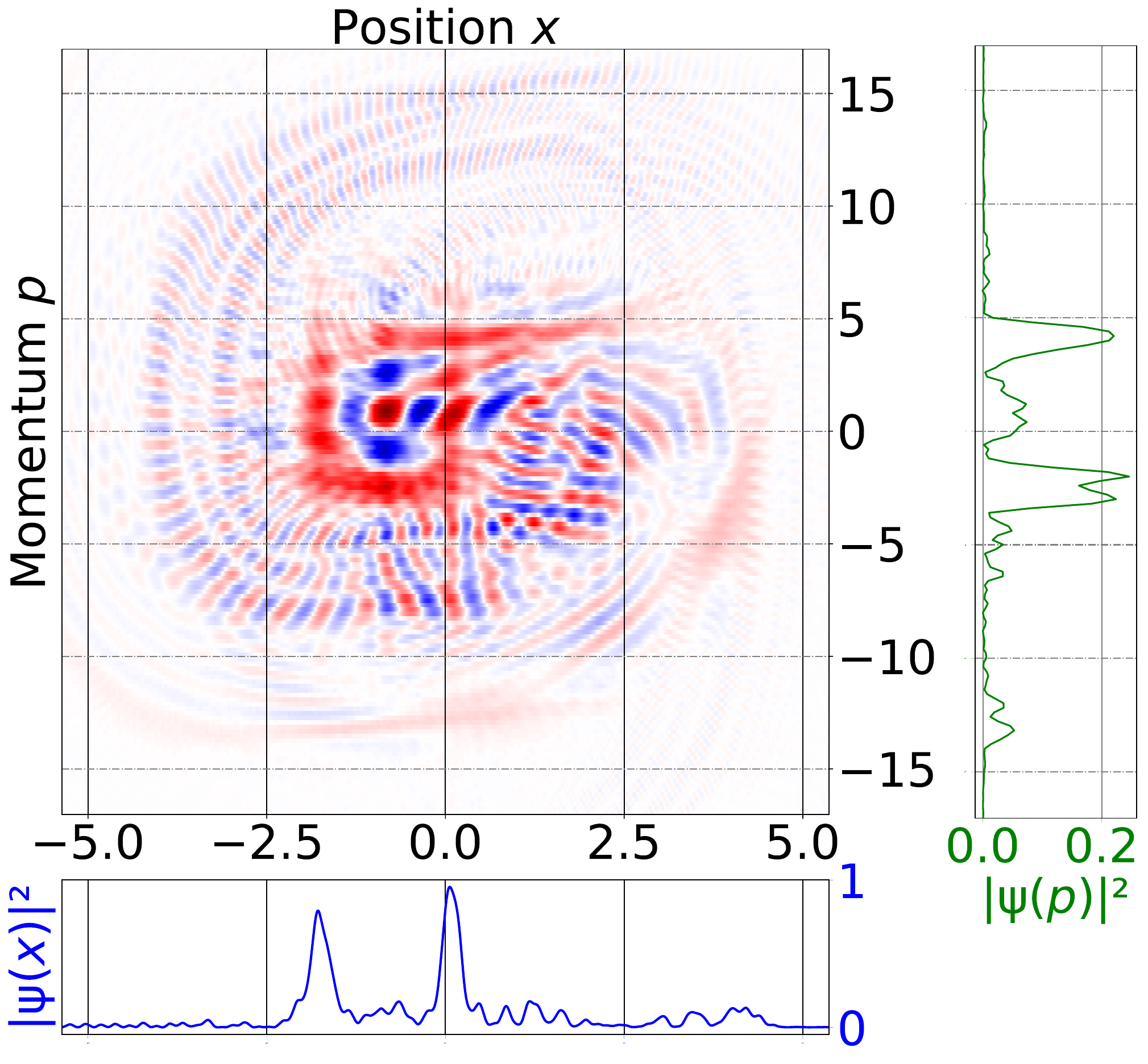}     
     \includegraphics[height=0.216\columnwidth]{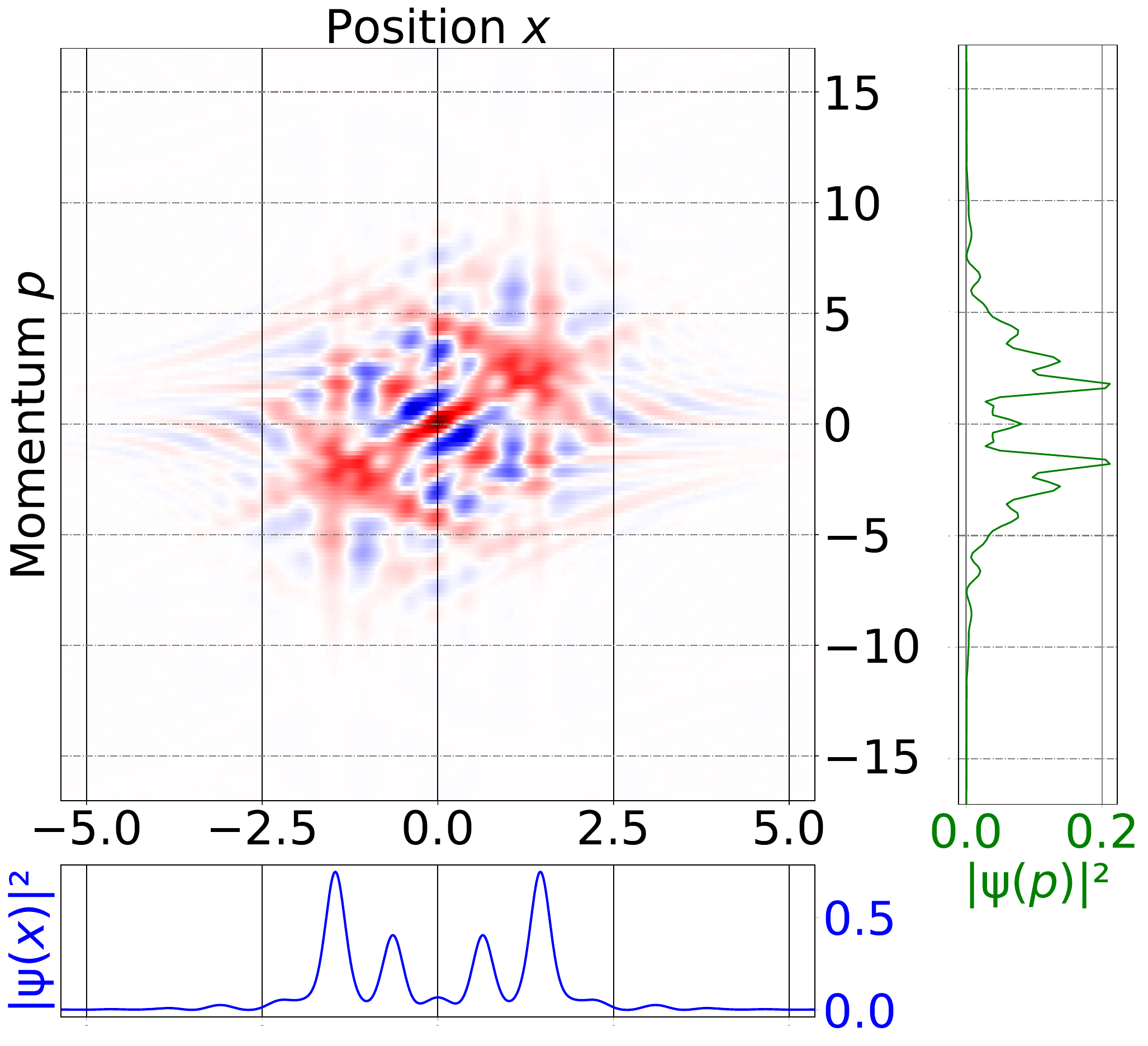}
     \includegraphics[height=0.216\columnwidth]{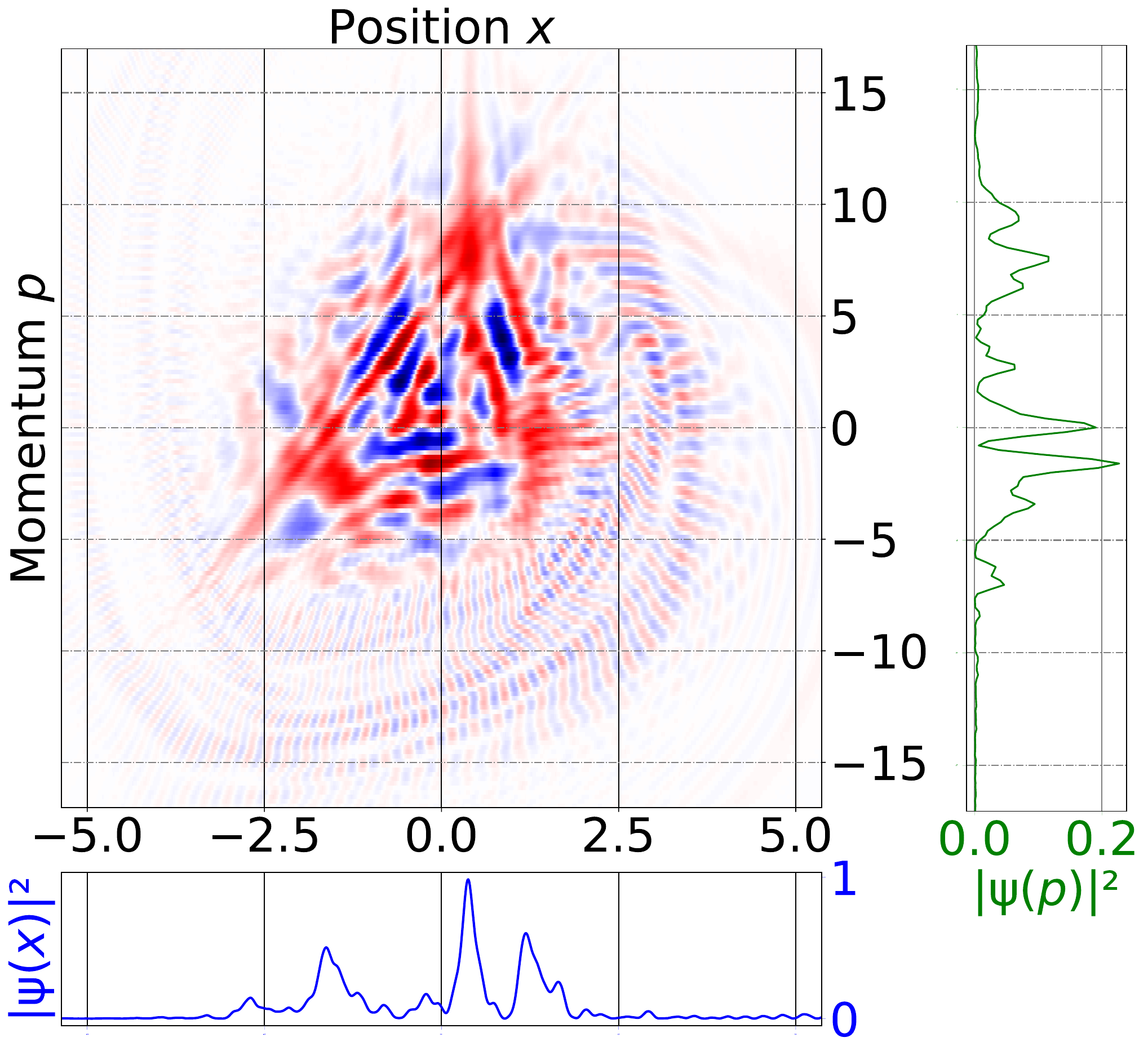}
     \put(-465,93){\rotatebox{0}{\emphLabel{A}}}
     \put(-348,93){\rotatebox{0}{\emphLabel{B}}}
     \put(-230,93){\rotatebox{0}{\emphLabel{C}}}
     \put(-110,93){\rotatebox{0}{\emphLabel{D}}}
     \caption{\emphCaption{Wigner distributions for NLSE systems~(\protect{\ref{eq:_NLSE}}) with
         varied order $\epsilon+1$, nonlinearity $\gamma$ and potential $V$ show formation of lines
         in \ps:} From an initial squeezed state of the form $\psi_0(x) = 0.43 \exp[-(x-x_0)^2/19]$,
       we show the time evolved state for \emphLabel{A} $\epsilon = 0.5, \gamma=40, V=0, x_0=0, t=1.5$,
       \emphLabel{B} $\epsilon = 0.5, \gamma=40, V=x^4/10, x_0=2, t=2.7$, \emphLabel{C}
       $\epsilon = 1, \gamma=40, V=0, x_0=0, t=1.38$, and \emphLabel{D}
       $\epsilon = 3, \gamma=40, V=x^4/10, x_0=2, t=2.0$.
       \label{fig:changeOrderEpsilon}}
 \end{minipage}
 \end{figure}
%\end{widetext}
%\vspace{\columnsep}
% \newpage
 
\section{REPULSIVE NLSE \label{subsec:Appendix_RepulsiveNLSE}}
 
\begin{figure}[h]
  \hspace{-0.2cm}
  \begin{minipage}[t]{0.98\columnwidth}
%%  "b" to have captions on the same line
    \includegraphics[height=0.216\columnwidth]{Figures/F_02__P__Exp2.0__NonLin0.0__Nx4096__Np512__X16__T6.pdf}
    \includegraphics[height=0.216\columnwidth]{Figures/F_02__00055__Wdist____NonLin0.0__eps2.0__V_0.1x4__Nx4096__X16__T0.5500.pdf}
    \includegraphics[height=0.216\columnwidth]{Figures/F_02__00176__Wdist____NonLin0.0__eps2.0__V_0.1x4__Nx4096__X16__T1.7600.pdf}
    \includegraphics[height=0.216\columnwidth]{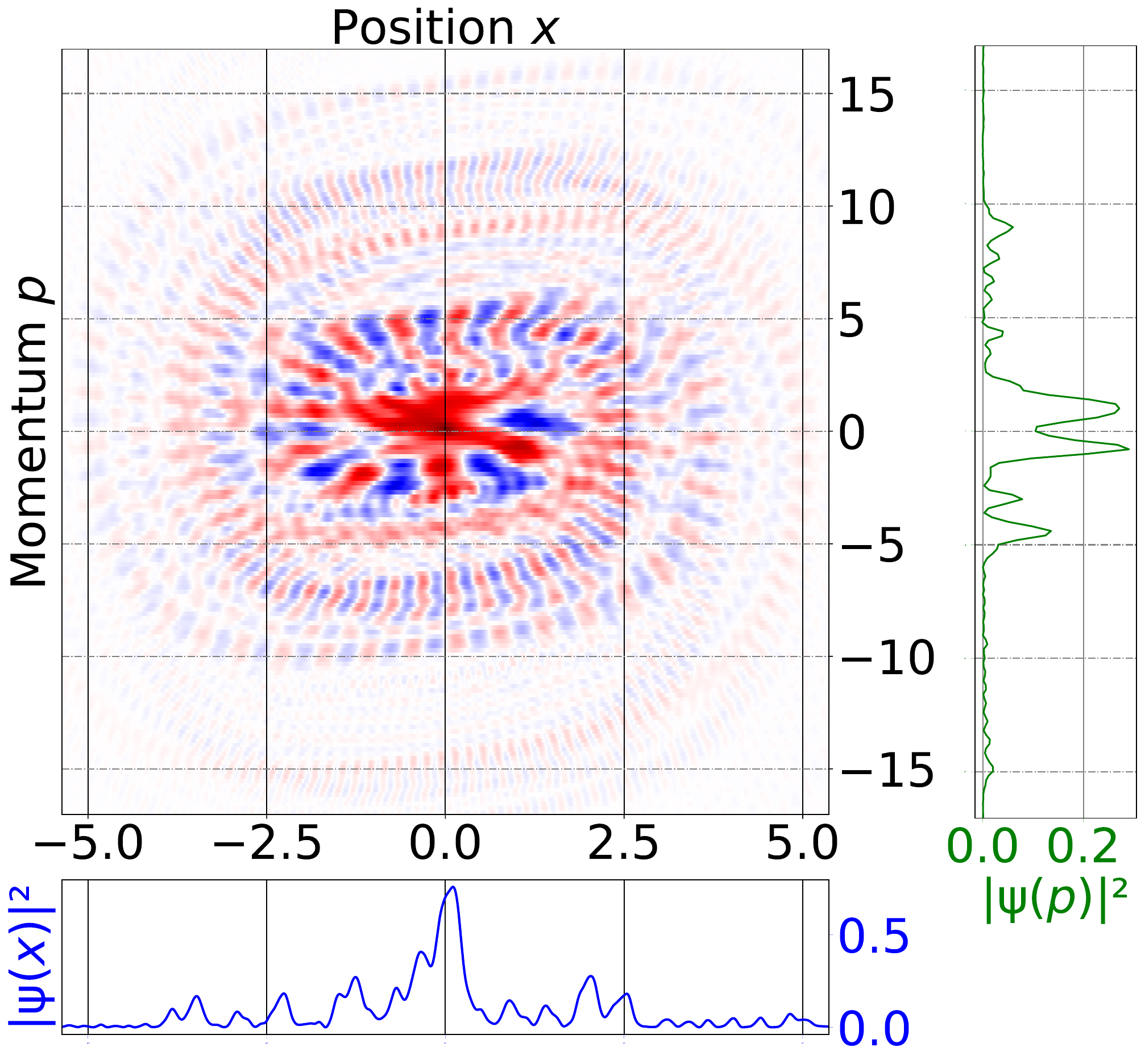}
    \put(-480,90){\rotatebox{0}{\textcolor{white}{\emphLabel{A}}}}
    \put(-477,34){\rotatebox{90}{\textcolor{white}{\boldmath{$|\psi(x,t)|^2$}}}}
    \put(-347,90){\rotatebox{0}{\emphLabel{B}}}
    \put(-229,90){\rotatebox{0}{\emphLabel{C}}}
    \put(-109,90){\rotatebox{0}{\emphLabel{D}}}
    \\
    \includegraphics[height=0.214\columnwidth]{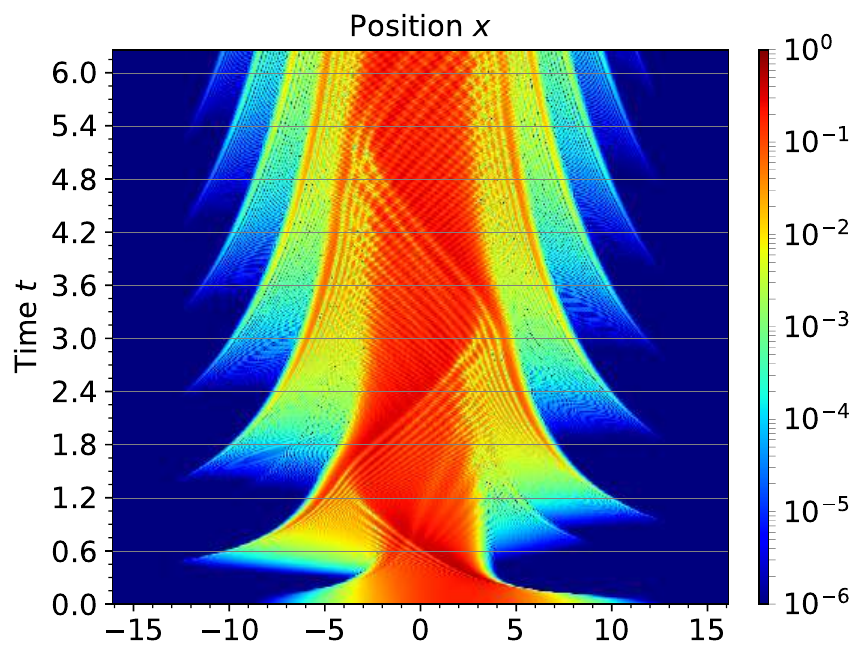}
    \includegraphics[height=0.214\columnwidth]{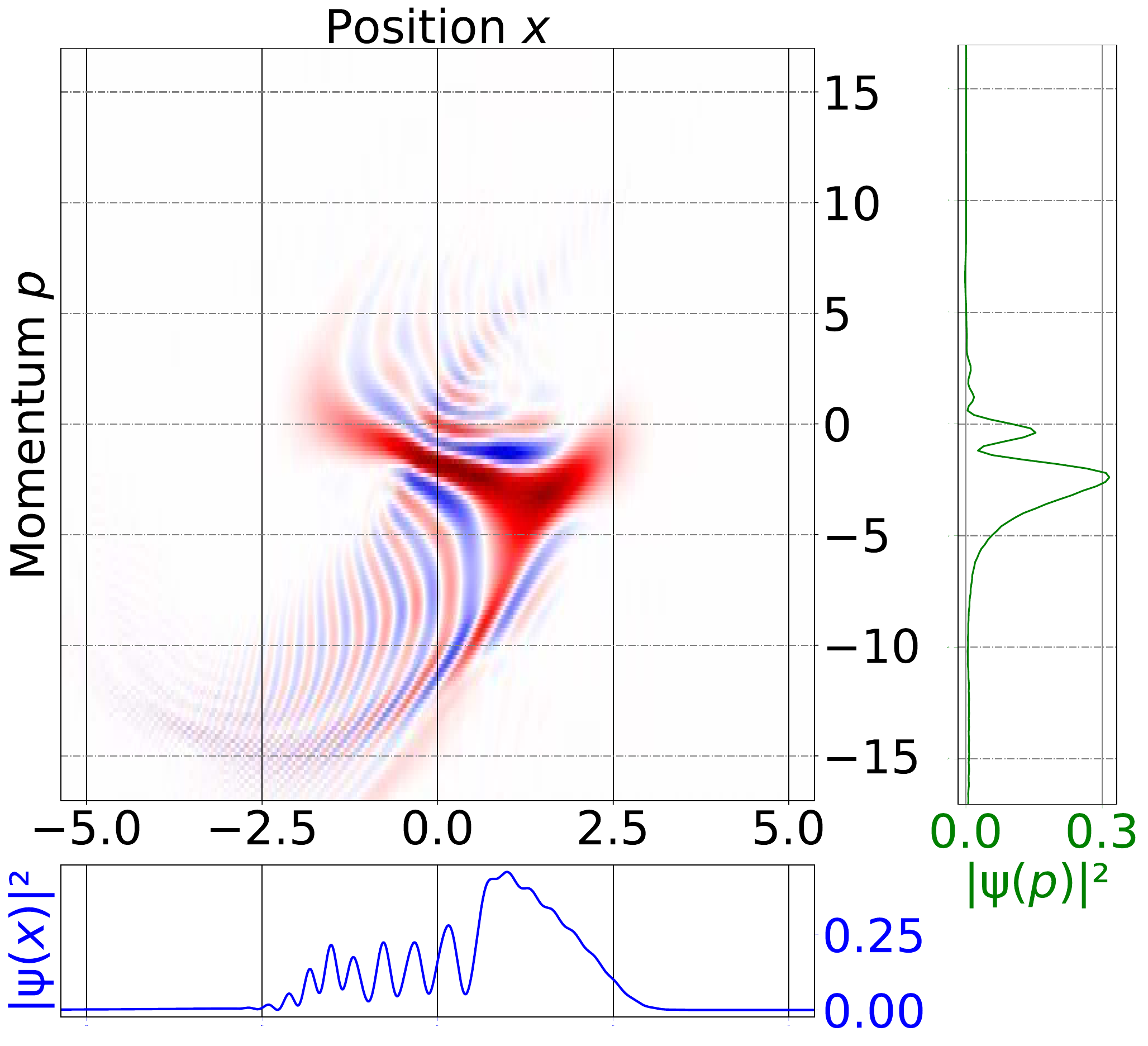}
    \includegraphics[height=0.214\columnwidth]{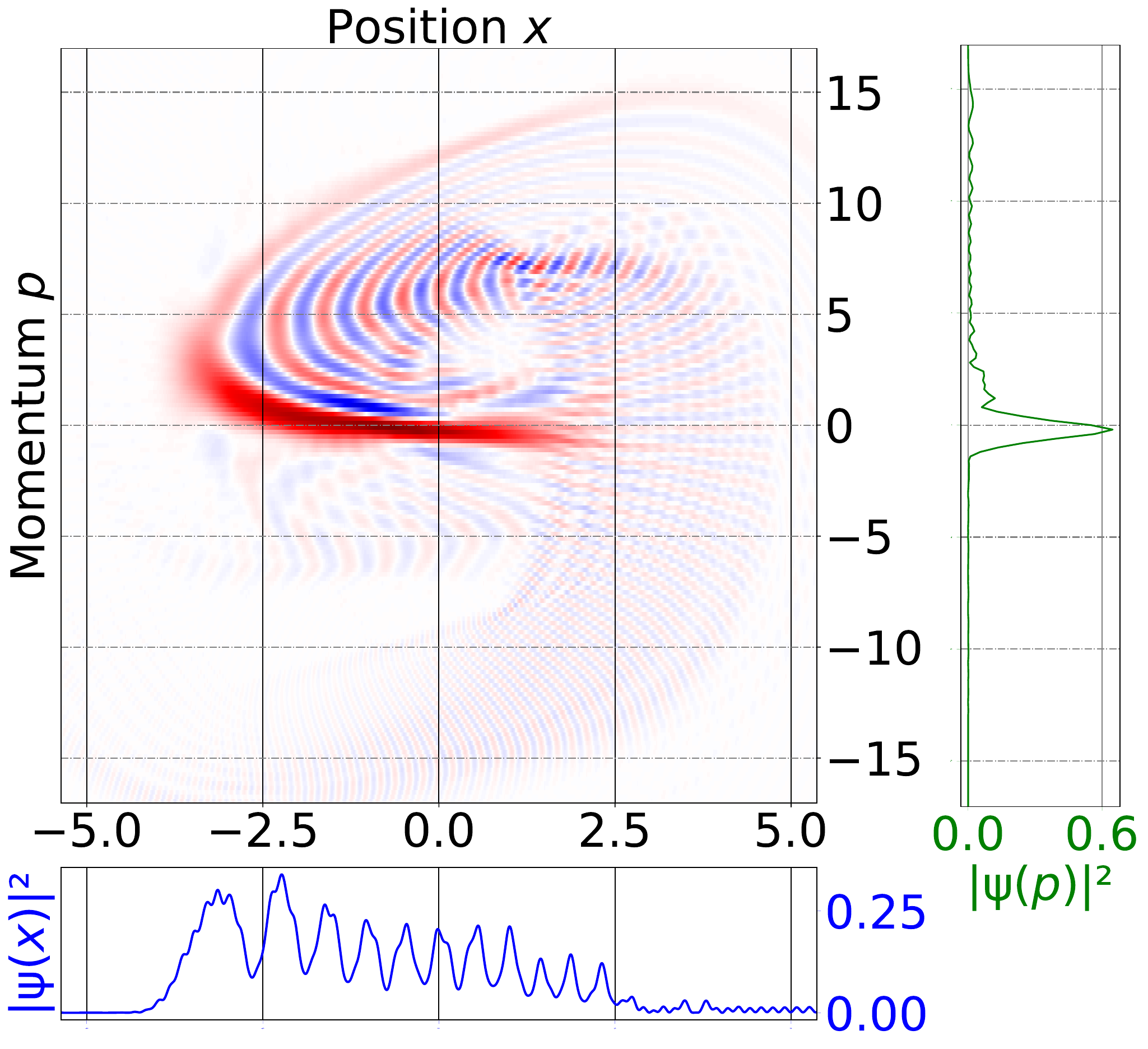}
    \includegraphics[height=0.214\columnwidth]{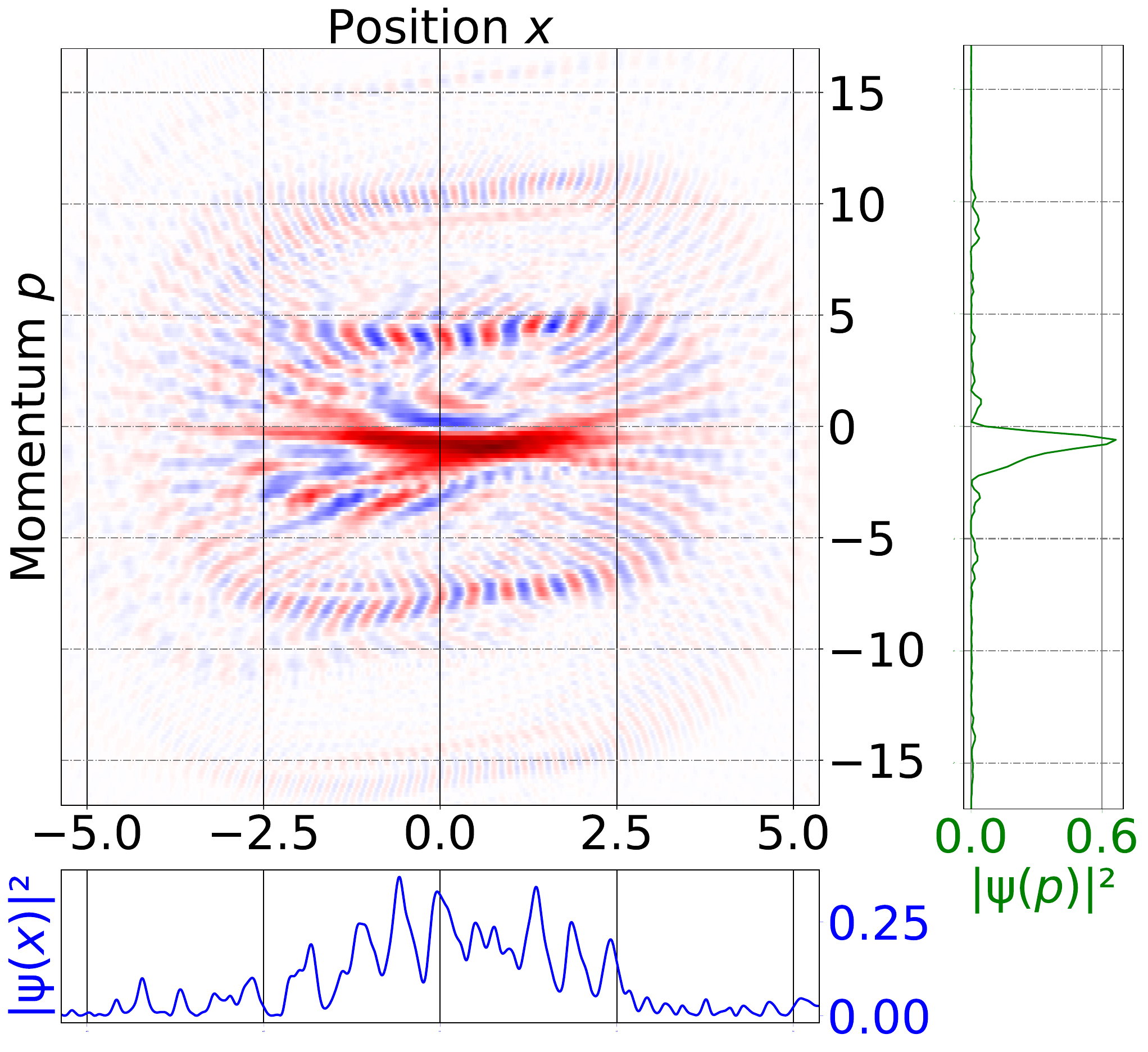}
    \put(-480,90){\rotatebox{0}{\textcolor{white}{\emphLabel{\boldmath{$\alpha$}}}}}
    \put(-478,34){\rotatebox{90}{\textcolor{white}{\boldmath{$|\psi(x,t)|^2$}}}}
    \put(-347,90){\rotatebox{0}{\emphLabel{\boldmath{$\beta$}}}}
    \put(-229,90){\rotatebox{0}{\emphLabel{\boldmath{$\gamma$}}}}
    \put(-109,90){\rotatebox{0}{\emphLabel{\boldmath{$\delta$}}}}
    \caption{ \emphCaption{Suppressed line formation in \ps for repulsive NLSE:} Top row: panels for
      linear \schr equation, ($V(x) = \frac{1}{10} x^4$, $\psi_0(x) = 0.43 \exp[-(x-2)^2/19]$,
      $\gamma =0$), copied over from Fig.~\ref{fig:WithPotential}. Bottom row with same parameters
      except for repulsive interaction~($\gamma=-50$ and $\epsilon=2$).  Evolved Wigner distribution
      at various times: \emphLabel{B} and {\boldmath{$\beta$}} at $t=0.55$, \emphLabel{C} and
      {\boldmath{$\gamma$}} at $t=1.76$, and \emphLabel{D} and {\boldmath{$\delta$}} at $t=4.61$.
      Attractive NLSEs ($\gamma > 0$) tend to display enhanced line formation, repulsive NLSEs
      ($\gamma < 0$) tend to display suppressed line formation .
        \label{fig:RepulsiveNLSE}}
\end{minipage}
\end{figure}

\clearpage

\section{Eyes of varying orders\label{sec:Appendix_Eyes}}
\begin{figure*}[h]
  \hspace{-0.2cm}
  \begin{minipage}[b]{0.89\columnwidth}
    %% "b" to have captions on the same line
    \includegraphics[width=\columnwidth]{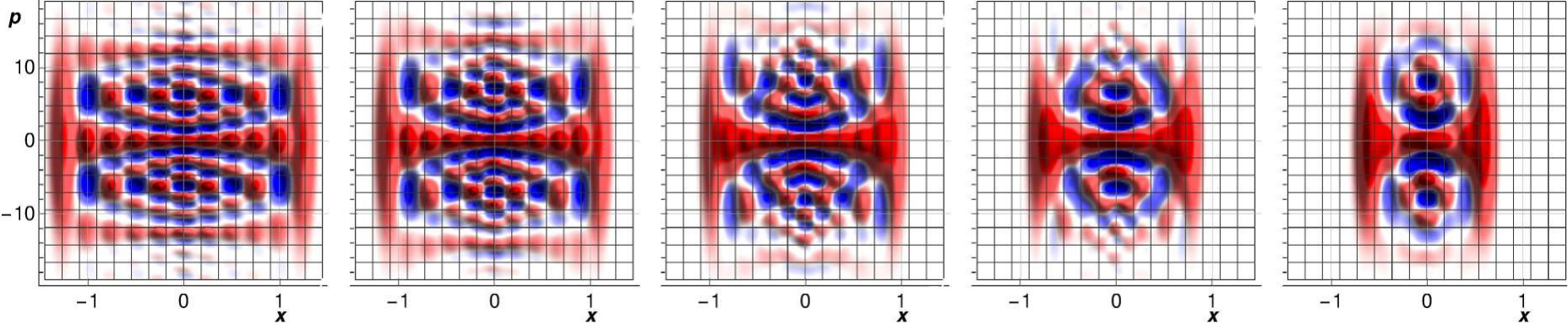}
    \\
    \includegraphics[width=\columnwidth]{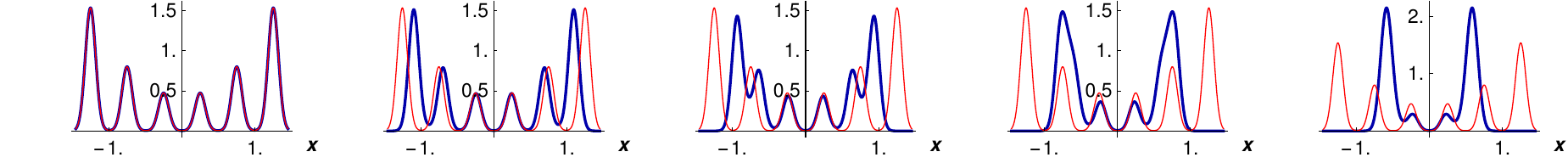}
  \end{minipage}
  \caption{\emphCaption{`Double eye' pattern in \ps: } \gss with a concave arrangement of an even
    number of peaks [see $P(x)$ in bottom row] yield a characteristic double eye interference pattern (with a negative [blue]
    center) with varying order of the number of concentric rings within each eye [see $W(x,p)$ in top row].
    Inter-peak phase differences are zero whereas distances are not constant. 
    \label{sfig:6eye}}
\end{figure*}

\begin{figure*}[h]
  \hspace{-0.2cm}
  \begin{minipage}[b]{0.89\columnwidth}
    %% "b" to have captions on the same line
    \includegraphics[width=\columnwidth]{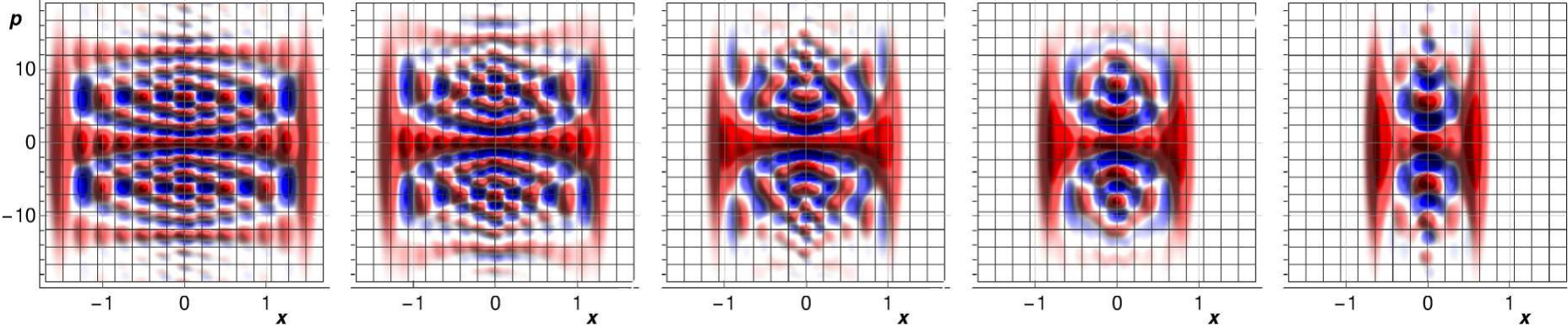}
    \\
    \includegraphics[width=\columnwidth]{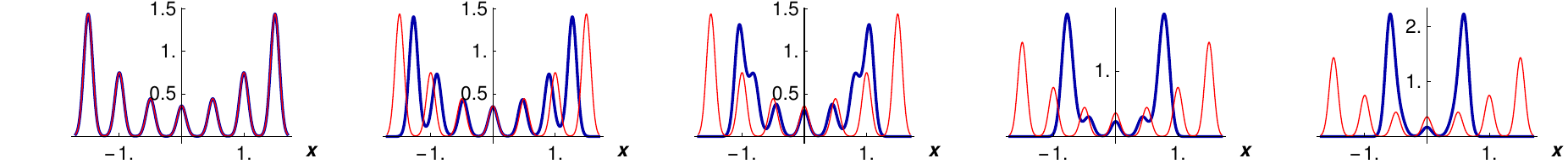}
  \end{minipage}
  \caption{\emphCaption{`Double eye' pattern in \ps: }Similar to Fig.~\ref{sfig:6eye},
    but for an odd number of peaks, yielding positive [red] centers in $W(x,p)$.
    Inter-peak phase differences are zero whereas distances are not constant.
    \label{sfig:7eye}}
\end{figure*}

\begin{figure*}[h]
  \hspace{-0.2cm}
  \begin{minipage}[b]{0.89\columnwidth}
    %% "b" to have captions on the same line
    \includegraphics[width=\columnwidth]{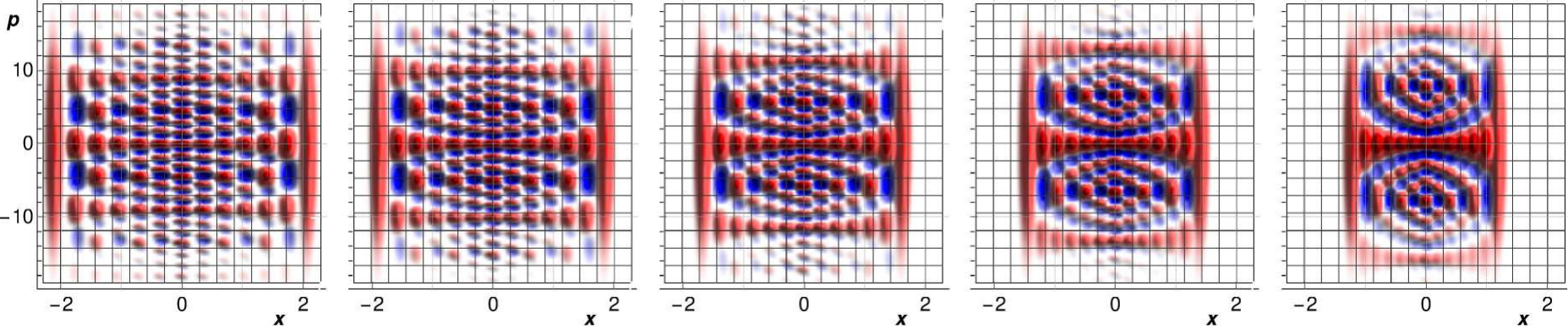}
    \\
    \includegraphics[width=\columnwidth]{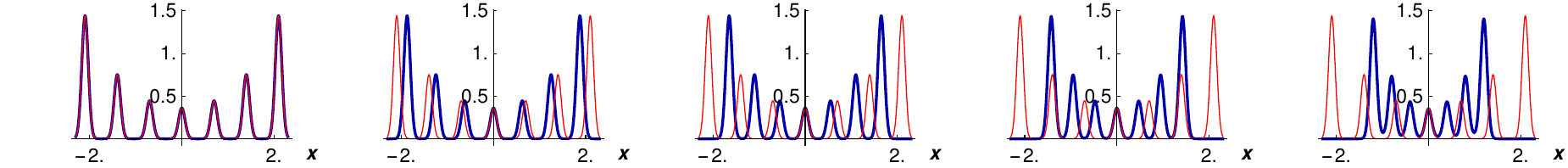}
  \end{minipage}
  \caption{\emphCaption{`Eye' pattern in \ps: }Similar to Fig.~\ref{sfig:7eye},
    but for peaks which are equidistant to each other. 
    \label{sfig:7eyeConcave}}
\end{figure*}

\begin{figure*}[h]
  \hspace{-0.2cm}
  \begin{minipage}[b]{0.89\columnwidth}
    %% "b" to have captions on the same line
    \includegraphics[width=\columnwidth]{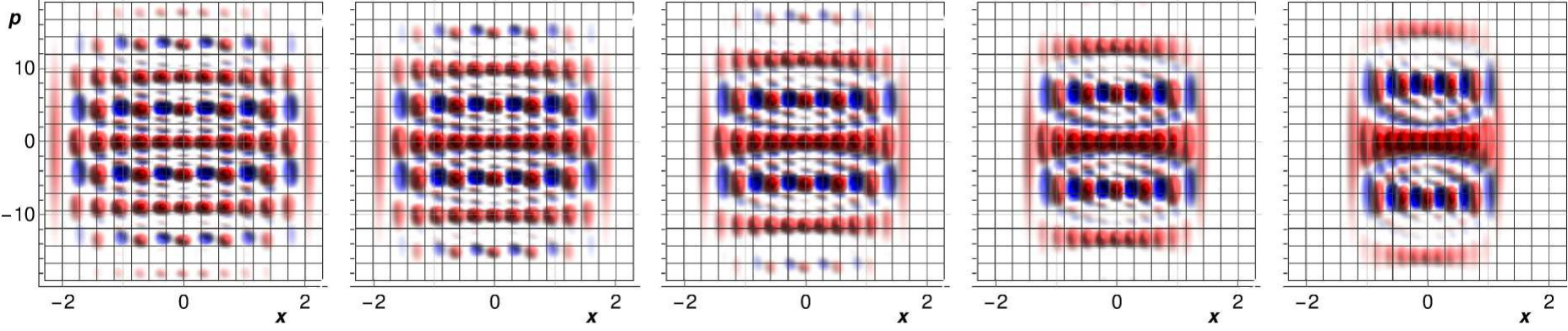}
    \\
    \includegraphics[width=\columnwidth]{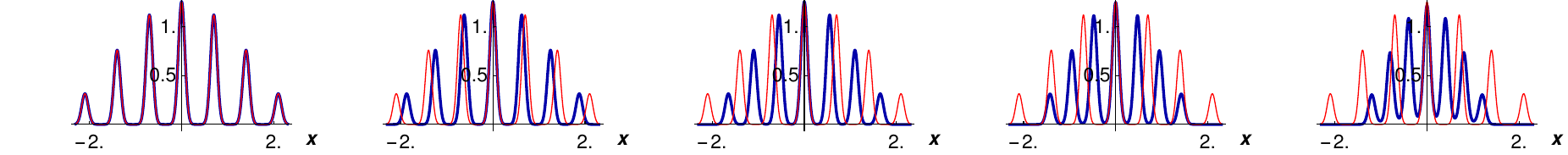}
  \end{minipage}
  \caption{\emphCaption{`Eye' pattern in \ps: }Similar to Fig.~\ref{sfig:7eyeConcave},
    but for peaks with a convex weighting distribution. In this convex case the formation of eye
    patterns is less `clean' than in the concave case of e.g. Fig.~\ref{sfig:7eyeConcave}.
    \label{sfig:7eyeConvex}}
\end{figure*}

\newpage
\section{Randomized momenta: single eyes and triangle lines\label{sec:Appendix_SingleEye}}
\begin{figure*}[h]
  \hspace{-0.2cm}
  \begin{minipage}[b]{0.98\columnwidth}
    %% "b" to have captions on the same line
    \includegraphics[width=\columnwidth]{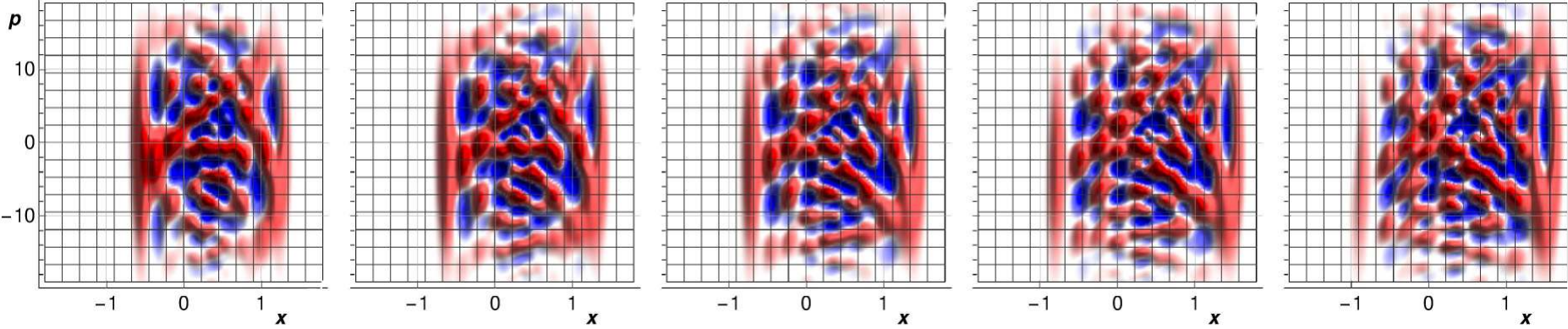}
    \\ \includegraphics[width=\columnwidth]{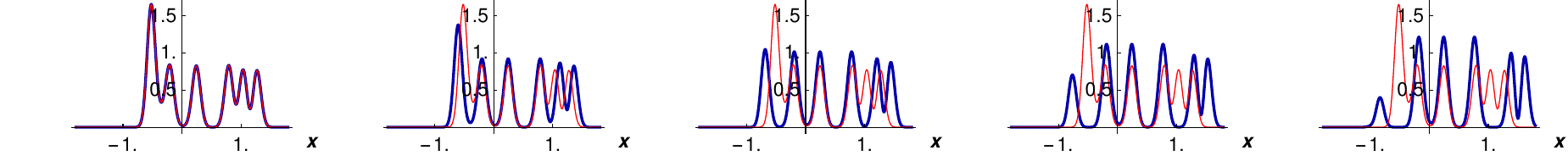}
    \\
\vspace{-2cm}    \includegraphics[width=\columnwidth]{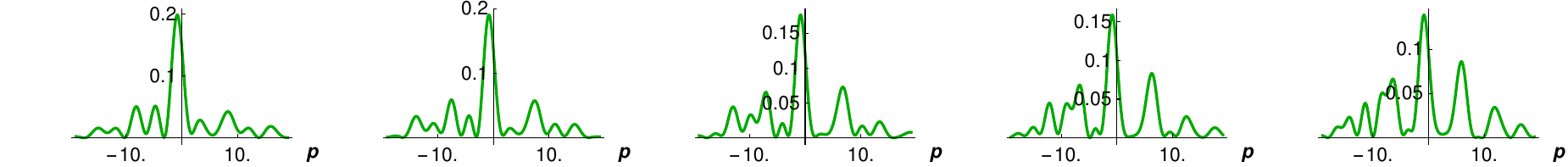}
    \put(-488,120){\rotatebox{0}{\emphLabel{A}}}
    \put(-387,120){\rotatebox{0}{\emphLabel{B}}}
    \put(-288,120){\rotatebox{0}{\emphLabel{C}}}
    \put(-189,120){\rotatebox{0}{\emphLabel{D}}}
    \put(-090,120){\rotatebox{0}{\emphLabel{E}}}
    \put(-488,35){\rotatebox{0}{\emphLabel{a}}}
    \put(-387,35){\rotatebox{0}{\emphLabel{b}}}
    \put(-288,35){\rotatebox{0}{\emphLabel{c}}}
    \put(-189,35){\rotatebox{0}{\emphLabel{d}}}
    \put(-090,35){\rotatebox{0}{\emphLabel{e}}}
  \end{minipage}
  \caption{\emphCaption{`Single eye' pattern in \ps: } Here, \gss for which not only the peak
    positions [see $P(x)$ in \emphLabel{A}--\emphLabel{E}] but also the mean momenta of the peaks
    are randomized [see $\tilde P(p)$ in \emphLabel{a}--\emphLabel{e} is not an even function]. This
    demonstrates that also single eye patterns can form [see $W(x,p)$ in \emphLabel{A} and
    \emphLabel{B}]. Triangle line arrangements, e.g. in panel \emphLabel{E}, resemble those in
    Fig.~\ref{fig:WithPotential}~\emphLabel{(d)} and
    Fig.~\ref{fig:changeOrderEpsilon}~\emphLabel{(D)}.
    \label{sfig:1eye}}
\end{figure*}
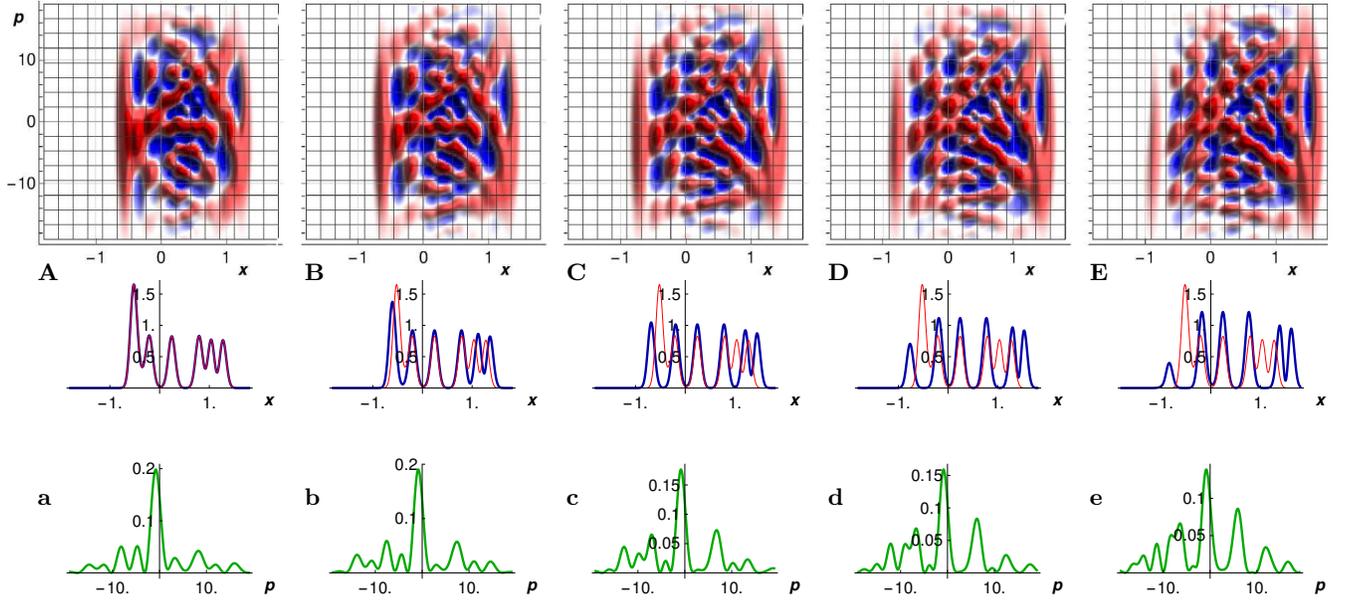

\clearpage

\section{Randomized phases\label{sec:Appendix_phiRand}}

\begin{figure*}[h]
  \hspace{-0.2cm}
  \begin{minipage}[b]{0.98\columnwidth}
    %% "b" to have captions on the same line
 \includegraphics[width=\columnwidth]{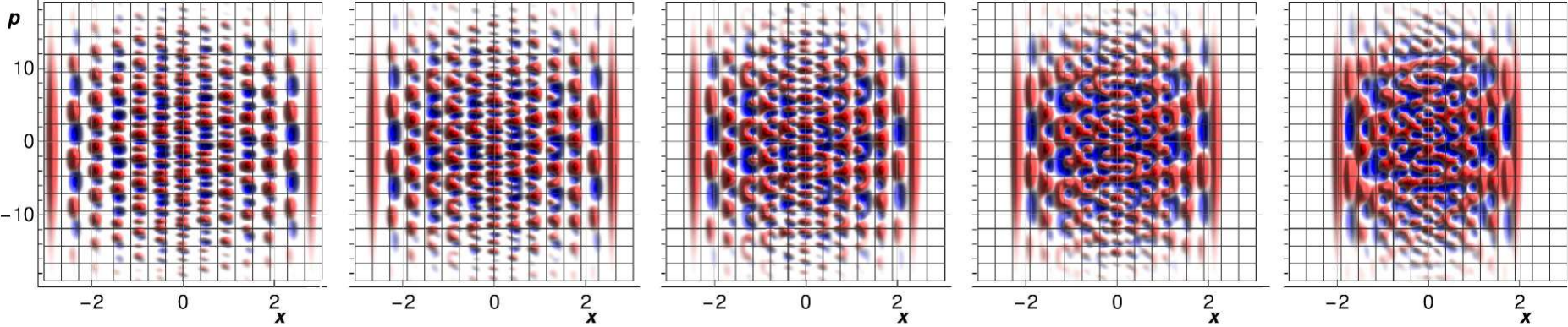}
 \\
\vspace{0.15cm} \includegraphics[width=\columnwidth]{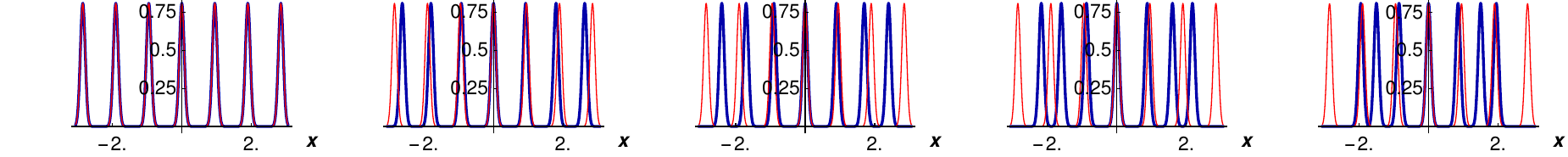}
    \\
\vspace{-2cm}    \includegraphics[width=\columnwidth]{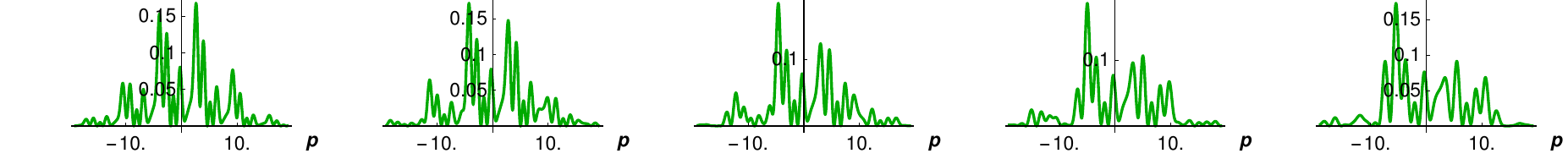}
    \put(-488,110){\rotatebox{0}{\emphLabel{A}}}
    \put(-387,110){\rotatebox{0}{\emphLabel{B}}}
    \put(-288,110){\rotatebox{0}{\emphLabel{C}}}
    \put(-189,110){\rotatebox{0}{\emphLabel{D}}}
    \put(-090,110){\rotatebox{0}{\emphLabel{E}}}
    \put(-488,35){\rotatebox{0}{\emphLabel{a}}}
    \put(-387,35){\rotatebox{0}{\emphLabel{b}}}
    \put(-288,35){\rotatebox{0}{\emphLabel{c}}}
    \put(-189,35){\rotatebox{0}{\emphLabel{d}}}
    \put(-090,35){\rotatebox{0}{\emphLabel{e}}}
  \end{minipage}
  \caption{\emphCaption{Comb-states with randomized phases form lines in \ps:} $P(x)$ in
    \emphLabel{A}--\emphLabel{E} has the same shape as in Fig.~\ref{fig:Qx_grid_Quadratic7} but
    since every peak carries a completely random phase the momentum distributions $\tilde P(p)$ in
    \emphLabel{a}--\emphLabel{e} are not even functions any more, yet, $W(x,p)$ forms lines in \ps.
    \label{fig:GridState_RandomPhases}}
\end{figure*}
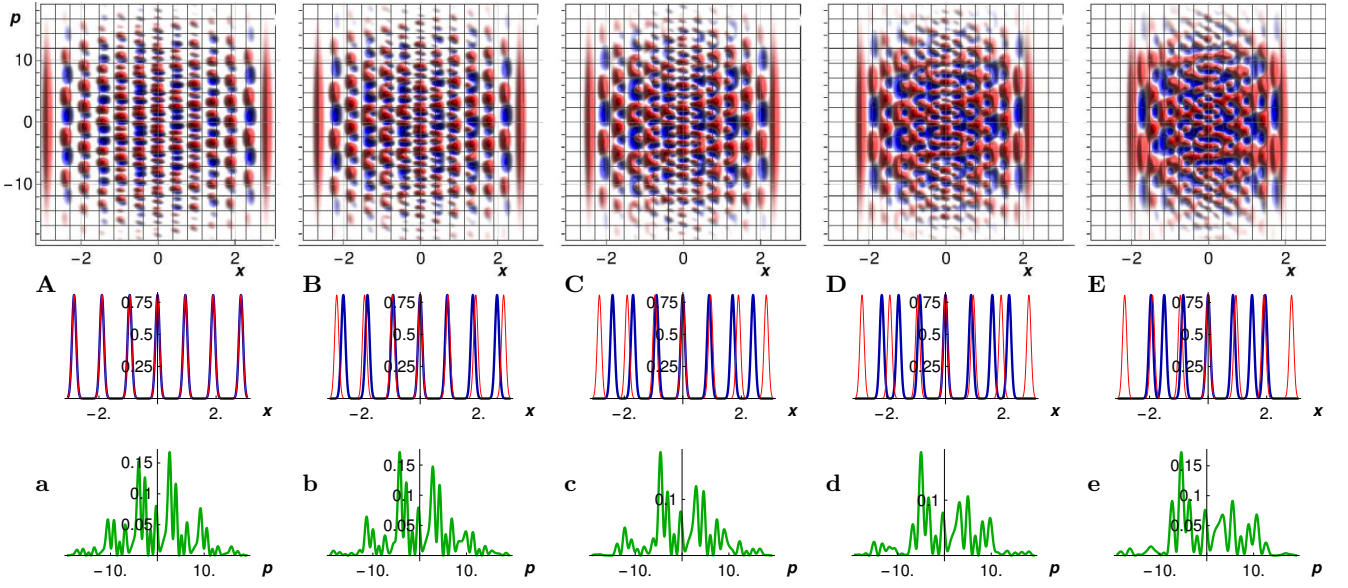

% \vspace{\columnsep}
% \twocolumngrid

% \newpage
% \input{Response_Report.tex}

\end{document}